\documentclass[12pt]{article}
\pdfoutput=1
\usepackage{jheppub}
\usepackage{ytableau}
\usepackage{comment}
\usepackage[vcentermath]{youngtab}
\usepackage{tikz}

\newcommand\be{\begin{equation}}
\newcommand\ee{\end{equation}}

\preprint{
RUP-23-17
}

\title{Large $N$ and large representations of Schur line defect correlators
}
\abstract{
We study the large $N$ and large representation limits of the Schur line defect correlators 
of the Wilson line operators transforming in the (anti)symmetric, hook and rectangular representations for $\mathcal{N}=4$ $U(N)$ super Yang-Mills theory. 
By means of the factorization property, the large $N$ correlators of the Wilson line operators in arbitrary representations can be exactly calculated in principle. 
In the large representation limit they turn out to be expressible in terms of certain infinite series such as Ramanujan's general theta functions 
and the $q$-analogues of multiple zeta values ($q$-MZVs). 
Several generating functions for combinatorial objects, including 
partitions with non-negative cranks and conjugacy classes of general linear groups over finite fields, 
emerge from the large $N$ correlators.
Also we find conjectured properties of the automorphy and the hook-length expansion satisfied by the large $N$ correlators. 
}
\author[a]{Yasuyuki Hatsuda}
\author[b]{and Tadashi Okazaki}

\emailAdd{yhatsuda@rikkyo.ac.jp, tokazaki@seu.edu.cn}
\affiliation[a]{Department of Physics, Rikkyo University, Toshima, Tokyo 171-8501, Japan}
\affiliation[b]{
Shing-Tung Yau Center of Southeast University,\\
Yifu Architecture Building, No.2 Sipailou, Xuanwu district, Nanjing, Jiangsu, 210096, China}
\begin{document}
%%%%%%%%%%%%%%%%%%%%%%%%%%%%%%%%%%%%%%%%%%%%
%%%%%%%%%%%%%%%%%%%%%%%%%%%%%%%%%%%%%%%%%%%%
\maketitle

%%%%%%%%%%%%%%%%%%%%%%%%%%%%%%%%%%%%%%%%%%%%%%%%%
%%%%%%%%%%%%%%%%%%%%%%%%%%%%%%%%%%%%%%%%%%%%%%%%%
\section{Introduction and summary}
\label{sec_intro}
%%%%%%%%%%%%%%%%%%%%%%%%%%%%%%%%%%%%%%%%%%%%%%%%%
%%%%%%%%%%%%%%%%%%%%%%%%%%%%%%%%%%%%%%%%%%%%%%%%%
The superconformal index \cite{Romelsberger:2005eg,Kinney:2005ej} of 4d $\mathcal{N}=2$ supersymmetric field theory 
which can be viewed as a supersymmetric partition function on $S^1\times S^{3}$ captures the protected states of the theory. 
It generally depends on three parameters 
coupled to the Cartan generators of the superconformal algebra as well as extra parameters coupled to the global charges. 
It admits the Schur index \cite{Gadde:2011ik,Gadde:2011uv}, that is a specialization of the index which only depends on one of the three parameters for the superconformal generators. 
For a 4d superconformal field theory of class $\mathcal{S}$ \cite{Gaiotto:2009we,Gaiotto:2009hg}, 
one can interpret it as a correlation function of a 2d topological field theory on a Riemann surface \cite{Gadde:2009kb,Gadde:2011ik}. 
It is observed in \cite{Beem:2013sza} that there is a correspondence between the protected subsector of $\mathcal{N}=2$ SCFT and vertex operator algebra (VOA) 
so that the unflavored Schur index is identified with the vacuum character of the associated VOA. 
In addition, closed-form expressions of the Schur indices are recently obtained 
in terms of the special functions endowed with nice elliptic and modular properties \cite{Beem:2021zvt,Pan:2021mrw,Hatsuda:2022xdv}. 
The Schur index can be decorated by the BPS line operators wrapping the $S^1$ and localized at a point along a great circle in the $S^3$ 
\cite{Dimofte:2011py,Gang:2012yr,Drukker:2015spa,Cordova:2016uwk,Neitzke:2017cxz,Gaiotto:2020vqj,Hatsuda:2023iwi,Guo:2023mkn}. 
It can be understood as a correlation function of the line operators which is independent of each distance of the adjacent line operators along the great circle. 
We call it the \textit{Schur line defect correlation function}. 

In this paper, we study the large $N$ limit of 
the Schur line defect correlation functions of the Wilson line operators transforming in various representations 
labeled by the Young diagrams for $\mathcal{N}=4$ $U(N)$ super Yang-Mills (SYM) theory. 
Such Wilson line operators indexed by the Young diagrams are conjectured to be holographically dual to 
the configurations of Type IIB string theory. 
For example, fundamental, symmetric and antisymmetric representations correspond to a fundamental string \cite{Maldacena:1998im,Rey:1998ik}, 
a D3-brane \cite{Drukker:2005kx,Gomis:2006sb,Gomis:2006im,Rodriguez-Gomez:2006fmx,Yamaguchi:2007ps} 
and a D5-brane \cite{Yamaguchi:2006tq,Gomis:2006sb,Rodriguez-Gomez:2006fmx,Hartnoll:2006hr} respectively. 
More general representations are conjectured to be holographically dual to a configuration 
with multiple D3- and D5-branes and bubbling geometries \cite{Gomis:2006sb,Yamaguchi:2006te,Lunin:2006xr,Okuda:2007kh,DHoker:2007mci,Okuda:2008px,Gomis:2008qa,Aguilera-Damia:2017znn}. 
In fact, it is shown \cite{Gang:2012yr,Hatsuda:2023iwi} that the large $N$ Schur line 2-point functions of the Wilson line operators in the fundamental representation 
and that of the Wilson line operators in the large (anti)symmetric representation precisely encodes the gravity spectra \cite{Faraggi:2011bb,Faraggi:2011ge},  
which fall into the representations of the 1d $\mathcal{N}=8$ superconformal group $OSp(4^*|4)$. 

Exact expressions for the large $N$ correlators of the Wilson line operators in arbitrary representations can be calculated 
by means of the factorization property \cite{Hatsuda:2023iwi} for the large $N$ correlators of the charged Wilson line operators as well as the relations of the symmetric functions. 
By taking the further limit where the number of boxes of the diagram becomes infinitely large, 
we obtain the large representation limit of the Schur line defect correlation functions which is expected to encode the fluctuation modes for the dual bubbling geometries. 
When the fugacities are specialized (i.e. in the unflavored limit or the half-BPS limit), 
several large $N$ and large representation correlators turn out to be expressible in terms of certain infinite series such as Ramanujan's general theta functions \cite{MR1117903} 
and the $q$-analogues of multiple zeta values ($q$-MZVs) \cite{schlesinger2001some,MR2069738,MR2111222,MR1992130,MR2341851,MR2322731,MR2843304,MR3141529,okounkov2014hilbert,MR3338962,MR3473421,MR3522085,milas2022generalized}. 

In addition, we find that the large $N$ 2-point functions of the Wilson line operators in the large hook representation 
and those in the large rectangular representation are given by the generating function \cite{MR3803977,MR4072556} for the partitions with non-negative cranks \cite{MR3077150,MR929094} and the generating functions \cite{MR109810,MR615131} for the conjugacy classes of general linear groups over finite fields respectively. 
According to the closed-form expressions, we evaluate the asymptotic growth of the states by applying the convolution theorem \cite{MR3043606}. 
It follows that the asymptotic degeneracy of the bubbling geometry with genus one surface dual to the large rectangular representation grows  
much faster than that for any $p$-branes \cite{Fubini:1972mf, Dethlefsen:1974dr,Strumia:1975rd,Alvarez:1991qs,Harms:1992jt}. 

Also we find several conjectured properties satisfied by the large $N$ Schur line defect correlation functions. 
Under the transformation $q\rightarrow q^{-1}$ (resp. $\mathfrak{q}\rightarrow \mathfrak{q}^{-1}$) and the conjugation of the Young diagrams associated with the Wilson line operators, 
the large $N$ normalized 2-point functions (resp. their half-BPS limit) are invariant up to the overall factor. 
In particular, the correlation function of the Wilson line operators in the representations labeled by a set of self-conjugate partitions enjoys the automorphy. 
Moreover, the large $N$ normalized correlator can be factorized to a certain polynomial with positive integer coefficients and hook-length denominator. 

%%%%%%%%%%%%%%%%%%%%%%%%%%%%%%%%%%%%
\subsection{Future works}
%%%%%%%%%%%%%%%%%%%%%%%%%%%%%%%%%%%%
Here we list several open problems for future research. 
\begin{itemize}

\item 
While we focus on the Wilson line operators in the hook and rectangular representations in this work, 
in the upcoming work \cite{HO23bubbling}, we plan to report the study of the excitations of the dual bubbling geometries containing higher genus surfaces by computing the large $N$ and large representation Schur line defect correlators of the Wilson line operators labeled by more general Young diagram with multiple parts. 

\item 
The determination of fully explicit formulas for 
general flavored large $N$ and large representation Schur line defect correlators is not all obvious whereas we find some of them. 

\item 
The nested sum expressions for the Schur line defect correlators \cite{Hatsuda:2023iwi} as well as for the Schur indices \cite{Hatsuda:2022xdv} indicate the close relation to 
the $q$-MZVs \cite{schlesinger2001some,MR2069738,MR2111222,MR1992130,MR2341851,MR2322731,MR2843304,MR3141529,okounkov2014hilbert,MR3338962,MR3473421,MR3522085,milas2022generalized}. While we find explicit relations between the large $N$ and large representation correlators of the Wilson line operators in the rectangular representation and the $q$-MZVs, the relation still deserve to be studied. 

\item 
It would be nice to give physical explanations or/and mathematical proofs of the conjectured properties satisfied by the large $N$ Schur line defect correlators. 
Also it would be interesting to explore further properties, e.g. elliptic and modular properties. 

\item We hope to report the detailed finite $N$ and finite representation corrections and the giant graviton expansions of the Schur line defect correlators, 
as discussed for various supersymmetric indices 
in e.g. \cite{Arai:2020qaj,Gaiotto:2021xce,Imamura:2021ytr,Murthy:2022ien,Lee:2022vig,Imamura:2022aua,Eniceicu:2023uvd,Beccaria:2023zjw}. 

\item 
The large $N$ and large representation correlators 
for other gauge theories, e.g. $\mathcal{N}=4$ SYM theory with different gauge groups and $\mathcal{N}=2$ SCFTs, will be also calculable 
by employing the similar factorization property.  

\end{itemize}

%%%%%%%%%%%%%%%%%%%%%%%%%%%%%%%%%%%%
\subsection{Structure}
%%%%%%%%%%%%%%%%%%%%%%%%%%%%%%%%%%%%
The organization of the paper is as follows. 
In section \ref{sec_largeNcorr} we review basic features of 
the large $N$ limit of the Schur line defect correlators of the Wilson line operators for $\mathcal{N}=4$ $U(N)$ SYM theory. 
It contains the factorization of the large $N$ correlators of the charged Wilson line operators and the holographic dual description of the Wilson line operators.   
In section \ref{sec_asymrep} we study the large $N$ correlators of the Wilson line operators in the (anti)symmetric representations. 
This generalizes the analysis in \cite{Gang:2012yr,Drukker:2015spa,Hatsuda:2023iwi} by examining more general correlators and their properties. 
In section \ref{sec_hook} we investigate the large $N$ correlators of the Wilson line operators in the hook representations. 
The generating function \cite{MR109810,MR615131} for the partitions with non-negative crank shows up in the large representation limit of the 2-point function. 
In section \ref{sec_rect} the large $N$ correlators of the Wilson line operators in the rectangular representations are examined. 
We find that in the large representation limit the 2-point functions
agree with the generating functions \cite{MR109810,MR615131} for the conjugacy classes of general linear groups over finite fields. 
In section \ref{sec_conj} we discuss several conjectured properties of the large $N$ Schur line defect correlators. 

%%%%%%%%%%%%%%%%%%%%%%%%%%%%%%%%%%%%
%%%%%%%%%%%%%%%%%%%%%%%%%%%%%%%%%%%%
\section{Large $N$ Schur line defect correlators}
\label{sec_largeNcorr}
%%%%%%%%%%%%%%%%%%%%%%%%%%%%%%%%%%%%
%%%%%%%%%%%%%%%%%%%%%%%%%%%%%%%%%%%%

%%%%%%%%%%%%%%%%%%%%%%%%%%%%%%%%%%%%
\subsection{Schur line defect correlators}
%%%%%%%%%%%%%%%%%%%%%%%%%%%%%%%%%%%%
The Schur line defect correlation function of the Wilson line operators transforming in the representations $\mathcal{R}_j$, 
$j=1,\cdots, k$ for $\mathcal{N}=4$ $U(N)$ SYM theory is given by \cite{Gang:2012yr}
\begin{align}
\label{sch_corr1}
&
\langle W_{\mathcal{R}_1}\cdots  W_{\mathcal{R}_k} \rangle^{U(N)}(t;q)
\nonumber\\
&=
\frac{1}{N!} 
\frac{(q;q)_{\infty}^{2N}}{(q^{\frac12}t^{\pm2};q)_{\infty}^N}
\oint 
\prod_{i=1}^N 
\frac{d\sigma_i}{2\pi i\sigma_i}
\prod_{i\neq j}
\frac{
\left(\frac{\sigma_i}{\sigma_j};q\right)_{\infty}
\left(q\frac{\sigma_i}{\sigma_j};q\right)_{\infty}
}
{
\left(q^{\frac12} t^2\frac{\sigma_i}{\sigma_j};q\right)_{\infty}
\left(q^{\frac12} t^{-2}\frac{\sigma_i}{\sigma_j};q\right)_{\infty}
}
\prod_{j=1}^k \chi_{\mathcal{R}_j}(\sigma), 
\end{align}
where $\chi_{\mathcal{R}_j}$ is the character of the representation $\mathcal{R}_j$ of the $U(N)$ gauge group. 
In the absence of the Wilson line operator, it reduces to the Schur index $\mathcal{I}^{U(N)}(t;q)$ realized as a certain supersymmetric partition function on $S^1\times S^3$. 
The Wilson line operator in the representation $\mathcal{R}_j$ wraps $S^1$ and localizes at a point on a great circle in $S^3$ \cite{Cordova:2016uwk}. 
Under the conformal map $S^1\times S^3$ $\rightarrow$ $\mathbb{R}^4$, it turns into a half-line, a ray emanating from the origin in $\mathbb{R}^4$. 
We define the normalized Schur line defect correlator by
\begin{align}
\langle \mathcal{W}_{\mathcal{R}_1}\cdots \mathcal{W}_{\mathcal{R}_k}\rangle^{U(N)}(t;q)
&:=
\frac{\langle W_{\mathcal{R}_1}\cdots  W_{\mathcal{R}_k} \rangle^{U(N)}(t;q)}
{\mathcal{I}^{U(N)}(t;q)}. 
\end{align}
The Schur line defect correlators which decorate the ordinary Schur index 
can be viewed as the ``index'' counting contributions of the BPS local operators due to the insertion of the Wilson line operators. 

The matrix integral (\ref{sch_corr1}) is invariant 
under an exchange of the characters with positive powers of gauge fugacities $\sigma_i$ and those with negative powers of $\sigma_i$. 
For example, the 2-point function satisfies 
\begin{align}
\label{sch_2pt}
\langle \mathcal{W}_{\lambda} \mathcal{W}_{\overline{\mu}}\rangle^{U(N)}(t;q)
&=\langle \mathcal{W}_{\mu} \mathcal{W}_{\overline{\lambda}}\rangle^{U(N)}(t;q),
\end{align}
where $\lambda$ and $\mu$ are the Young diagrams labeling the representations of the Wilson line operators
and the notation $\overline{\mu}$ stands for the representation labeled by the diagram $\mu$ for which the character is a symmetric function of the negative powers of gauge fugacities. 

According to the Gauss law, the correlator (\ref{sch_corr1}) does not vanish  
if the sum of the weights of the partitions associated with the characters with positive powers of gauge fugacities
is equal to that of the weights of the partitions for which the characters have negative powers of gauge fugacities. 
For example, the 2-point function (\ref{sch_2pt}) is non-trivial when $|\lambda|$ $=$ $|\mu|$. 

When the two Young diagrams are equal, $\lambda=\mu$, a pair of two half-lines conformally maps to a straight superconformal line along $\mathbb{R}$ in $\mathbb{R}^4$ 
so that the $q$-series expansion of the 2-point function of the Wilson line operators starts with $1+\cdots$ and contains positive powers of $q$.  

%%%%%%%%%%%%%%%%%%%%%%%%%%%%%%%%%%%%
\subsection{Half-BPS limit}
%%%%%%%%%%%%%%%%%%%%%%%%%%%%%%%%%%%%
The Schur indices of $\mathcal{N}=4$ SYM theories admit the special limit of fugacities 
keeping $\mathfrak{q}:=q^{\frac12}t^2$ being finite and setting $q$ to $0$ \cite{Hatsuda:2022xdv}. 
For $U(N)$ SYM this results in 
\begin{align}
\mathcal{I}^{U(N)}_{\textrm{$\frac12$BPS}} (\mathfrak{q})
&=\prod_{n=1}^{N}\frac{1}{1-\mathfrak{q}^n}, 
\end{align}
which enumerates the half-BPS local operators \cite{Corley:2001zk}, 
which correspond to partitions whose length is no greater than $N$. 
In this limit, the matrix integral (\ref{sch_corr1}) reduces to 
\begin{align}
\label{h_corr}
\langle W_{\mathcal{R}_1}\cdots W_{\mathcal{R}_k}\rangle^{U(N)}_{\textrm{$\frac12$BPS}} (\mathfrak{q})
&=\frac{1}{N!}\oint \prod_{i=1}^{N}
\frac{d\sigma_i}{2\pi i\sigma_i}
\frac{\prod_{i\neq j} (1-\frac{\sigma_i}{\sigma_j})}
{\prod_{i,j}(1-\mathfrak{q}\frac{\sigma_i}{\sigma_j})}
\prod_{j=1}^k \chi_{\mathcal{R}_j}(\sigma). 
\end{align}
Since the orthogonal basis of this integral measure is the Hall-Littlewood function, for the two-point function of the Wilson line operators indexed by the two Young diagrams $\lambda$ and $\mu$, 
the half-BPS limit is shown to be given by \cite{Hatsuda:2023iwi}
\begin{align}
\label{h_2pt_lambda_mu}
\langle W_{\lambda} W_{\overline{\mu}}\rangle^{U(N)}_{\textrm{$\frac12$BPS}} (\mathfrak{q})
&=\sum_{\nu}
\frac{K_{\lambda\nu}(\mathfrak{q}) K_{\mu\nu} (\mathfrak{q})}
{\prod_{n=1}^{N-l(\nu)} (1-\mathfrak{q}^n) \prod_{j\ge 1}\prod_{n=1}^{m_j(\nu)} (1-\mathfrak{q}^n)}, 
\end{align}
where $K_{\lambda\nu}(\mathfrak{q})$ is the Kostka-Foulkes polynomial \cite{MR1354144} 
and $m_j(\nu)$ is the multiplicity of the partition $\nu$. 
For the antisymmetric representation, we have
\begin{align}
P_{(1^m)}(\sigma;\mathfrak{q})&=s_{(1^m)}(\sigma), 
\end{align}
where $P_{\lambda}(\sigma;\mathfrak{q})$ is the Hall-Littlewood function.
Therefore if setting $\lambda=\mu=(1^m)$ in \eqref{h_2pt_lambda_mu}, we obtain the exact result
\begin{align}
\label{h_2pt_asym}
\langle W_{(1^m)} W_{\overline{(1^m)}}\rangle^{U(N)}_{\textrm{$\frac12$BPS}} (\mathfrak{q})
=\prod_{n=1}^{N-m} \frac{1}{1-\mathfrak{q}^n} \prod_{n=1}^{m}\frac{1}{1-\mathfrak{q}^n}.
\end{align}
Similarly, the two-point function $\langle W_{\lambda} W_{\overline{(1^m)}}\rangle^{U(N)}_{\textrm{$\frac12$BPS}} (\mathfrak{q})$ can be evaluated in a compact form
\begin{align}
\langle W_{\lambda} W_{\overline{(1^m)}}\rangle^{U(N)}_{\textrm{$\frac12$BPS}} (\mathfrak{q})
=K_{\lambda, (1^m)}(\mathfrak{q})\langle W_{(1^m)} W_{\overline{(1^m)}}\rangle^{U(N)}_{\textrm{$\frac12$BPS}} (\mathfrak{q}).
\end{align}
In particular, for $\lambda=(m)$, we have
\begin{align}
\langle W_{(m)} W_{\overline{(1^m)}}\rangle^{U(N)}_{\textrm{$\frac12$BPS}} (\mathfrak{q})
&=\mathfrak{q}^{\frac{m(m-1)}{2}} \langle W_{(1^m)} W_{\overline{(1^m)}}\rangle^{U(N)}_{\textrm{$\frac12$BPS}} (\mathfrak{q}). 
\end{align}

For the symmetric representation, the Hall-Littlewood function are expanded by the sum of the Schur functions with the hook representations \cite{MR1354144}
\begin{align}
P_{(m)}(\sigma;\mathfrak{q})
=\sum_{r=0}^{m-1}(-\mathfrak{q})^r s_{(m-r,1^r)}.
\end{align}
We can constrain relations of the correlators from the norm and the orthogonality of the Hall-Littlewood function. Using it, one finds the relations
\begin{align}
\label{hook_relation1}
\sum_{r_1=0}^{m-1}
\sum_{r_2=0}^{m-1}
(-\mathfrak{q})^{r_1+r_2}
\langle W_{(m-r_1, 1^{r_1})}W_{\overline{(m-r_2,1^{r_2})}}\rangle^{U(N)}_{\textrm{$\frac12$BPS}}(\mathfrak{q})
&=\frac{1}{1-\mathfrak{q}}
\prod_{n=1}^{N-1}\frac{1}{1-\mathfrak{q}^n}, \\
\sum_{r=0}^{m-1} (-\mathfrak{q})^r \langle W_{(m-r, 1^{r})}W_{\overline{(1^m)}}\rangle^{U(N)}_{\textrm{$\frac12$BPS}}(\mathfrak{q})&=0 \qquad (m >1).
\end{align}
In the following sections, we will implicitly use these relations to check our evaluation. 

%%%%%%%%%%%%%%%%%%%%%%%%%%%%%%%%%%%%
\subsection{Charged Wilson line operators}
%%%%%%%%%%%%%%%%%%%%%%%%%%%%%%%%%%%%
%large N charge n
Let $W_n$ be the Wilson line operator of charge $n$ 
which is described by the power sum symmetric function $p_n$. 
In the large $N$ limit, the normalized 2-point function of $W_n$ and $W_{-n}$ is given by \cite{Hatsuda:2023iwi}
\begin{align}
\label{f+n_-n}
\langle \mathcal{W}_{n} \mathcal{W}_{-n}\rangle^{U(\infty)}(t;q)
&=\frac{n (1-q^n)}{(1-q^{\frac{n}{2}}t^{2n}) (1-q^{\frac{n}{2}}t^{-2n})}. 
\end{align}
Turning off the flavored fugacity $t$, we get
\begin{align}
\label{u+n_-n}
\langle \mathcal{W}_{n} \mathcal{W}_{-n}\rangle^{U(\infty)}(q)
&=\frac{n(1+q^{\frac{n}{2}})}{(1-q^{\frac{n}{2}})}. 
\end{align}
In the half-BPS limit, we find
\begin{align}
\label{h+n_-n}
\langle \mathcal{W}_{n} \mathcal{W}_{-n}\rangle^{U(\infty)}_{\textrm{$\frac12$BPS}}(\mathfrak{q}) 
&=\frac{n}{1-\mathfrak{q}^n}. 
\end{align}
We have
\begin{align}
\label{trans_1}
\langle \mathcal{W}_{n} \mathcal{W}_{-n}\rangle^{U(\infty)}(t;q^{-1})&=
-\langle \mathcal{W}_{n} \mathcal{W}_{-n}\rangle^{U(\infty)}(t;q), \\
\label{trans_2}
\langle \mathcal{W}_{n} \mathcal{W}_{-n}\rangle^{U(\infty)}_{\textrm{$\frac12$BPS}}(\mathfrak{q}^{-1})
&=-\mathfrak{q}^n 
\langle \mathcal{W}_{n} \mathcal{W}_{-n}\rangle^{U(\infty)}_{\textrm{$\frac12$BPS}}(\mathfrak{q}). 
\end{align}
For $n=1$ they are the normalized 2-point functions of the Wilson line operators in the fundamental representation. 
One finds
%unflavored correlators rank1
\begin{align}
\label{f+1_-1}
\langle \mathcal{W}_{\tiny \yng(1)}\mathcal{W}_{\overline{\tiny \yng(1)}}\rangle^{U(\infty)}(t;q)
&=\frac{1-q}{(1-q^{\frac12}t^2) (1-q^{\frac12}t^{-2})}, \\
\label{u+1_-1}
\langle \mathcal{W}_{\tiny \yng(1)}\mathcal{W}_{\overline{\tiny \yng(1)}}\rangle^{U(\infty)}(q)
&=\frac{1+q^{\frac12}}{1-q^{\frac12}}, \\
\label{h+n_-n}
\langle \mathcal{W}_{\tiny \yng(1)}\mathcal{W}_{\overline{\tiny \yng(1)}}\rangle_{\textrm{$\frac12$BPS}}^{U(\infty)}(\mathfrak{q})
&=\frac{1}{1-\mathfrak{q}}. 
\end{align}

%%%%%%%%%%%%%%%%%%%%%%%%%%%%%%%%%%%%
\subsection{Factorization}
%%%%%%%%%%%%%%%%%%%%%%%%%%%%%%%%%%%%
%large N
It follows that 
in the large $N$ limit, the Schur line defect even-point functions of the charged Wilson line operators have the following factorization 
\cite{Hatsuda:2023iwi} 
\begin{align}
\label{fac1}
\langle 
\prod_{j=1}^{k}
(\mathcal{W}_{n_j}\mathcal{W}_{-n_j})^{m_j}
\rangle^{U(\infty)}
&=\prod_{j=1}^{k}m_j ! 
\Bigl(
\langle \mathcal{W}_{n_j} \mathcal{W}_{-n_j}\rangle^{U(\infty)}
\Bigr)^{m_j}. 
\end{align}
On the other hand, the large $N$ odd-point functions of the charged Wilson line operators vanish. 
In the following sections, we present the large $N$ correlation functions of various Wilson line operators 
by making use of the factorization (\ref{fac1}). 

According to the relations (\ref{trans_1}), (\ref{trans_2}) and the factorization (\ref{fac1}), 
the large $N$ normalized correlation function of the charged Wilson line operators enjoys the automorphy
\begin{align}
\langle 
\prod_{j=1}^{k}
(\mathcal{W}_{n_j}\mathcal{W}_{-n_j})^{m_j}
\rangle^{U(\infty)}(t;q^{-1})
&=(-1)^{k}
\langle 
\prod_{j=1}^{k}
(\mathcal{W}_{n_j}\mathcal{W}_{-n_j})^{m_j}
\rangle^{U(\infty)}(t;q), \\
\langle 
\prod_{j=1}^{k}
(\mathcal{W}_{n_j}\mathcal{W}_{-n_j})^{m_j}
\rangle^{U(\infty)}_{\textrm{$\frac12$BPS}}(\mathfrak{q}^{-1})
&=(-1)^{k} \mathfrak{q}^{\sum_{j=1}^{k}n_j}
\langle 
\prod_{j=1}^{k}
(\mathcal{W}_{n_j}\mathcal{W}_{-n_j})^{m_j}
\rangle^{U(\infty)}_{\textrm{$\frac12$BPS}}(\mathfrak{q}). 
\end{align}

%%%%%%%%%%%%%%%%%%%%%%%%%%%%%%%%%%%%
\subsection{Holographic dual brane configuration}
%%%%%%%%%%%%%%%%%%%%%%%%%%%%%%%%%%%%
In later sections, we examine the large $N$ limit of the Schur line defect correlation functions 
which play a significant role in the study of the AdS/CFT correspondence for the line operator.  
Let us briefly review the holographic dual brane configuration of the Wilson line operators in $\mathcal{N}=4$ $U(N)$ SYM theory. 
Suppose that $N$ D3-branes are supported on $(x^0,x^1,x^2,x^3)$ in Type IIB string theory on $\mathbb{R}^{1,9}$. 
The low-energy effective description of the D3-branes is $\mathcal{N}=4$ $U(N)$ SYM theory on $(x^0,x^1,x^2,x^3)$.  

When a fundamental string on $(x^0,x^9)$ is added, one finds the half-BPS Wilson line operator 
in the fundamental representation for $\mathcal{N}=4$ $U(N)$ SYM theory \cite{Maldacena:1998im,Rey:1998ik}. 
When $m$ fundamental strings end on the $N$ D3-branes and on another D3-brane, 
we obtain the half-BPS Wilson line operator in the rank-$m$ symmetric representation 
\cite{Drukker:2005kx,Gomis:2006sb,Gomis:2006im,Rodriguez-Gomez:2006fmx,Yamaguchi:2007ps}. 
Without breaking further supersymmetry, the configuration also admit a D5-brane on $(x^0,$ $x^4,$ $x^5,$ $x^6,$ $x^7,$ $x^8)$. 
When $m$ fundamental strings ending on the $N$ D3-branes and a D5-brane, 
the Wilson line operator in the rank-$m$ antisymmetric representation is introduced \cite{Yamaguchi:2006tq,Gomis:2006sb,Rodriguez-Gomez:2006fmx,Hartnoll:2006hr}. 
The brane setup is summarized as follows: 
\begin{align}
\label{brane_conf}
\begin{array}{c|cccccccccc}
&0&1&2&3&4&5&6&7&8&9 \\ \hline
\textrm{D3}&\circ&\circ&\circ&\circ&&&&&&\\
\textrm{F1}&\circ&&&&&&&&&\circ\\
\textrm{D5}&\circ&&&&\circ&\circ&\circ&\circ&\circ&\\
\end{array}
\end{align}
Here $\circ$ denotes the directions in which branes are extended. 
The configuration preserves $SO(1,2)$ $\times$ $SO(3)$ $\times$ $SO(5)$ global symmetry, 
the bosonic subgroup of the 1d $\mathcal{N}=8$ superconformal group $OSp(4^*|4)$. 

\begin{figure}
\begin{center}
\includegraphics[width=7cm]{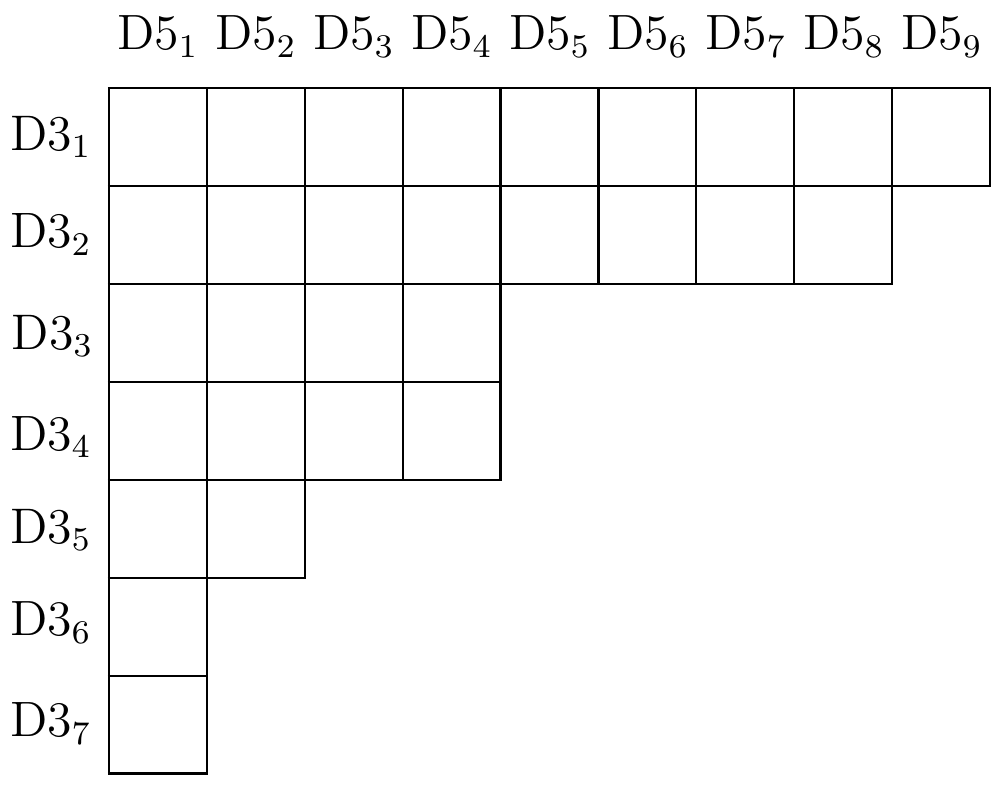}
\caption{
The Young diagram $\lambda=(9,8,4,4,2,1,1)$ labeling the representation of the Wilson line operator. 
The holographic dual brane configuration is realized 
as $7$ D3-branes with fundamental string charges $\lambda=(9,8,4,4,2,1,1)$ or 
that as $9$ D5-branes with fundamental string charges $\lambda'=(7,5,4,4,2,2,2,2,1)$. 
}
\label{figyoungbrane}
\end{center}
\end{figure}

More generally, it is argued in \cite{Gomis:2006sb} that 
the Wilson line operator in the representation labeled by the Young diagram 
$\lambda=(\lambda_1,\cdots,\lambda_r)$ 
can be realized in the holographic dual gravitational description in terms of 
a configuration of $r$ coincident D3-branes 
$(\textrm{D3}_1,\cdots, \textrm{D3}_r)$ in $AdS_5\times S^5$ 
where D3$_i$ is the $i$-th D3-brane that carries $\lambda_i$ units of fundamental string charge dissolved in it. 
Alternatively, it is also shown to be described in terms of 
a configuration of $\lambda_1$ coincident D5-branes 
$(\textrm{D5}_1,\cdots, \textrm{D5}_{\lambda_r})$ in $AdS_5\times S^5$ 
where D5$_j$ is the $j$-th D5-brane that carries $\lambda_j'$ units of fundamental string charge dissolved in it 
See Figure \ref{figyoungbrane} for an example with the Young diagram $(9,8,4,4,2,1,1)$. 

In the near horizon limit of the brane configuration (\ref{brane_conf}),  
the resulting Type IIB supergravity solution contains $AdS_2$ factor due to the one-dimensional conformal group $SO(1,2)$, 
$S^2$ factor corresponding to the $SO(3)$ and $S^4$ factor associated to the $SO(5)$. 
It takes the form \cite{DHoker:2007mci}
\begin{align}
AdS_2\times S^2\times S^4\times \Sigma, 
\end{align}
where $\Sigma$ is a two-dimensional surface, over which the $AdS_2\times S^2\times S^4$ part is warped. 
It has the induced metric \cite{DHoker:2007mci}
\begin{align}
ds^2&=f_1^2 ds_{AdS_2}^2+f_2^2 ds_{S^2}^2+f_4 ds_{S^4}^2+ds_{\Sigma}^2, 
\end{align}
where $f_1$, $f_2$, $f_4$ and $ds_{\Sigma}^2$ are real functions on $\Sigma$. 
When $\Sigma$ is chosen as a hyperelliptic Riemann surface of genus $g$ with boundary, 
the bubbling solutions are parametrized by two harmonic functions $h_1, h_2$ on $\Sigma$ as well as $\Sigma$ 
in such a way that $g$ is the number of parts of the Young diagram \cite{DHoker:2007mci}. 

The fundamental strings wrapping $AdS_2$ in the global $AdS_5$ can meet the boundary of the $AdS_5$ at two points 
so that they should correspond to the superconformal lines which map to a pair of half-lines localized at the north pole and the south pole of $S^3$. 
Accordingly, the fluctuation modes of the gravity dual configuration is expected to be calculated from the gauge theory side 
as the 2-point functions of the corresponding Wilson line operators in $\mathcal{N}=4$ $U(N)$ SYM theory. 
In fact, it is confirmed in \cite{Gang:2012yr,Hatsuda:2023iwi} 
\footnote{See \cite{Gang:2012yr} for the fundamental representation and the large antisymmetric representation 
and \cite{Hatsuda:2023iwi} for arbitrary antisymmetric and symmetric representations.} 
that the large $N$ normalized 2-point function of the Wilson line operator in the fundamental representation 
and that in the large (anti)symmetric representation precisely agree with the gravity indices 
encoding the corresponding gravity dual spectra \cite{Faraggi:2011bb,Faraggi:2011ge}. 

The calculation of the fluctuation modes for more general geometries from the gravity side seems a rather non-trivial problem. 
To our knowledge, they have not yet been calculated in the literature. 
However, in the following sections, we will address them from gauge theory side by computing the large $N$ Schur line defect correlators 
of the Wilson line operators in the representations labeled by the Young diagrams with hook shape and rectangular shape. 

%%%%%%%%%%%%%%%%%%%%%%%%%%%%%%%%%%%%
%%%%%%%%%%%%%%%%%%%%%%%%%%%%%%%%%%%%
\section{(Anti)Symmetric representations}
\label{sec_asymrep}
%%%%%%%%%%%%%%%%%%%%%%%%%%%%%%%%%%%%
%%%%%%%%%%%%%%%%%%%%%%%%%%%%%%%%%%%%
In this section, we begin by examining the large $N$ correlation functions of the Wilson line operators 
$W_{(1^m)}$ (resp. $W_{(m)}$) 
transforming in the rank-$m$ antisymmetric (resp. symmetric) representations labeled by the Young diagrams $(1^m)$ (resp. $(m)$). 
While the exact closed-form expression for the large $N$ 2-point function of the two Wilson line operators 
$W_{(1^m)}$'s (resp. $W_{(m)}$'s) is given in \cite{Hatsuda:2023iwi}, 
several new properties and more general correlation functions are discussed. 

The Wilson line operator $W_{(1^m)}$ (resp. $W_{(m)}$) is associated with the elementary symmetric function $e_m$
(resp. complete homogeneous symmetric function $h_m$). 
According to Newton's identities
\begin{align}
\label{Newton1}
mh_m&=\sum_{i=1}^{m}h_{m-i}p_i,\\
\label{Newton2}
me_m&=\sum_{i=1}^{m} (-1)^{i-1}e_{m-i}p_i, 
\end{align}
as well as the factorization property (\ref{fac1}), 
the large $N$ correlation functions of the Wilson line operators in the (anti)symmetric representations 
can be expressed in terms of the 2-point function (\ref{f+n_-n}) of the charged Wilson line operators.  

%%%%%%%%%%%%%%%%%%%%%%%%%%%%%%%%%%%%
\subsection{Rank-$2$ representations}
%%%%%%%%%%%%%%%%%%%%%%%%%%%%%%%%%%%%
The large $N$ normalized 2-point functions of the Wilson line operators in the rank-$2$ (anti)symmetric representations are given by 
\begin{align}
\label{f2}
\langle \mathcal{W}_{\tiny \yng(2)}\mathcal{W}_{\overline{\tiny \yng(2)}}\rangle^{U(\infty)}
&=\frac12 \langle \mathcal{W}_{1} \mathcal{W}_{-1}\rangle^{U(\infty)}
+\frac14 \langle \mathcal{W}_{2} \mathcal{W}_{-2}\rangle^{U(\infty)}
, \\
\label{f2&1_1}
\langle \mathcal{W}_{\tiny \yng(2)}\mathcal{W}_{\overline{\tiny \yng(1,1)}}\rangle^{U(\infty)}
&=\frac12 \langle \mathcal{W}_{1} \mathcal{W}_{-1}\rangle^{U(\infty)}
-\frac14 \langle \mathcal{W}_{2} \mathcal{W}_{-2}\rangle^{U(\infty)}. 
\end{align}
The two correlators (\ref{f2}) and (\ref{f2&1_1}) are related by
\begin{align}
\langle \mathcal{W}_{\tiny \yng(2)}\mathcal{W}_{\overline{\tiny \yng(2)}}\rangle^{U(\infty)}(t;q^{-1})
=
\langle \mathcal{W}_{\tiny \yng(2)}\mathcal{W}_{\overline{\tiny \yng(1,1)}}\rangle^{U(\infty)}(t;q). 
\end{align}

%%%%%%%%%%%%%%%%%%%%%%%%%%%%%%%%%%%%
\subsubsection{Unflavored limit}
%%%%%%%%%%%%%%%%%%%%%%%%%%%%%%%%%%%%
%unflavored correlators rank2
The large $N$ unflavored Schur line defect 2-point functions of the Wilson line operators 
transforming in the rank-$2$ (anti)symmetric representations are  
\begin{align}
\label{u2&2}
\langle \mathcal{W}_{\tiny \yng(2)}\mathcal{W}_{\overline{\tiny \yng(2)}}\rangle^{U(\infty)}(q)
&=\langle \mathcal{W}_{\tiny \yng(1,1)}\mathcal{W}_{\overline{\tiny \yng(1,1)}}\rangle^{U(\infty)}(q)
=\frac{1+q^{\frac12}+2q}{(1-q^{\frac12})(1-q)}
\nonumber\\
&=\sum_{n\ge0}a_{\{(2), (2)\}}(n) q^{\frac{n}{2}}, \\
\label{u2&1_1}
\langle \mathcal{W}_{\tiny \yng(2)}\mathcal{W}_{\overline{\tiny \yng(1,1)}}\rangle^{U(\infty)}(q)
&=\frac{q^{\frac12} (2+q^{\frac12}+q)}{(1-q^{\frac12}) (1-q)}
\nonumber\\
&=\sum_{n\ge0}a_{\{(2), (1^2)\}}(n) q^{\frac{n}{2}}. 
\end{align}
Here the coefficients 
\begin{align}
a_{\{(2), (2)\}}(n)&=\sum_{n\ge0} \frac{4n+1+(-1)^{n}}{2}
\end{align}
and 
\begin{align}
a_{\{(2), (1^2)\}}&=\sum_{n\ge 0}\frac{4n+3+(-1)^n}{2} 
\end{align}
are the numbers congruent to $1$ or $2 \mod 4$ 
and the numbers congruent to $2$ or $3\mod 4$ respectively. 

%%%%%%%%%%%%%%%%%%%%%%%%%%%%%%%%%%%%
\subsubsection{Half-BPS limit}
%%%%%%%%%%%%%%%%%%%%%%%%%%%%%%%%%%%%
%1/2BPS correlators rank2
The half-BPS limit of the large $N$ 2-point functions for the rank-$2$ (anti)symmetric representations is given by 
\begin{align}
\label{h2&2}
\langle \mathcal{W}_{\tiny \yng(2)}\mathcal{W}_{\overline{\tiny \yng(2)}}\rangle^{U(\infty)}_{\textrm{$\frac12$BPS}}(\mathfrak{q})
&=\langle \mathcal{W}_{\tiny \yng(1,1)}\mathcal{W}_{\overline{\tiny \yng(1,1)}}\rangle^{U(\infty)}_{\textrm{$\frac12$BPS}}(\mathfrak{q})
=\frac{1}{(1-\mathfrak{q}) (1-\mathfrak{q}^2)}, \\
\label{h2&1_1}
\langle \mathcal{W}_{\tiny \yng(2)}\mathcal{W}_{\overline{\tiny \yng(1,1)}}\rangle^{U(\infty)}_{\textrm{$\frac12$BPS}}(\mathfrak{q})
&=\frac{\mathfrak{q}}{(1-\mathfrak{q}) (1-\mathfrak{q}^2)}. 
\end{align}
We have
\begin{align}
\langle \mathcal{W}_{\tiny \yng(2)}\mathcal{W}_{\overline{\tiny \yng(2)}}\rangle^{U(\infty)}_{\textrm{$\frac12$BPS}}(\mathfrak{q}^{-1})
&=\mathfrak{q}^2 \langle \mathcal{W}_{\tiny \yng(2)}\mathcal{W}_{\overline{\tiny \yng(1,1)}}\rangle^{U(\infty)}_{\textrm{$\frac12$BPS}}(\mathfrak{q}). 
\end{align}

%%%%%%%%%%%%%%%%%%%%%%%%%%%%%%%%%%%%
\subsection{Rank-$3$ representations}
%%%%%%%%%%%%%%%%%%%%%%%%%%%%%%%%%%%%
The large $N$ normalized 2-point functions of the Wilson line operators in the rank-$3$ (anti)symmetric representations are given by
\begin{align}
\label{f3&3}
&
\langle \mathcal{W}_{\tiny \yng(3)}\mathcal{W}_{\overline{\tiny \yng(3)}}\rangle^{U(\infty)}
\nonumber\\
&=\frac16 {\langle \mathcal{W}_{1}\mathcal{W}_{-1}\rangle^{U(\infty)}}^3
+\frac14 \langle \mathcal{W}_{1}\mathcal{W}_{-1}\rangle^{U(\infty)} \langle \mathcal{W}_{2}\mathcal{W}_{-2}\rangle^{U(\infty)}
+\frac19 \langle \mathcal{W}_{3}\mathcal{W}_{-3}\rangle^{U(\infty)}, \\
\label{f3&1_1_1}
&
\langle \mathcal{W}_{\tiny \yng(3)}\mathcal{W}_{\overline{\tiny \yng(1,1,1)}}\rangle^{U(\infty)}
\nonumber\\
&=\frac16 {\langle \mathcal{W}_{1}\mathcal{W}_{-1}\rangle^{U(\infty)}}^3
-\frac14 {\langle \mathcal{W}_{1}\mathcal{W}_{-1}\rangle^{U(\infty)}} {\langle \mathcal{W}_{2}\mathcal{W}_{-2}\rangle^{U(\infty)}}
+\frac19 {\langle \mathcal{W}_{3}\mathcal{W}_{-3}\rangle^{U(\infty)}}. 
\end{align}
It is shown that  
\begin{align}
\langle \mathcal{W}_{\tiny \yng(3)}\mathcal{W}_{\overline{\tiny \yng(3)}}\rangle^{U(\infty)}(t;q^{-1})
&=-\langle \mathcal{W}_{\tiny \yng(3)}\mathcal{W}_{\overline{\tiny \yng(1,1,1)}}\rangle^{U(\infty)}(t;q). 
\end{align}

%%%%%%%%%%%%%%%%%%%%%%%%%%%%%%%%%%%%
\subsubsection{Unflavored limit}
%%%%%%%%%%%%%%%%%%%%%%%%%%%%%%%%%%%%
%unflavored correlators rank3
The large $N$ unflavored 2-point functions of the Wilson line operators in the rank-$3$ (anti)symmetric representations are 
\begin{align}
\label{u3&3}
\langle \mathcal{W}_{\tiny \yng(3)}\mathcal{W}_{\overline{\tiny \yng(3)}}\rangle^{U(\infty)}(q)
&=\langle \mathcal{W}_{\tiny \yng(1,1,1)}\mathcal{W}_{\overline{\tiny \yng(1,1,1)}}\rangle^{U(\infty)}(q)
=\frac{1+q^{\frac12}+2q+3q^{\frac32}+q^2}{(1-q^{\frac12}) (1-q) (1-q^{\frac32})}, \\
\label{u3&1_1_1}
\langle \mathcal{W}_{\tiny \yng(3)}\mathcal{W}_{\overline{\tiny \yng(1,1,1)}}\rangle^{U(\infty)}(q)
&=\frac{q(1+3q^{\frac12}+2q+q^{\frac32}+q^2)}{(1-q^{\frac12}) (1-q) (1-q^{\frac32})}. 
\end{align}

%%%%%%%%%%%%%%%%%%%%%%%%%%%%%%%%%%%%
\subsubsection{Half-BPS limit}
%%%%%%%%%%%%%%%%%%%%%%%%%%%%%%%%%%%%
%1/2BPS correlators rank3
For the rank-$3$ (anti)symmetric representations the half-BPS limit of the large $N$ normalized 2-point functions of the Wilson line operators is  
\begin{align}
\label{h3&3}
\langle \mathcal{W}_{\tiny \yng(3)}\mathcal{W}_{\overline{\tiny \yng(3)}}\rangle^{U(\infty)}_{\textrm{$\frac12$BPS}}(\mathfrak{q})
&=\langle \mathcal{W}_{\tiny \yng(1,1,1)}\mathcal{W}_{\overline{\tiny \yng(1,1,1)}}\rangle^{U(\infty)}_{\textrm{$\frac12$BPS}}(\mathfrak{q})
=\frac{1}{(1-\mathfrak{q}) (1-\mathfrak{q}^2) (1-\mathfrak{q}^3)}, \\
\label{h3&1_1_1}
\langle \mathcal{W}_{\tiny \yng(3)}\mathcal{W}_{\overline{\tiny \yng(1,1,1)}}\rangle^{U(\infty)}_{\textrm{$\frac12$BPS}}(\mathfrak{q})
&=\frac{\mathfrak{q}^3}{(1-\mathfrak{q}) (1-\mathfrak{q}^2) (1-\mathfrak{q}^3)}. 
\end{align}
They obey
\begin{align}
\langle \mathcal{W}_{\tiny \yng(3)}\mathcal{W}_{\overline{\tiny \yng(3)}}\rangle^{U(\infty)}_{\textrm{$\frac12$BPS}}(\mathfrak{q}^{-1})
&=-\mathfrak{q}^{3}
\langle \mathcal{W}_{\tiny \yng(3)}\mathcal{W}_{\overline{\tiny \yng(1,1,1)}}\rangle^{U(\infty)}_{\textrm{$\frac12$BPS}}(\mathfrak{q}). 
\end{align}

%%%%%%%%%%%%%%%%%%%%%%%%%%%%%%%%%%%%
\subsection{Rank-$4$ representations}
%%%%%%%%%%%%%%%%%%%%%%%%%%%%%%%%%%%%
The large $N$ normalized 2-point function of the Wilson line operators in the rank-$4$ (anti)symmetric representations 
can be expanded with respect to the large $N$ 2-point functions of the charged Wilson line operators 
\begin{align}
\label{f4&4}
&\langle \mathcal{W}_{\tiny \yng(4)}\mathcal{W}_{\overline{\tiny \yng(4)}}\rangle^{U(\infty)}(q)
=\langle \mathcal{W}_{\tiny \yng(1,1,1,1)}\mathcal{W}_{\overline{\tiny \yng(1,1,1,1)}}\rangle^{U(\infty)}(q)
\nonumber\\
&=\frac{1}{24}{\langle W_{1}W_{-1}\rangle^{U(\infty)}}^{4}
+\frac{1}{8}{\langle W_{1}W_{-1}\rangle^{U(\infty)}}^{2}{\langle W_{2}W_{-2}\rangle^{U(\infty)}}
\nonumber\\
&+\frac{1}{32}{\langle W_{2}W_{-2}\rangle^{U(\infty)}}^{2}
+\frac{1}{9}{\langle W_{1}W_{-1}\rangle^{U(\infty)}}{\langle W_{3}W_{-3}\rangle^{U(\infty)}}
+\frac{1}{16}{\langle W_{4}W_{-4}\rangle^{U(\infty)}}
, \\
&
\label{f4&1111}
\langle \mathcal{W}_{\tiny \yng(4)}\mathcal{W}_{\overline{\tiny \yng(1,1,1,1)}}\rangle^{U(\infty)}(q)
\nonumber\\
&=\frac{1}{24}{\langle W_{1}W_{-1}\rangle^{U(\infty)}}^{4}
-\frac{1}{8}{\langle W_{1}W_{-1}\rangle^{U(\infty)}}^{2}{\langle W_{2}W_{-2}\rangle^{U(\infty)}}
\nonumber\\
&+\frac{1}{32}{\langle W_{2}W_{-2}\rangle^{U(\infty)}}^{2}
+\frac{1}{9}{\langle W_{1}W_{-1}\rangle^{U(\infty)}}{\langle W_{3}W_{-3}\rangle^{U(\infty)}}
-\frac{1}{16}{\langle W_{4}W_{-4}\rangle^{U(\infty)}}. 
\end{align}
We have 
\begin{align}
\langle \mathcal{W}_{\tiny \yng(4)}\mathcal{W}_{\overline{\tiny \yng(4)}}\rangle^{U(\infty)}(t;q^{-1})
&=
\langle \mathcal{W}_{\tiny \yng(4)}\mathcal{W}_{\overline{\tiny \yng(1,1,1,1)}}\rangle^{U(\infty)}(t;q). 
\end{align}

%%%%%%%%%%%%%%%%%%%%%%%%%%%%%%%%%%%%
\subsubsection{Unflavored limit}
%%%%%%%%%%%%%%%%%%%%%%%%%%%%%%%%%%%%
%unflavored correlators rank4
The unflavored 2-point functions of the Wilson line operators in the rank-$4$ (anti)symmetric representations are given by
\begin{align}
\label{u4&4}
\langle \mathcal{W}_{\tiny \yng(4)}\mathcal{W}_{\overline{\tiny \yng(4)}}\rangle^{U(\infty)}(q)
&=\langle \mathcal{W}_{\tiny \yng(1,1,1,1)}\mathcal{W}_{\overline{\tiny \yng(1,1,1,1)}}\rangle^{U(\infty)}(q)
=\frac{1+q^{\frac12}+2q+3q^{\frac32}+5q^2+2q^{\frac52}+2q^3}{(1-q^{\frac12}) (1-q) (1-q^{\frac32}) (1-q^2)}, \\
\label{u4&1_1_1_1}
\langle \mathcal{W}_{\tiny \yng(4)}\mathcal{W}_{\overline{\tiny \yng(1,1,1,1)}}\rangle^{U(\infty)}(q)
&=\frac{q^2(2+2q^{\frac12}+5q+3q^{\frac32}+2q^2+q^{\frac52}+q^3)}{(1-q^{\frac12}) (1-q) (1-q^{\frac32}) (1-q^{2})}. 
\end{align}

%%%%%%%%%%%%%%%%%%%%%%%%%%%%%%%%%%%%
\subsubsection{Half-BPS limit}
%%%%%%%%%%%%%%%%%%%%%%%%%%%%%%%%%%%%
%1/2BPS correlators rank4
Also we find
\begin{align}
\label{h4&4}
\langle \mathcal{W}_{\tiny \yng(4)}\mathcal{W}_{\overline{\tiny \yng(4)}}\rangle^{U(\infty)}_{\textrm{$\frac12$BPS}}(\mathfrak{q})
&=\langle \mathcal{W}_{\tiny \yng(1,1,1,1)}\mathcal{W}_{\overline{\tiny \yng(1,1,1,1)}}\rangle^{U(\infty)}_{\textrm{$\frac12$BPS}}(\mathfrak{q})
=\frac{1}{(1-\mathfrak{q}) (1-\mathfrak{q}^2) (1-\mathfrak{q}^3) (1-\mathfrak{q}^4)}, \\
\label{h4&1_1_1}
\langle \mathcal{W}_{\tiny \yng(4)}\mathcal{W}_{\overline{\tiny \yng(1,1,1,1)}}\rangle^{U(\infty)}_{\textrm{$\frac12$BPS}}(\mathfrak{q})
&=\frac{\mathfrak{q}^6}{(1-\mathfrak{q}) (1-\mathfrak{q}^2) (1-\mathfrak{q}^3) (1-\mathfrak{q}^4)}. 
\end{align}
They satisfy the relation
\begin{align}
\langle \mathcal{W}_{\tiny \yng(4)}\mathcal{W}_{\overline{\tiny \yng(4)}}\rangle_{\textrm{$\frac12$BPS}}^{U(\infty)}(\mathfrak{q}^{-1})
&=
\mathfrak{q}^4\langle \mathcal{W}_{\tiny \yng(4)}\mathcal{W}_{\overline{\tiny \yng(1,1,1,1)}}\rangle_{\textrm{$\frac12$BPS}}^{U(\infty)}(\mathfrak{q}). 
\end{align}

%%%%%%%%%%%%%%%%%%%%%%%%%%%%%%%%%%%%
\subsection{Rank-$m$ representations}
%%%%%%%%%%%%%%%%%%%%%%%%%%%%%%%%%%%%
For general rank-$m$ (anti)symmetric representations, 
the large $N$ normalized 2-point functions of the Wilson line operators are given by \cite{Hatsuda:2023iwi}
\begin{align}
\label{fm&m}
&\langle \mathcal{W}_{(m)} \mathcal{W}_{\overline{(m)}}\rangle^{U(\infty)}(t;q)
=\langle \mathcal{W}_{(1^m)} \mathcal{W}_{\overline{(1^m)}}\rangle^{U(\infty)}(t;q)
\nonumber\\
&=\sum_{n=0}^{m}
\frac{1}{(q^{\frac12}t^2;q^{\frac12}t^2)_{n} (q^{\frac12}t^{-2};q^{\frac12}t^{-2})_{m-n}}
-\sum_{n=0}^{m-1}
\frac{1}{(q^{\frac12}t^2;q^{\frac12}t^2)_{n} (q^{\frac12}t^{-2};q^{\frac12}t^{-2})_{m-n-1}}
\nonumber\\
&=\sum_{n=0}^{m}
\frac{q^{\frac{m-n}{2}}t^{-2(m-n)}}
{(q^{\frac12}t^2;q^{\frac12}t^2)_{n} (q^{\frac12}t^{-2};q^{\frac12}t^{-2})_{m-n}}. 
\end{align}
Also we get
\begin{align}
\label{fm&1^m}
\langle \mathcal{W}_{(m)} \mathcal{W}_{\overline{(1)^m}}\rangle^{U(\infty)}(t;q)
&=\sum_{n=0}^{m}
\frac{q^{\frac14 (m^2+2n(n+1)-m(2n+1))} t^{m(2n-m+1)}}
{(q^{\frac12}t^2;q^{\frac12}t^2)_{m-n} (q^{\frac12}t^{-2};q^{\frac12}t^{-2})_{n}}. 
\end{align}
We have
\begin{align}
\langle \mathcal{W}_{(m)} \mathcal{W}_{\overline{(m)}}\rangle^{U(\infty)}(t;q^{-1})
&=(-1)^m \langle \mathcal{W}_{(m)} \mathcal{W}_{\overline{(1)^m}}\rangle^{U(\infty)}(t;q). 
\end{align}

%%%%%%%%%%%%%%%%%%%%%%%%%%%%%%%%%%%%
\subsubsection{Unflavored limit}
%%%%%%%%%%%%%%%%%%%%%%%%%%%%%%%%%%%%
In the unflavored limit, the large $N$ 2-point functions (\ref{fm&m}) and (\ref{fm&1^m}) reduce to
\begin{align}
\label{um&m}
\langle \mathcal{W}_{(m)} \mathcal{W}_{\overline{(m)}}\rangle^{U(\infty)}(q)
&=\langle \mathcal{W}_{(1^m)} \mathcal{W}_{\overline{(1^m)}}\rangle^{U(\infty)}(q)
=\sum_{n=0}^{\infty}
\frac{q^{\frac{m-n}{2}}}{(q^{\frac12};q^{\frac12})_{n} (q^{\frac12};q^{\frac12})_{m-n}}, \\
\label{um&1^m}
\langle \mathcal{W}_{(m)} \mathcal{W}_{\overline{(1^m)}}\rangle^{U(\infty)}(q)
&=\sum_{n=0}^{\infty}
\frac{q^{\frac14 (m^2+2n(n+1)-m(2n+1))}}{(q^{\frac12};q^{\frac12})_{n} (q^{\frac12};q^{\frac12})_{m-n}}. 
\end{align}
They can be written as
\begin{align}
\langle \mathcal{W}_{(m)} \mathcal{W}_{\overline{(m)}}\rangle^{U(\infty)}(q)
&=\langle \mathcal{W}_{(1^m)} \mathcal{W}_{\overline{(1^m)}}\rangle^{U(\infty)}(q)
=G_{\{(m),(m)\}}(q)\prod_{n=1}^{m}\frac{1}{1-q^{\frac{n}{2}}}, \\
\langle \mathcal{W}_{(m)} \mathcal{W}_{\overline{(1^{m})}}\rangle^{U(\infty)}(q)
&=q^{\frac{m(m-1)}{2}}G_{\{(m),(m)\}}(q^{-1})\prod_{n=1}^{m}\frac{1}{1-q^{\frac{n}{2}}}, 
\end{align}
where $G_{\{(m) (m)\}}(q)$ is a polynomial in $q$ with positive coefficients. 
So we have
\begin{align}
\langle \mathcal{W}_{(m)} \mathcal{W}_{\overline{(m)}}\rangle^{U(\infty)}(q^{-1})
&=(-1)^m \langle \mathcal{W}_{(m)} \mathcal{W}_{\overline{(1^{m})}}\rangle^{U(\infty)}(q). 
\end{align}
It follows that 
\begin{align}
G_{\{(m),(m)\}}(q)&\equiv \prod_{n=1}^{\infty}\frac{1}{(1-q^{\frac{n}{2}})}
\mod q^{\frac{m+1}{2}}, 
\end{align}
where the notation $f(q)\equiv g(q)\mod q^{\frac{m}{2}}$ implies that 
the coefficients of $q^{\frac{k}{2}}$ in $f(q)$ and $g(q)$ are the same for $k=0,1,\cdots, m-1$. 

In the large representation limit $m\rightarrow \infty$, 
the large $N$ normalized unflavored 2-point function 
of the symmetric Wilson line operators or equivalently that of the antisymmetric Wilson line operators becomes
\begin{align}
&
\lim_{m\rightarrow \infty}
\langle \mathcal{W}_{(m)} \mathcal{W}_{\overline{(m)}}\rangle^{U(\infty)}(q)
=
\lim_{m\rightarrow \infty}
\langle \mathcal{W}_{(1^m)} \mathcal{W}_{\overline{(1^m)}}\rangle^{U(\infty)}(q)
= \prod_{n=1}^{\infty}\frac{1}{(1-q^{\frac{n}{2}})^2}
\nonumber\\
&=\sum_{n=0}^{\infty}a_{\{(\infty),(\infty) \}}^{(G)}(n)q^{\frac{n}{2}}
\nonumber\\
&=1+2q^{\frac12}+5q+10q^{\frac32}+20q^2+36q^{\frac52}+65q^3+110q^{\frac72}+185q^4+300q^{\frac92}+481q^5+\cdots.
\end{align}
The degeneracy $a_{\{(\infty),(\infty) \}}^{(G)}(n)$ is equal to the number of double partitions of $n$ \cite{MR1442260}. 
The asymptotic degeneracy is given by the Meinardus theorem \cite{MR62781}
\begin{align}
a_{\{(\infty),(\infty) \}}^{(G)}(n)
&\sim 
\frac{1}{4\cdot 3^{\frac34} n^{\frac54}} \exp\left(
\frac{2}{\sqrt{3}}\pi n^{\frac12}
\right). 
\end{align}
The actual values and the analytic values are 
\begin{align}
\begin{array}{c|c|c}
n&a_{\{(\infty),(\infty) \}}^{(G)}(n)&a_{\{(\infty),(\infty)\} \textrm{asym.}}^{(G)}(n)\\ \hline 
10&481&591.694 \\
100&1.84365\times 10^{12}&1.97043\times 10^{12} \\
1000&1.26159\times 10^{45}&1.28849\times 10^{45} \\
10000&3.81778\times 10^{151}&3.84335\times 10^{151} \\
\end{array}. 
\end{align}

As discussed in \cite{Hatsuda:2023iwi}, the full large $N$ unflavored 2-point function is given by
\begin{align}
\lim_{m\rightarrow \infty}
\langle W_{(m)} W_{\overline{(m)}}\rangle^{U(\infty)}(q)
&=
\lim_{m\rightarrow \infty}
\langle W_{(1^m)} W_{\overline{(1^m)}}\rangle^{U(\infty)}(q)
=\prod_{n=1}^{\infty} 
\frac{1-q^n}{(1-q^{\frac{n}{2}})^4}. 
\end{align}
This is identified with the generating function for the plane partition diamonds of unrestricted length \cite{MR4370530}. 
On the other hand, the large $N$ 2-point function of the symmetric Wilson line operator 
and the antisymmetric Wilson line operator starts from $q^{\frac{m(m-1)}{4}}$. Therefore it
does not allow for a non-trivial large representation limit 
\begin{align}
\lim_{m\rightarrow\infty}
\langle \mathcal{W}_{(m)} \mathcal{W}_{\overline{(1^m)}}\rangle^{U(\infty)}(q)&=0. 
\end{align}

The large $N$ unflavored higher-point functions of (anti)symmetric Wilson line operators 
vanish in the large representation limit $m\rightarrow \infty$ for $|q|<1$. 
However, we find that the difference of the 4-point functions of the Wilson line operators with different ranks are 
\begin{align}
\label{um&m&m&m_diff}
&
\langle \mathcal{W}_{(m)}\mathcal{W}_{(m)}\mathcal{W}_{\overline{(m)}}\mathcal{W}_{\overline{(m)}}\rangle^{U(\infty)}(q)
-\langle \mathcal{W}_{(m-1)}\mathcal{W}_{(m-1)}\mathcal{W}_{\overline{(m-1)}}\mathcal{W}_{\overline{(m-1)}}\rangle^{U(\infty)}(q)
\nonumber\\
&\equiv \prod_{n=1}^{\infty}\frac{1}{(1-q^{\frac{n}{2}})^8}\mod q^{\frac{m+1}{2}}. 
\end{align}

%%%%%%%%%%%%%%%%%%%%%%%%%%%%%%%%%%%%
\subsubsection{Half-BPS limit}
%%%%%%%%%%%%%%%%%%%%%%%%%%%%%%%%%%%%
For general rank-$m$ (anti)symmetric representations we find
\begin{align}
\label{hm&m}
\langle \mathcal{W}_{(m)}\mathcal{W}_{\overline{(m)}}\rangle^{U(\infty)}_{\textrm{$\frac12$BPS}}(\mathfrak{q})
&=\langle \mathcal{W}_{(1^m)}\mathcal{W}_{\overline{(1^m)}}\rangle^{U(\infty)}_{\textrm{$\frac12$BPS}}(\mathfrak{q})
=\prod_{n=1}^{m} \frac{1}{1-\mathfrak{q}^n}, \\
\label{m&1^m}
\langle \mathcal{W}_{(m)}\mathcal{W}_{\overline{(1^m)}}\rangle^{U(\infty)}_{\textrm{$\frac12$BPS}}(\mathfrak{q})
&=\mathfrak{q}^{\frac{m(m-1)}{2}} \prod_{n=1}^{m} \frac{1}{1-\mathfrak{q}^n}. 
\end{align}
The expression (\ref{hm&m}) implies that the half-BPS local operators of dimension $n$ which 
additionally appear at a junction of rank-$m$ (anti)symmetric Wilson line operators are one-to-one with partitions of $n$ into at most $m$ parts. 
It follows that
\begin{align}
\langle \mathcal{W}_{(m)}\mathcal{W}_{\overline{m}}\rangle^{U(\infty)}_{\textrm{$\frac12$BPS}}(\mathfrak{q}^{-1})
&=(-\mathfrak{q})^m \langle \mathcal{W}_{(m)}\mathcal{W}_{\overline{(1^m)}}\rangle^{U(\infty)}_{\textrm{$\frac12$BPS}}(\mathfrak{q}). 
\end{align}

In the large representation limit $m\rightarrow \infty$, we find
\begin{align}
\lim_{m\rightarrow \infty}
\langle \mathcal{W}_{(m)}\mathcal{W}_{\overline{(m)}}\rangle^{U(\infty)}_{\textrm{$\frac12$BPS}}(\mathfrak{q})
&=\prod_{n=1}^{\infty}\frac{1}{1-\mathfrak{q}^n}, \\
\lim_{m\rightarrow \infty}
\langle \mathcal{W}_{(m)}\mathcal{W}_{\overline{(1)^m}}\rangle^{U(\infty)}_{\textrm{$\frac12$BPS}}(\mathfrak{q})
&=0. 
\end{align} 

The haif-BPS limit of the large $N$ higher-point functions vanishes for $|\mathfrak{q}|<1$, 
however, certain linear combinations of the correlation functions can be finite.  
For example the half-BPS limit of the large $N$ 4-point functions of (anti)symmetric Wilson line operators obeys
\begin{align}
\label{hm&m&m&m_diff}
&
\langle \mathcal{W}_{(m)}\mathcal{W}_{(m)}\mathcal{W}_{\overline{(m)}}\mathcal{W}_{\overline{(m)}}\rangle^{U(\infty)}_{\textrm{$\frac12$BPS}}(\mathfrak{q})
-\langle \mathcal{W}_{(m-1)}\mathcal{W}_{(m-1)}\mathcal{W}_{\overline{(m-1)}}\mathcal{W}_{\overline{(m-1)}}\rangle^{U(\infty)}_{\textrm{$\frac12$BPS}}(\mathfrak{q})
\nonumber\\
&\equiv \prod_{n=1}^{\infty}\frac{1}{(1-\mathfrak{q})^4}\mod \mathfrak{q}^{m+1}. 
\end{align}

%%%%%%%%%%%%%%%%%%%%%%%%%%%%%%%%%%%%
%%%%%%%%%%%%%%%%%%%%%%%%%%%%%%%%%%%%
\section{Hook representations}
\label{sec_hook}
%%%%%%%%%%%%%%%%%%%%%%%%%%%%%%%%%%%%
%%%%%%%%%%%%%%%%%%%%%%%%%%%%%%%%%%%%
Now we study the Schur line defect correlators of the Wilson line operators in the representations labeled by the Young diagrams of the hook shapes. 
Note that the power sum symmetric function can be decomposed with respect to the Schur functions indexed by the Young diagrams of the hook shapes
\begin{align}
\label{pn_hook}
p_n&=\sum_{k+l=n} (-1)^l s_{(k,1^l)}. 
\end{align}
From the relation (\ref{pn_hook}) as well as the factorization (\ref{fac1}), 
the large $N$ correlation functions of the Wilson line operators in the  hook representations can be expressed in terms of the large $N$ 2-point functions of the charged Wilson line operators. 

%%%%%%%%%%%%%%%%%%%%%%%%%%%%%%%%%%%%
\subsection{$(m,1)$ and $(2,1^{m-1})$}
%%%%%%%%%%%%%%%%%%%%%%%%%%%%%%%%%%%%
%1/2BPS correlators (m,1)
Consider the Wilson line operator in the hook representation labeled by the partition $\lambda$ $=$ $(m,1)$ and its conjugate $(2,1^{m-1})$ with $m>2$. 

For example, for $m=3$ we have 
\begin{align}
&
\langle \mathcal{W}_{\tiny \yng(3,1)} \mathcal{W}_{\overline{\tiny \yng(3,1)}}\rangle^{U(\infty)}
=\langle \mathcal{W}_{\tiny \yng(2,1,1)} \mathcal{W}_{\overline{\tiny \yng(2,1,1)}}\rangle^{U(\infty)}
\nonumber\\
&=\frac38 {\langle \mathcal{W}_{1}\mathcal{W}_{-1}\rangle^{U(\infty)}}^{4}
+\frac18  {\langle \mathcal{W}_{1}\mathcal{W}_{-1}\rangle^{U(\infty)}}^{2}
{\langle \mathcal{W}_{2}\mathcal{W}_{-2}\rangle^{U(\infty)}}
\nonumber\\
&+\frac{1}{32} {\langle \mathcal{W}_{2}\mathcal{W}_{-2}\rangle^{U(\infty)}}^{2}
+\frac{1}{16} {\langle \mathcal{W}_{4}\mathcal{W}_{-4}\rangle^{U(\infty)}}, \\
&
\langle \mathcal{W}_{\tiny \yng(3,1)} \mathcal{W}_{\overline{\tiny \yng(2,1,1)}}\rangle^{U(\infty)}
\nonumber\\
&=\frac38 {\langle \mathcal{W}_{1}\mathcal{W}_{-1}\rangle^{U(\infty)}}^{4}
-\frac18  {\langle \mathcal{W}_{1}\mathcal{W}_{-1}\rangle^{U(\infty)}}^{2}
{\langle \mathcal{W}_{2}\mathcal{W}_{-2}\rangle^{U(\infty)}}
\nonumber\\
&+\frac{1}{32} {\langle \mathcal{W}_{2}\mathcal{W}_{-2}\rangle^{U(\infty)}}^{2}
-\frac{1}{16} {\langle \mathcal{W}_{4}\mathcal{W}_{-4}\rangle^{U(\infty)}}. 
\end{align}
The Wilson line operators in the hook representations $\tiny \yng(3,1)$ and $\tiny \yng(2,1,1)$ also 
have correlators with those in the (anti)symmetric representations
\begin{align}
&
\langle \mathcal{W}_{\tiny \yng(3,1)} \mathcal{W}_{\overline{\tiny \yng(4)}}\rangle^{U(\infty)}
=\langle \mathcal{W}_{\tiny \yng(2,1,1)} \mathcal{W}_{\overline{\tiny \yng(1,1,1,1)}}\rangle^{U(\infty)}
\nonumber\\
&=\frac{1}{8} {\langle \mathcal{W}_{1}\mathcal{W}_{-1}\rangle^{U(\infty)}}^{4}
+\frac{1}{8}  {\langle \mathcal{W}_{1}\mathcal{W}_{-1}\rangle^{U(\infty)}}^{2}
{\langle \mathcal{W}_{2}\mathcal{W}_{-2}\rangle^{U(\infty)}}
\nonumber\\
&-\frac{1}{32} {\langle \mathcal{W}_{2}\mathcal{W}_{-2}\rangle^{U(\infty)}}^{2}
-\frac{1}{16} {\langle \mathcal{W}_{4}\mathcal{W}_{-4}\rangle^{U(\infty)}}, \\
&
\langle \mathcal{W}_{\tiny \yng(3,1)} \mathcal{W}_{\overline{\tiny \yng(1,1,1,1)}}\rangle^{U(\infty)}
=\langle \mathcal{W}_{\tiny \yng(2,1,1)} \mathcal{W}_{\overline{\tiny \yng(4)}}\rangle^{U(\infty)}
\nonumber\\
&=\frac{1}{8} {\langle \mathcal{W}_{1}\mathcal{W}_{-1}\rangle^{U(\infty)}}^{4}
-\frac{1}{8}  {\langle \mathcal{W}_{1}\mathcal{W}_{-1}\rangle^{U(\infty)}}^{2}
{\langle \mathcal{W}_{2}\mathcal{W}_{-2}\rangle^{U(\infty)}}
\nonumber\\
&-\frac{1}{32} {\langle \mathcal{W}_{2}\mathcal{W}_{-2}\rangle^{U(\infty)}}^{2}
+\frac{1}{16} {\langle \mathcal{W}_{4}\mathcal{W}_{-4}\rangle^{U(\infty)}}.
\end{align}
They are related by
\begin{align}
\langle \mathcal{W}_{\tiny \yng(3,1)} \mathcal{W}_{\overline{\tiny \yng(3,1)}}\rangle^{U(\infty)}(t;q^{-1})
&=\langle \mathcal{W}_{\tiny \yng(3,1)} \mathcal{W}_{\overline{\tiny \yng(2,1,1)}}\rangle^{U(\infty)}(t;q), \\
\langle \mathcal{W}_{\tiny \yng(3,1)} \mathcal{W}_{\tiny \yng(4)}\rangle^{U(\infty)}(t;q^{-1})
&=\langle \mathcal{W}_{\tiny \yng(3,1)} \mathcal{W}_{\tiny \yng(1,1,1,1)}\rangle^{U(\infty)}(t;q). 
\end{align}

For $m=4$ we have 
\begin{align}
&
\langle \mathcal{W}_{\tiny \yng(4,1)} \mathcal{W}_{\overline{\tiny \yng(4,1)}}\rangle^{U(\infty)}
=\langle \mathcal{W}_{\tiny \yng(2,1,1,1)} \mathcal{W}_{\overline{\tiny \yng(2,1,1,1)}}\rangle^{U(\infty)}
\nonumber\\
&=\frac{2}{15} {\langle \mathcal{W}_{1}\mathcal{W}_{-1}\rangle^{U(\infty)}}^{5}
+\frac16  {\langle \mathcal{W}_{1}\mathcal{W}_{-1}\rangle^{U(\infty)}}^{3}
{\langle \mathcal{W}_{2}\mathcal{W}_{-2}\rangle^{U(\infty)}}
\nonumber\\
&+\frac{1}{18} {\langle \mathcal{W}_{1}\mathcal{W}_{-1}\rangle^{U(\infty)}}^{2}
{\langle \mathcal{W}_{3}\mathcal{W}_{-3}\rangle^{U(\infty)}}
+\frac{1}{36} {\langle \mathcal{W}_{2}\mathcal{W}_{-2}\rangle^{U(\infty)}}
{\langle \mathcal{W}_{3}\mathcal{W}_{-3}\rangle^{U(\infty)}}
+\frac{1}{25} {\langle \mathcal{W}_{5}\mathcal{W}_{-5}\rangle^{U(\infty)}}, \\
&
\langle \mathcal{W}_{\tiny \yng(4,1)} \mathcal{W}_{\overline{\tiny \yng(2,1,1,1)}}\rangle^{U(\infty)}
\nonumber\\
&=\frac{2}{15} {\langle \mathcal{W}_{1}\mathcal{W}_{-1}\rangle^{U(\infty)}}^{5}
-\frac16  {\langle \mathcal{W}_{1}\mathcal{W}_{-1}\rangle^{U(\infty)}}^{3}
{\langle \mathcal{W}_{2}\mathcal{W}_{-2}\rangle^{U(\infty)}}
\nonumber\\
&+\frac{1}{18} {\langle \mathcal{W}_{1}\mathcal{W}_{-1}\rangle^{U(\infty)}}^{2}
{\langle \mathcal{W}_{3}\mathcal{W}_{-3}\rangle^{U(\infty)}}
-\frac{1}{36} {\langle \mathcal{W}_{2}\mathcal{W}_{-2}\rangle^{U(\infty)}}
{\langle \mathcal{W}_{3}\mathcal{W}_{-3}\rangle^{U(\infty)}}
+\frac{1}{25} {\langle \mathcal{W}_{5}\mathcal{W}_{-5}\rangle^{U(\infty)}}. 
\end{align}
Also we have the correlation functions of the Wilson line operators in the hook representations $\tiny \yng(4,1)$, $\tiny \yng(2,1,1,1)$ 
and those in the (anti)symmetric representations 
\begin{align}
&
\langle \mathcal{W}_{\tiny \yng(4,1)} \mathcal{W}_{\overline{\tiny \yng(5)}}\rangle^{U(\infty)}
=\langle \mathcal{W}_{\tiny \yng(2,1,1,1)} \mathcal{W}_{\overline{\tiny \yng(1,1,1,1,1)}}\rangle^{U(\infty)}
\nonumber\\
&=\frac{1}{30} {\langle \mathcal{W}_{1}\mathcal{W}_{-1}\rangle^{U(\infty)}}^{5}
+\frac{1}{12}  {\langle \mathcal{W}_{1}\mathcal{W}_{-1}\rangle^{U(\infty)}}^{3}
{\langle \mathcal{W}_{2}\mathcal{W}_{-2}\rangle^{U(\infty)}}
\nonumber\\
&+\frac{1}{18} {\langle \mathcal{W}_{1}\mathcal{W}_{-1}\rangle^{U(\infty)}}^{2}
{\langle \mathcal{W}_{3}\mathcal{W}_{-3}\rangle^{U(\infty)}}
-\frac{1}{36} {\langle \mathcal{W}_{2}\mathcal{W}_{-2}\rangle^{U(\infty)}}
{\langle \mathcal{W}_{3}\mathcal{W}_{-3}\rangle^{U(\infty)}}
-\frac{1}{25}
{\langle \mathcal{W}_{5}\mathcal{W}_{-5}\rangle^{U(\infty)}}, \\
&
\langle \mathcal{W}_{\tiny \yng(4,1)} \mathcal{W}_{\overline{\tiny \yng(1,1,1,1,1)}}\rangle^{U(\infty)}
=\langle \mathcal{W}_{\tiny \yng(2,1,1,1)} \mathcal{W}_{\overline{\tiny \yng(5)}}\rangle^{U(\infty)}
\nonumber\\
&=\frac{1}{30} {\langle \mathcal{W}_{1}\mathcal{W}_{-1}\rangle^{U(\infty)}}^{5}
-\frac{1}{12}  {\langle \mathcal{W}_{1}\mathcal{W}_{-1}\rangle^{U(\infty)}}^{3}
{\langle \mathcal{W}_{2}\mathcal{W}_{-2}\rangle^{U(\infty)}}
\nonumber\\
&+\frac{1}{18} {\langle \mathcal{W}_{1}\mathcal{W}_{-1}\rangle^{U(\infty)}}^{2}
{\langle \mathcal{W}_{3}\mathcal{W}_{-3}\rangle^{U(\infty)}}
+\frac{1}{36} {\langle \mathcal{W}_{2}\mathcal{W}_{-2}\rangle^{U(\infty)}}
{\langle \mathcal{W}_{3}\mathcal{W}_{-3}\rangle^{U(\infty)}}
-\frac{1}{25}
{\langle \mathcal{W}_{5}\mathcal{W}_{-5}\rangle^{U(\infty)}}. 
\end{align}
It follows that
\begin{align}
\langle \mathcal{W}_{\tiny \yng(4,1)} \mathcal{W}_{\overline{\tiny \yng(4,1)}}\rangle^{U(\infty)}(t;q^{-1})
&=- \langle \mathcal{W}_{\tiny \yng(4,1)} \mathcal{W}_{\overline{\tiny \yng(2,1,1,1)}}\rangle^{U(\infty)}(t;q), \\
\langle \mathcal{W}_{\tiny \yng(4,1)} \mathcal{W}_{\overline{\tiny \yng(5)}}\rangle^{U(\infty)}(t;q^{-1})
&=- \langle \mathcal{W}_{\tiny \yng(4,1)} \mathcal{W}_{\overline{\tiny \yng(1,1,1,1,1)}}\rangle^{U(\infty)}(t;q). 
\end{align}

%%%%%%%%%%%%%%%%%%%%%%%%%%%%%%%%%%%%
\subsubsection{Unflavored limit}
%%%%%%%%%%%%%%%%%%%%%%%%%%%%%%%%%%%%
%unflavored correlators (3,1)
For the partition $\tiny \yng(3,1)$ and its conjugate $\tiny \yng(2,1,1)$, 
the large $N$ unflavored 2-point functions are evaluated as
\begin{align}
\label{u31&31}
&
\langle \mathcal{W}_{\tiny \yng(3,1)} \mathcal{W}_{\overline{\tiny \yng(3,1)}}\rangle^{U(\infty)}
=\langle \mathcal{W}_{\tiny \yng(2,1,1)} \mathcal{W}_{\overline{\tiny \yng(2,1,1)}}\rangle^{U(\infty)}
\nonumber\\
&=\frac{1+2q^{\frac12}+7q+10q^{\frac32}+12q^2+8q^{\frac52}+6q^3+2q^{\frac72}}
{(1-q^{\frac12})^2 (1-q) (1-q^2)}, \\
\label{u31&211}
&
\langle \mathcal{W}_{\tiny \yng(3,1)} \mathcal{W}_{\overline{\tiny \yng(2,1,1)}}\rangle^{U(\infty)}
=\frac{q^{\frac12} (2+6q^{\frac12}+8q+12q^{\frac32}+10q^2+7q^{\frac52}+2q^3+q^{\frac72})}
{(1-q^{\frac12}) (1-q) (1-q^2)}. 
\end{align}
The large $N$ correlation functions with the rank-$4$ (anti)symmetric Wilson line operators are
\begin{align}
\label{u31&4}
&
\langle \mathcal{W}_{\tiny \yng(3,1)} \mathcal{W}_{\tiny \yng(4)}\rangle^{U(\infty)}(q)
=\langle \mathcal{W}_{\tiny \yng(2,1,1)} \mathcal{W}_{\tiny \yng(1,1,1,1)}\rangle^{U(\infty)}(q)
\nonumber\\
&=\frac{q^{\frac12} (2+2q^{\frac12}+4q+3q^{\frac32}+4q^2+q^{\frac52})}
{(1-q^{\frac12})^2 (1-q) (1-q^2)}, \\
&
\label{u31&1111}
\langle \mathcal{W}_{\tiny \yng(3,1)} \mathcal{W}_{\tiny \yng(1,1,1,1)}\rangle^{U(\infty)}(q)
=\langle \mathcal{W}_{\tiny \yng(2,1,1)} \mathcal{W}_{\tiny \yng(4)}\rangle^{U(\infty)}(q)
\nonumber\\
&=\frac{q(1+4q^{\frac12}+3q+4q^{\frac32}+2q^2+2q^{\frac52})}
{(1-q^{\frac12})^2 (1-q) (1-q^{2})}. 
\end{align}

%unflavored correlators (4,1)
For the partition $\tiny \yng(4,1)$ and its conjugate $\tiny \yng(2,1,1,1)$, we obtain
\begin{align}
\label{u41&41}
&
\langle \mathcal{W}_{\tiny \yng(4,1)} \mathcal{W}_{\overline{\tiny \yng(4,1)}}\rangle^{U(\infty)}
=\langle \mathcal{W}_{\tiny \yng(2,1,1,1)} \mathcal{W}_{\overline{\tiny \yng(2,1,1,1)}}\rangle^{U(\infty)}
\nonumber\\
&=\frac{1+2q^{\frac12}+7q+13q^{\frac32}+20q^2+24q^{\frac52}+22q^3+19q^{\frac72}
+13q^4+6q^{\frac92}+q^5}
{(1-q^{\frac12})^2 (1-q) (1-q^{\frac32}) (1-q^{\frac52})}, \\
\label{u31&211}
&
\langle \mathcal{W}_{\tiny \yng(4,1)} \mathcal{W}_{\overline{\tiny \yng(2,1,1,1)}}\rangle^{U(\infty)}
\nonumber\\
&=\frac{q (1+6q^{\frac12}+13q+19q^{\frac32}+22q^2+24q^{\frac52}+20q^3+13q^{\frac72}+7q^4+2q^{\frac92}+q^5)}
{(1-q^{\frac12})^2 (1-q) (1-q^{\frac32}) (1-q^{\frac52})}. 
\end{align}
Also we find the correlators with the rank-$5$ (anti)symmetric Wilson line operators
\begin{align}
\label{u41&5}
&
\langle \mathcal{W}_{\tiny \yng(4,1)} \mathcal{W}_{\tiny \yng(5)}\rangle^{U(\infty)}(q)
=\langle \mathcal{W}_{\tiny \yng(2,1,1,1)} \mathcal{W}_{\tiny \yng(1,1,1,1,1)}\rangle^{U(\infty)}(q)
\nonumber\\
&=\frac{2 q^{\frac12} (1+q^{\frac12}+2q+3q^{\frac32}+3q^2+3q^{\frac52}+2q^{3}+q^{\frac72})}
{(1-q^{\frac12})^2 (1-q) (1-q^{\frac32}) (1-q^{\frac52})}, \\
&
\label{u41&11111}
\langle \mathcal{W}_{\tiny \yng(4,1)} \mathcal{W}_{\tiny \yng(1,1,1,1,1)}\rangle^{U(\infty)}(q)
=\langle \mathcal{W}_{\tiny \yng(2,1,1,1)} \mathcal{W}_{\tiny \yng(5)}\rangle^{U(\infty)}(q)
\nonumber\\
&=\frac{2q^2(1+2q^{\frac12}+3q+3q^{\frac32}+3q^2+2q^{\frac52}+q^3+q^{\frac92})}
{(1-q^{\frac12})^2 (1-q) (1-q^{\frac32}) (1-q^{\frac52})}. 
\end{align}

%unflavored correlators (m,1)
For general $m$, we can write 
\begin{align}
&
\langle \mathcal{W}_{(m,1)} \mathcal{W}_{\overline{(m,1)}}\rangle^{U(\infty)}(q)
=\langle \mathcal{W}_{(2,1^{m-1})} \mathcal{W}_{\overline{(2,1^{m-1})}}\rangle^{U(\infty)}(q)
\nonumber\\
&=\frac{G_{\{(m,1), (m,1)\}}(q)}
{(1-q^{\frac12}) (1-q^{\frac{m+1}{2}})}
\prod_{n=1}^{m-1}
\frac{1}{1-q^{\frac{n}{2}}}, \\
&
\langle \mathcal{W}_{(m,1)} \mathcal{W}_{\overline{(m+1)}}\rangle^{U(\infty)}(q)
=\langle \mathcal{W}_{(2,1^{m-1})} \mathcal{W}_{\overline{(1^{m+1})}}\rangle^{U(\infty)}(q)
\nonumber\\
&=\frac{G_{\{(m,1), (m+1)\}}(q)}
{(1-q^{\frac12}) (1-q^{\frac{m+1}{2}})}
\prod_{n=1}^{m-1}
\frac{1}{1-q^{\frac{n}{2}}}, 
\end{align}
where $G_{\{(m,1), (m,1)\}}(q)$ and $G_{\{(m,1), (m+1)\}}$ are polynomials in $q$ with positive integer coefficients. 
We find that 
\begin{align}
G_{\{(m,1), (m,1)\}}(q)&\equiv \frac{1+3q}{1-q^{\frac12}} \prod_{n=1}^{\infty}\frac{1}{1-q^{\frac{n}{2}}}\mod q^{\frac{m}{2}}, \\
G_{\{(m,1), (m+1)\}}(q)&\equiv 2q^{\frac12} \prod_{n=1}^{\infty}\frac{1}{1-q^{\frac{n}{2}}}\mod q^{\frac{m+1}{2}}. 
\end{align}

In the large $m$ limit, we obtain 
\begin{align}
\label{large_um1&m1}
\lim_{m\rightarrow \infty} 
\langle \mathcal{W}_{(m,1)} \mathcal{W}_{\overline{(m,1)}}\rangle^{U(\infty)}(q)
&=\frac{1+3q}{(1-q^{\frac12})^2} 
\prod_{n=1}^{\infty} \frac{1}{(1-q^{\frac{n}{2}})^2}, \\
\label{large_um1&m}
\lim_{m\rightarrow \infty} 
\langle \mathcal{W}_{(m,1)} \mathcal{W}_{\overline{(m+1)}}\rangle^{U(\infty)}(q)
&=\frac{2q^{\frac12}}{1-q^{\frac12}} 
\prod_{n=1}^{\infty} \frac{1}{(1-q^{\frac{n}{2}})^2}. 
\end{align}

%%%%%%%%%%%%%%%%%%%%%%%%%%%%%%%%%%%%
\subsubsection{Half-BPS limit}
%%%%%%%%%%%%%%%%%%%%%%%%%%%%%%%%%%%%

%1/2BPS correlators (3,1)
The half-BPS limit of the 2-point functions of the Wilson line operators associated with the partition  
$\tiny \yng(3,1)$ and its conjugate $\tiny \yng(2,1,1)$ is evaluated as
\begin{align}
\label{h31&31}
&
\langle \mathcal{W}_{\tiny \yng(3,1)} \mathcal{W}_{\overline{\tiny \yng(3,1)}}\rangle^{U(\infty)}_{\textrm{$\frac12$BPS}}(\mathfrak{q})
=\langle \mathcal{W}_{\tiny \yng(2,1,1)} \mathcal{W}_{\overline{\tiny \yng(2,1,1)}}\rangle^{U(\infty)}_{\textrm{$\frac12$BPS}}(\mathfrak{q})
\nonumber\\
&=\frac{1+\mathfrak{q}^2+\mathfrak{q}^3}{(1-\mathfrak{q})^2 (1-\mathfrak{q}^2) (1-\mathfrak{q}^4)}, \\
\label{h31&211}
&
\langle \mathcal{W}_{\tiny \yng(3,1)} \mathcal{W}_{\overline{\tiny \yng(4)}}\rangle^{U(\infty)}_{\textrm{$\frac12$BPS}}(\mathfrak{q})
=\langle \mathcal{W}_{\tiny \yng(2,1,1)} \mathcal{W}_{\overline{\tiny \yng(1,1,1,1)}}\rangle^{U(\infty)}_{\textrm{$\frac12$BPS}}(\mathfrak{q})
\nonumber\\
&=\frac{\mathfrak{q}}{(1-\mathfrak{q})^2 (1-\mathfrak{q}^2) (1-\mathfrak{q}^4)}. 
\end{align}

%1/2BPS correlators (4,1)
For the partition $\tiny \yng(4,1)$ and its conjugate $\tiny \yng(2,1,1,1)$, we obtain
\begin{align}
\label{h41&41}
&
\langle \mathcal{W}_{\tiny \yng(4,1)} \mathcal{W}_{\overline{\tiny \yng(4,1)}}\rangle^{U(\infty)}_{\textrm{$\frac12$BPS}}(\mathfrak{q})
=\langle \mathcal{W}_{\tiny \yng(2,1,1,1)} \mathcal{W}_{\overline{\tiny \yng(2,1,1,1)}}\rangle^{U(\infty)}_{\textrm{$\frac12$BPS}}(\mathfrak{q})
\nonumber\\
&=\frac{1+\mathfrak{q}^2+\mathfrak{q}^3+\mathfrak{q}^4}
{(1-\mathfrak{q})^2 (1-\mathfrak{q}^2) (1-\mathfrak{q}^3) (1-\mathfrak{q}^5)}, \\
\label{h41&2111}
&
\langle \mathcal{W}_{\tiny \yng(4,1)} \mathcal{W}_{\overline{\tiny \yng(5)}}\rangle^{U(\infty)}_{\textrm{$\frac12$BPS}}(\mathfrak{q})
\nonumber\\
&=\frac{\mathfrak{q}}{(1-\mathfrak{q})^2 (1-\mathfrak{q}^2) (1-\mathfrak{q}^3) (1-\mathfrak{q}^5)}. 
\end{align}

The half-BPS limit of the large $N$ 2-point function of the Wilson line operators is given by
\begin{align}
\label{hm1&m1}
\langle \mathcal{W}_{(m,1)} \mathcal{W}_{\overline{(m,1)}}\rangle^{U(\infty)}_{\textrm{$\frac12$BPS}}(\mathfrak{q})
&=\langle \mathcal{W}_{(2,1^{m-1})} \mathcal{W}_{\overline{(2,1^{m-1})}}\rangle^{U(\infty)}_{\textrm{$\frac12$BPS}}(\mathfrak{q})
\nonumber\\
&=
\frac{1-\mathfrak{q}+\mathfrak{q}^2-\mathfrak{q}^{m+1}}{(1-\mathfrak{q})^2 (1-\mathfrak{q}^{m+1})}
\prod_{n=1}^{m-1}
\frac{1}{1-\mathfrak{q}^n}, 
\end{align}
\begin{align}
\label{hm1&21^m-1}
\langle \mathcal{W}_{(m,1)} \mathcal{W}_{\overline{(2,1^{m-1})}}\rangle^{U(\infty)}_{\textrm{$\frac12$BPS}}(\mathfrak{q})
&=
\mathfrak{q}^{\frac{(m-2) (m-1)}{2}}
\frac{1-\mathfrak{q}^{m-1}+\mathfrak{q}^m-\mathfrak{q}^{m+1}}
{(1-\mathfrak{q})^2 (1-\mathfrak{q}^{m+1})}
\prod_{n=1}^{m-1}
\frac{1}{1-\mathfrak{q}^{n}}. 
\end{align}
The correlation functions (\ref{hm1&m1}) and (\ref{hm1&21^m-1}) are related under the transformation $\mathfrak{q}$ $\rightarrow$ $\mathfrak{q}^{-1}$
\begin{align}
\langle \mathcal{W}_{(m,1)} \mathcal{W}_{\overline{(m,1)}}\rangle^{U(\infty)}_{\textrm{$\frac12$BPS}}(\mathfrak{q}^{-1})
&=(-\mathfrak{q})^{m+1}
\langle \mathcal{W}_{(m,1)} \mathcal{W}_{\overline{(2,1^{m-1})}}\rangle^{U(\infty)}_{\textrm{$\frac12$BPS}}(\mathfrak{q}). 
\end{align}
In the large representation limit $m\rightarrow \infty$, we find 
\begin{align}
\label{large_hm1&m1}
\lim_{m\rightarrow \infty}
\langle \mathcal{W}_{(m,1)} \mathcal{W}_{\overline{(m,1)}}\rangle^{U(\infty)}_{\textrm{$\frac12$BPS}}(\mathfrak{q})
&=\frac{1-\mathfrak{q}+\mathfrak{q}^2}{(1-\mathfrak{q})^2}
\prod_{n=1}^{\infty}\frac{1}{1-\mathfrak{q}^n}, \\
\lim_{m\rightarrow \infty}
\langle \mathcal{W}_{(m,1)} \mathcal{W}_{\overline{(2,1^{m-1})}}\rangle^{U(\infty)}_{\textrm{$\frac12$BPS}}(\mathfrak{q})
&=0. 
\end{align}

They also have non-trivial correlation functions with the (anti)symmetric Wilson line operators. 
We find 
\begin{align}
\label{hm1&m}
\langle \mathcal{W}_{(m,1)} \mathcal{W}_{\overline{(m+1)}}\rangle^{U(\infty)}_{\textrm{$\frac12$BPS}}(\mathfrak{q})
&=\frac{\mathfrak{q}}{(1-\mathfrak{q}) (1-\mathfrak{q}^{m+1})}
\prod_{n=1}^{m-1}\frac{1}{1-\mathfrak{q}^{n}}, \\
\label{hm1&1^m}
\langle \mathcal{W}_{(m,1)} \mathcal{W}_{\overline{(1^{m+1})}}\rangle^{U(\infty)}_{\textrm{$\frac12$BPS}}(\mathfrak{q})
&=\frac{\mathfrak{q}^{\frac{m(m-1)}{2}}}{(1-\mathfrak{q}) (1-\mathfrak{q}^{m+1})}
\prod_{n=1}^{m-1}\frac{1}{1-\mathfrak{q}^{n}}. 
\end{align}

In the large representation limit $m\rightarrow \infty$, we obtain
\begin{align}
\label{large_hm1&m}
&\lim_{m\rightarrow \infty}
\langle \mathcal{W}_{(m,1)} \mathcal{W}_{\overline{(m+1)}}\rangle^{U(\infty)}_{\textrm{$\frac12$BPS}}(\mathfrak{q})
=\frac{\mathfrak{q}}{1-\mathfrak{q}}\prod_{n=1}^{\infty}
\frac{1}{1-\mathfrak{q}^n}
\nonumber\\
&=\sum_{n=1}^{\infty} a_{\{(\infty,1),(\infty+1)\}}^{(H)}(n) \mathfrak{q}^{n}
\nonumber\\
&=\mathfrak{q}+2\mathfrak{q}^2+4\mathfrak{q}^3+7\mathfrak{q}^4+12\mathfrak{q}^5
+19\mathfrak{q}^6+30\mathfrak{q}^7+45\mathfrak{q}^8+67\mathfrak{q}^9+97\mathfrak{q}^{10}+\cdots
\end{align}
and
\begin{align}
&\lim_{m\rightarrow \infty}
\langle \mathcal{W}_{(m,1)} \mathcal{W}_{\overline{(1^{m+1})}}\rangle^{U(\infty)}_{\textrm{$\frac12$BPS}}(\mathfrak{q})
=0. 
\end{align}
The function (\ref{large_hm1&m}) is known as the generating function for the sums of the numbers of partitions  
or that for the numbers of $1$'s appearing in all the partitions of $n$ \cite{MR1442260}. 
This implies that 
\begin{align}
a_{\{(\infty,1), (\infty+1)\}}^{(H)}(n)&=\sum_{k=0}^{n-1}p(k). 
\end{align}

The asymptotic growth of the coefficient $a_{\{(\infty,1), (\infty+1)\}}^{(H)}(n)$ is given by \cite{MR3181090}
\begin{align}
a_{\{(\infty,1), (\infty+1)\}}^{(H)}(n)&\sim 
\frac{1}{2\sqrt{2}\pi n^{\frac12}}
\left(
1-\frac{13\sqrt{2}\pi}{48\sqrt{3}n^{\frac12}}
\right)
\exp\left(
\sqrt{ \frac{2}{3}} \pi n^{\frac12}
\right). 
\end{align}

The actual coefficients and the analytic values are given by
\begin{align}
\begin{array}{c|c|c}
n&a_{\{ (\infty,1), (\infty+1) \}}^{(H)}(n)&a_{\{ (\infty,1), (\infty+1) \} \textrm{asym.}}^{(H)}(n)\\ \hline 
10&97&92.5503 \\
100&1.45242\times 10^{9}&1.44538\times 10^{9} \\
1000&5.88705\times 10^{32}&5.88442\times 10^{32} \\
10000&2.81293\times 10^{108}&2.81281\times 10^{108} \\
\end{array}. 
\end{align}

\subsubsection{Reconstructing flavored correlators}
In the previous two subsections, we found analytic expressions in the large representation limit of the large $N$ correlators for the unflavored case and the half-BPS case. To interpolate these two limits smoothly, we can guess analytic expressions for the flavored case. For example, from \eqref{large_um1&m1} and \eqref{large_hm1&m1}, we find
\begin{align}
&\lim_{m \to \infty} \langle \mathcal{W}_{(m,1)} \mathcal{W}_{\overline{(m,1)}}\rangle^{U(\infty)}(t;q)\\
&=\frac{1-(t^2+t^{-2})q^{\frac{1}{2}}+(t^2+t^{-2})^2 q-3(t^2+t^{-2})q^{\frac{3}{2}}+3q^2}{(1-q^{\frac{1}{2}}t^2)^2(1-q^{\frac{1}{2}}t^{-2})^2}
\prod_{n=1}^\infty \frac{1}{(1-q^{\frac{n}{2}}t^{2n})(1-q^{\frac{n}{2}}t^{-2n})}.\notag
\end{align}
Once we obtain such an analytic form, we can easily check its validity by comparing the $q$-series from the factorization property. 
Similarly, from \eqref{large_um1&m} and \eqref{large_hm1&m}, we find
\begin{align}
\lim_{m \to \infty} \langle \mathcal{W}_{(m,1)} \mathcal{W}_{\overline{(m+1)}}\rangle^{U(\infty)}(t;q)
=\frac{(t^2+t^{-2})q^{\frac{1}{2}}-2q}{(1-q^{\frac{1}{2}}t^2)(1-q^{\frac{1}{2}}t^{-2})}
\prod_{n=1}^\infty \frac{1}{(1-q^{\frac{n}{2}}t^{2n})(1-q^{\frac{n}{2}}t^{-2n})}.
\end{align}

%%%%%%%%%%%%%%%%%%%%%%%%%%%%%%%%%%%%
\subsection{$(m,1^2)$ and $(3,1^{m-1})$}
%%%%%%%%%%%%%%%%%%%%%%%%%%%%%%%%%%%%
%1/2BPS correlators (m,1^2)
Next consider the Wilson line operators in the hook representations labeled by 
the partition $\lambda$ $=$ $(m,1^2)$ and its conjugate $\lambda'$ $=$ $(3,1^{m-1})$ for $m>3$. 

For $m=4$ we have
\begin{align}
&
\langle \mathcal{W}_{\tiny \yng(4,1,1)} \mathcal{W}_{\overline{\tiny \yng(4,1,1)}}\rangle^{U(\infty)}
\nonumber\\
&=\frac{5}{36} {\langle \mathcal{W}_{1} \mathcal{W}_{-1}\rangle^{U(\infty)}}^{6}
+\frac{1}{24} {\langle \mathcal{W}_{1} \mathcal{W}_{-1}\rangle^{U(\infty)}}^{4} {\langle \mathcal{W}_{2} \mathcal{W}_{-2}\rangle^{U(\infty)}}
\nonumber\\
&
+\frac{1}{16} {\langle \mathcal{W}_{1} \mathcal{W}_{-1}\rangle^{U(\infty)}}^{2}  {\langle \mathcal{W}_{2} \mathcal{W}_{-2}\rangle^{U(\infty)}}^{2}
+\frac{1}{96} {\langle \mathcal{W}_{2} \mathcal{W}_{-2}\rangle^{U(\infty)}}^{3}
+\frac{1}{54} {\langle \mathcal{W}_{1} \mathcal{W}_{-1}\rangle^{U(\infty)}}^{3} {\langle \mathcal{W}_{3} \mathcal{W}_{-3}\rangle^{U(\infty)}}
\nonumber\\
&+\frac{1}{36} {\langle \mathcal{W}_{1} \mathcal{W}_{-1}\rangle^{U(\infty)}}
{\langle \mathcal{W}_{2} \mathcal{W}_{-2}\rangle^{U(\infty)}}
{\langle \mathcal{W}_{3} \mathcal{W}_{-3}\rangle^{U(\infty)}}
+\frac{1}{162}{\langle \mathcal{W}_{3} \mathcal{W}_{-3}\rangle^{U(\infty)}}^2
+\frac{1}{36}{\langle \mathcal{W}_{6} \mathcal{W}_{-6}\rangle^{U(\infty)}}, \\
&
\langle \mathcal{W}_{\tiny \yng(4,1,1)} \mathcal{W}_{\overline{\tiny \yng(3,1,1,1)}}\rangle^{U(\infty)}
\nonumber\\
&=\frac{5}{36} {\langle \mathcal{W}_{1} \mathcal{W}_{-1}\rangle^{U(\infty)}}^{6}
-\frac{1}{24} {\langle \mathcal{W}_{1} \mathcal{W}_{-1}\rangle^{U(\infty)}}^{4} {\langle \mathcal{W}_{2} \mathcal{W}_{-2}\rangle^{U(\infty)}}
\nonumber\\
&+\frac{1}{16} {\langle \mathcal{W}_{1} \mathcal{W}_{-1}\rangle^{U(\infty)}}^{2}  {\langle \mathcal{W}_{2} \mathcal{W}_{-2}\rangle^{U(\infty)}}^{2}
-\frac{1}{96} {\langle \mathcal{W}_{2} \mathcal{W}_{-2}\rangle^{U(\infty)}}^{3}
+\frac{1}{54} {\langle \mathcal{W}_{1} \mathcal{W}_{-1}\rangle^{U(\infty)}}^{3} {\langle \mathcal{W}_{3} \mathcal{W}_{-3}\rangle^{U(\infty)}}
\nonumber\\
&
-\frac{1}{36} {\langle \mathcal{W}_{1} \mathcal{W}_{-1}\rangle^{U(\infty)}}
{\langle \mathcal{W}_{2} \mathcal{W}_{-2}\rangle^{U(\infty)}}
{\langle \mathcal{W}_{3} \mathcal{W}_{-3}\rangle^{U(\infty)}}
+\frac{1}{162}{\langle \mathcal{W}_{3} \mathcal{W}_{-3}\rangle^{U(\infty)}}^2
-\frac{1}{36}{\langle \mathcal{W}_{6} \mathcal{W}_{-6}\rangle^{U(\infty)}}. 
\end{align}
The large $N$ normalized correlators obey the transformation law
\begin{align}
\langle \mathcal{W}_{\tiny \yng(4,1,1)} \mathcal{W}_{\overline{\tiny \yng(4,1,1)}}\rangle^{U(\infty)}(t;q^{-1})
&=\langle \mathcal{W}_{\tiny \yng(4,1,1)} \mathcal{W}_{\overline{\tiny \yng(3,1,1,1)}}\rangle^{U(\infty)}(t;q). 
\end{align}

The Wilson line operators in these hook representations also have correlation functions 
with those in the (anti)symmetric representations. 
It follows that 
\begin{align}
&
\langle \mathcal{W}_{\tiny \yng(4,1,1)} \mathcal{W}_{\overline{\tiny \yng(6)}}\rangle^{U(\infty)}
\nonumber\\
&=\frac{1}{72} {\langle \mathcal{W}_{1} \mathcal{W}_{-1}\rangle^{U(\infty)}}^{6}
+\frac{1}{48} {\langle \mathcal{W}_{1} \mathcal{W}_{-1}\rangle^{U(\infty)}}^{4}{\langle \mathcal{W}_{2} \mathcal{W}_{-2}\rangle^{U(\infty)}}
-\frac{1}{32} {\langle \mathcal{W}_{1} \mathcal{W}_{-1}\rangle^{U(\infty)}}^{2}{\langle \mathcal{W}_{2} \mathcal{W}_{-2}\rangle^{U(\infty)}}^{2}
\nonumber\\
&
-\frac{1}{192} {\langle \mathcal{W}_{2} \mathcal{W}_{-2}\rangle^{U(\infty)}}^{3}
+\frac{1}{54} {\langle \mathcal{W}_{1} \mathcal{W}_{-1}\rangle^{U(\infty)}}^{3}{\langle \mathcal{W}_{3} \mathcal{W}_{-3}\rangle^{U(\infty)}}
\nonumber\\
&-\frac{1}{36} {\langle \mathcal{W}_{1} \mathcal{W}_{-1}\rangle^{U(\infty)}}
{\langle \mathcal{W}_{2} \mathcal{W}_{-2}\rangle^{U(\infty)}}
{\langle \mathcal{W}_{3} \mathcal{W}_{-3}\rangle^{U(\infty)}}
+\frac{1}{162}{\langle \mathcal{W}_{3} \mathcal{W}_{-3}\rangle^{U(\infty)}}^2
+\frac{1}{36}{\langle \mathcal{W}_{6} \mathcal{W}_{-6}\rangle^{U(\infty)}}, \\
&
\langle \mathcal{W}_{\tiny \yng(4,1,1)} \mathcal{W}_{\overline{\tiny \yng(1,1,1,1,1,1)}}\rangle^{U(\infty)}
\nonumber\\
&=\frac{1}{72} {\langle \mathcal{W}_{1} \mathcal{W}_{-1}\rangle^{U(\infty)}}^{6}
-\frac{1}{48} {\langle \mathcal{W}_{1} \mathcal{W}_{-1}\rangle^{U(\infty)}}^{4}{\langle \mathcal{W}_{2} \mathcal{W}_{-2}\rangle^{U(\infty)}}
-\frac{1}{32} {\langle \mathcal{W}_{1} \mathcal{W}_{-1}\rangle^{U(\infty)}}^{2}{\langle \mathcal{W}_{2} \mathcal{W}_{-2}\rangle^{U(\infty)}}^{2}
\nonumber\\
&
+\frac{1}{192} {\langle \mathcal{W}_{2} \mathcal{W}_{-2}\rangle^{U(\infty)}}^{3}
+\frac{1}{54} {\langle \mathcal{W}_{1} \mathcal{W}_{-1}\rangle^{U(\infty)}}^{3}{\langle \mathcal{W}_{3} \mathcal{W}_{-3}\rangle^{U(\infty)}}
\nonumber\\
&+\frac{1}{36} {\langle \mathcal{W}_{1} \mathcal{W}_{-1}\rangle^{U(\infty)}}
{\langle \mathcal{W}_{2} \mathcal{W}_{-2}\rangle^{U(\infty)}}
{\langle \mathcal{W}_{3} \mathcal{W}_{-3}\rangle^{U(\infty)}}
+\frac{1}{162}{\langle \mathcal{W}_{3} \mathcal{W}_{-3}\rangle^{U(\infty)}}^2
-\frac{1}{36}{\langle \mathcal{W}_{6} \mathcal{W}_{-6}\rangle^{U(\infty)}}. 
\end{align}

We find that 
they are related by the transformation $q$ $\rightarrow$ $q^{-1}$
\begin{align}
\langle \mathcal{W}_{\tiny \yng(4,1,1)} \mathcal{W}_{\overline{\tiny \yng(6)}}\rangle^{U(\infty)}(t;q^{-1})
&=
\langle \mathcal{W}_{\tiny \yng(4,1,1)} \mathcal{W}_{\overline{\tiny \yng(1,1,1,1,1,1)}}\rangle^{U(\infty)}(t;q). 
\end{align}

%%%%%%%%%%%%%%%%%%%%%%%%%%%%%%%%%%%%
\subsubsection{Unflavored limit}
%%%%%%%%%%%%%%%%%%%%%%%%%%%%%%%%%%%%
The large $N$ unflavored 2-point functions of the Wilson line operators associated with the partitions $\tiny \yng(4,1,1)$ and $\tiny \yng(3,1,1,1)$ are 
\begin{align}
\label{u411&411}
&
\langle \mathcal{W}_{\tiny \yng(4,1,1)} \mathcal{W}_{\overline{\tiny \yng(4,1,1)}}\rangle^{U(\infty)}(q)
=\langle \mathcal{W}_{\tiny \yng(3,1,1,1)} \mathcal{W}_{\overline{\tiny \yng(3,1,1,1)}}\rangle^{U(\infty)}(q)
\nonumber\\
&=\frac{1}
{(1-q^{\frac12})^2 (1-q)^2 (1-q^{\frac32}) (1-q^{3})}
\Bigl(
1+2q^{\frac12}+9q+23q^{\frac32}+45q^2+67q^{\frac52}+91q^3
\nonumber\\
&+95q^{\frac72}+98q^4+83q^{\frac92}+64q^5+36q^{\frac{11}{2}}+18q^6+6q^{\frac{13}{2}}+2q^7
\Bigr), \\
\label{u411&3111}
&
\langle \mathcal{W}_{\tiny \yng(4,1,1)} \mathcal{W}_{\overline{\tiny \yng(3,1,1,1)}}\rangle^{U(\infty)}(q)
\nonumber\\
&=\frac{q^{\frac12}}
{(1-q^{\frac12})^2 (1-q)^2 (1-q^{\frac32}) (1-q^{3})}
\Bigl(
2+6q^{\frac12}+18q+36q^{\frac32}+64q^2+83q^{\frac52}
\nonumber\\
&+98q^{3}+95q^{\frac72}+91q^4+67q^{\frac92}+45q^5+23q^{\frac{11}{2}}+9q^6+2q^{\frac{13}{2}}+q^7
\Bigr). 
\end{align}
Also we obtain
\begin{align}
\label{u411&6}
&
\langle \mathcal{W}_{\tiny \yng(4,1,1)} \mathcal{W}_{\overline{\tiny \yng(6)}}\rangle^{U(\infty)}(q)
=\langle \mathcal{W}_{\tiny \yng(3,1,1,1)} \mathcal{W}_{\overline{\tiny \yng(1,1,1,1,1,1)}}\rangle^{U(\infty)}(q)
\nonumber\\
&=\frac{q}
{(1-q^{\frac12})^2 (1-q)^2 (1-q^{\frac32}) (1-q^{3})}
\Bigl(
1+4q^{\frac12}+5q+9q^{\frac32}+10q^2+11q^{\frac52}+8q^3
\nonumber\\
&+10q^{\frac72}+4q^4+2q^{\frac92}
\Bigr), \\
&
\label{u411&111111}
\langle \mathcal{W}_{\tiny \yng(4,1,1)} \mathcal{W}_{\overline{\tiny \yng(1,1,1,1,1,1)}}\rangle^{U(\infty)}(q)
=\langle \mathcal{W}_{\tiny \yng(3,1,1,1)} \mathcal{W}_{\overline{\tiny \yng(6)}}\rangle^{U(\infty)}(q)
\nonumber\\
&=\frac{q^2}
{(1-q^{\frac12})^2 (1-q)^2 (1-q^{\frac32}) (1-q^{3})}
\Bigl(
2+4q^{\frac12}+10q+8q^{\frac32}+11q^2+10q^{\frac52}+9q^3
\nonumber\\
&+5q^{\frac72}+4q^4+q^{\frac92}
\Bigr). 
\end{align}

For general $m$, the correlators can be written as
\begin{align}
&\langle \mathcal{W}_{(m,1^2)} \mathcal{W}_{\overline{(m,1^2)}}\rangle^{U(\infty)}(q)
=\langle \mathcal{W}_{(3,1^{m-1})} \mathcal{W}_{\overline{(3,1^{m-1})}}\rangle^{U(\infty)}(q)
\nonumber\\
&=\frac{G_{\{(m,1^2), (m,1^2)\}} (q)}
{(1-q^{\frac12}) (1-q) (1-q^{\frac{m+2}{2}})}
\prod_{n=1}^{m-1}\frac{1}{1-q^{\frac{n}{2}}}, \\
&\langle \mathcal{W}_{(m,1^2)} \mathcal{W}_{\overline{(m+2)}}\rangle^{U(\infty)}(q)
=\langle \mathcal{W}_{(3,1^{m-1})} \mathcal{W}_{\overline{(1^{m+2})}}\rangle^{U(\infty)}(q)
\nonumber\\
&=\frac{G_{\{(m,1^2), (m+2)\}} (q)}
{(1-q^{\frac12}) (1-q) (1-q^{\frac{m+2}{2}})}
\prod_{n=1}^{m-1}\frac{1}{1-q^{\frac{n}{2}}}, 
\end{align}
where $G_{\{(m,1^2) (m,1^2)\}}(q)$ and $G_{\{(m,1^2) (m+2)\}}(q)$ are polynomials in $q$ with positive integer coefficients. 
It follows that 
\begin{align}
G_{\{(m,1^2) (m,1^2)\}}(q)
&\equiv \frac{1+4q+6q^{\frac32}+5q^3}{(1-q^{\frac12}) (1-q)}
\prod_{n=1}^{\infty}\frac{1}{1-q^{\frac{n}{2}}}\mod q^{\frac{m}{2}}, \\
G_{\{(m,1^2) (m+2)\}}(q)
&\equiv q(1+3q^{\frac12})
\prod_{n=1}^{\infty}\frac{1}{1-q^{\frac{n}{2}}}\mod q^{\frac{m+2}{2}}. 
\end{align}

In the large $m$ limit, we find
\begin{align}
\label{large_um11&m11}
\lim_{m\rightarrow \infty}
\langle \mathcal{W}_{(m,1^2)} \mathcal{W}_{\overline{(m,1^2)}}\rangle^{U(\infty)}(q)
&=\frac{1+4q+6q^{\frac32}+5q^3}
{(1-q^{\frac12})^2 (1-q)^2}
\prod_{n=1}^{\infty}\frac{1}{(1-q^{\frac{n}{2}})^2}, \\
\label{large_um11&mp2}
\lim_{m\rightarrow \infty}
\langle \mathcal{W}_{(m,1^2)} \mathcal{W}_{\overline{(m+2)}}\rangle^{U(\infty)}(q)
&=\frac{q(1+3q^{\frac12})}{(1-q^{\frac12}) (1-q)}
\prod_{n=1}^{\infty}\frac{1}{(1-q^{\frac{n}{2}})^2}. 
\end{align}

%%%%%%%%%%%%%%%%%%%%%%%%%%%%%%%%%%%%
\subsubsection{Half-BPS limit}
%%%%%%%%%%%%%%%%%%%%%%%%%%%%%%%%%%%%
The half-BPS limit of the large $N$ 2-point functions is given by
\begin{align}
\label{hm11&m11}
&
\langle \mathcal{W}_{(m,1^2)} \mathcal{W}_{\overline{(m,1^2)}}\rangle^{U(\infty)}_{\textrm{$\frac12$BPS}}(\mathfrak{q})
=\langle \mathcal{W}_{(3,1^{m-1})} \mathcal{W}_{\overline{(3,1^{m-1})}}\rangle^{U(\infty)}_{\textrm{$\frac12$BPS}}(\mathfrak{q})
\nonumber\\
&=\frac{1-\mathfrak{q}+2\mathfrak{q}^3-\mathfrak{q}^4-\mathfrak{q}^5+\mathfrak{q}^6
-\mathfrak{q}^{m+1}-\mathfrak{q}^{m+2}+\mathfrak{q}^{m+3}-\mathfrak{q}^{m+5}+\mathfrak{q}^{2m+3}}
{(1-\mathfrak{q})^2 (1-\mathfrak{q}^2)^2 (1-\mathfrak{q}^{m+2})}
\prod_{n=1}^{m-1}
\frac{1}{1-\mathfrak{q}^n}, \\
\label{hm11&31^m-1}
&
\langle \mathcal{W}_{(m,1^2)} \mathcal{W}_{\overline{(3,1^{m-1})}}\rangle^{U(\infty)}_{\textrm{$\frac12$BPS}}(\mathfrak{q})
\nonumber\\
&=
\frac{1-\mathfrak{q}^{m-2}+\mathfrak{q}^m-\mathfrak{q}^{m+1}-\mathfrak{q}^{m+2}
+\mathfrak{q}^{2m-3}-\mathfrak{q}^{2m-2}-\mathfrak{q}^{2m-1}+2\mathfrak{q}^{2m}-\mathfrak{q}^{2m+2}+\mathfrak{q}^{2m+3}
}
{
(1-\mathfrak{q})^2 (1-\mathfrak{q}^2)^2 (1-\mathfrak{q}^{m+2})
}
\nonumber\\
&\times 
\mathfrak{q}^{\frac{(m-2) (m-3)}{2}}
\prod_{n=1}^{m-1}\frac{1}{1-\mathfrak{q}^n}. 
\end{align}
The correlation functions (\ref{hm11&m11}) and (\ref{hm11&31^m-1}) are related by the transformation $\mathfrak{q}$ $\rightarrow$ $\mathfrak{q}^{-1}$
\begin{align}
\langle \mathcal{W}_{(m,1^2)} \mathcal{W}_{\overline{(m,1^2)}}\rangle^{U(\infty)}_{\textrm{$\frac12$BPS}}(\mathfrak{q}^{-1})
&=(-\mathfrak{q})^{m+2} 
\langle \mathcal{W}_{(m,1^2)} \mathcal{W}_{\overline{(3,1^{m-1})}}\rangle^{U(\infty)}_{\textrm{$\frac12$BPS}}(\mathfrak{q}). 
\end{align}
In the large representation limit $m\rightarrow \infty$, we obtain
\begin{align}
\label{large_hm11&m11}
\lim_{m\rightarrow \infty}
\langle \mathcal{W}_{(m,1^2)} \mathcal{W}_{\overline{(m,1^2)}}\rangle^{U(\infty)}_{\textrm{$\frac12$BPS}}(\mathfrak{q})
&=\frac{1-\mathfrak{q}+2\mathfrak{q}^3-\mathfrak{q}^4-\mathfrak{q}^5+\mathfrak{q}^6}
{(1-\mathfrak{q})^2 (1-\mathfrak{q}^2)^2}
\prod_{n=1}^{\infty}\frac{1}{1-\mathfrak{q}^n}, \\
\label{large_hm11&31^m-1}
\lim_{m\rightarrow \infty}
\langle \mathcal{W}_{(m,1^2)} \mathcal{W}_{\overline{(3,1^{m-1})}}\rangle^{U(\infty)}_{\textrm{$\frac12$BPS}}(\mathfrak{q})&=0. 
\end{align}

Also they have non-trivial correlators with the Wilson line operators 
labeled by the partitions $(m)$, $(1^{m})$, $(m+1,1)$ and $(2,1^{m})$
\begin{align}
\label{hm11&mp2}
&
\langle \mathcal{W}_{(m,1^2)} \mathcal{W}_{\overline{(m+2)}}\rangle^{U(\infty)}_{\textrm{$\frac12$BPS}}(\mathfrak{q})
=\langle \mathcal{W}_{(3,1^{m-1})} \mathcal{W}_{\overline{(1^{m+2})}}\rangle^{U(\infty)}_{\textrm{$\frac12$BPS}}(\mathfrak{q})
\nonumber\\
&=\frac{\mathfrak{q}^3}{(1-\mathfrak{q}) (1-\mathfrak{q}^2) (1-\mathfrak{q}^{m+2})}
\prod_{n=1}^{m-1}\frac{1}{1-\mathfrak{q}^n}, \\
&
\langle \mathcal{W}_{(m,1^2)} \mathcal{W}_{\overline{(1^{m+2})}}\rangle^{U(\infty)}_{\textrm{$\frac12$BPS}}(\mathfrak{q})
=\langle \mathcal{W}_{(3,1^{m-1})} \mathcal{W}_{\overline{(m+2)}}\rangle^{U(\infty)}_{\textrm{$\frac12$BPS}}(\mathfrak{q})
\nonumber\\
&=\frac{\mathfrak{q}^{\frac{m(m-1)}{2}}}{(1-\mathfrak{q}) (1-\mathfrak{q}^2) (1-\mathfrak{q}^{m+2})}
\prod_{n=1}^{m-1}\frac{1}{1-\mathfrak{q}^n}
, \\
&
\langle \mathcal{W}_{(m,1^2)} \mathcal{W}_{\overline{(m+1,1)}}\rangle^{U(\infty)}_{\textrm{$\frac12$BPS}}(\mathfrak{q})
=\langle \mathcal{W}_{(3,1^{m-1})} \mathcal{W}_{\overline{(2,1^{m})}}\rangle^{U(\infty)}_{\textrm{$\frac12$BPS}}(\mathfrak{q})
\nonumber\\
&=\frac{\mathfrak{q} (1-\mathfrak{q}^2+\mathfrak{q}^3-\mathfrak{q}^{m+2})}
{(1-\mathfrak{q})^2 (1-\mathfrak{q}^2) (1-\mathfrak{q}^{m+2})}
\prod_{n=1}^{m-1}\frac{1}{1-\mathfrak{q}^n}, \\
&
\langle \mathcal{W}_{(m,1^2)} \mathcal{W}_{\overline{(2,1^{m})}}\rangle^{U(\infty)}_{\textrm{$\frac12$BPS}}(\mathfrak{q})
=\langle \mathcal{W}_{(3,1^{m-1})} \mathcal{W}_{\overline{(m+1,1)}}\rangle^{U(\infty)}_{\textrm{$\frac12$BPS}}(\mathfrak{q})
\nonumber\\
&=\mathfrak{q}^{\frac{(m-1) (m-2)}{2}} \frac{1-\mathfrak{q}^{m-1}+\mathfrak{q}^{m}-\mathfrak{q}^{m+2}}
{(1-\mathfrak{q})^2 (1-\mathfrak{q}^2) (1-\mathfrak{q}^{m+2})}
\prod_{n=1}^{m-1}\frac{1}{1-\mathfrak{q}^n}. 
\end{align}
They obey
\begin{align}
\langle \mathcal{W}_{(m,1^2)} \mathcal{W}_{\overline{(m+2)}}\rangle^{U(\infty)}_{\textrm{$\frac12$BPS}}(\mathfrak{q}^{-1})
&=(-\mathfrak{q})^{m+2}
\langle \mathcal{W}_{(m,1^2)} \mathcal{W}_{\overline{(1^{m+2})}}\rangle^{U(\infty)}_{\textrm{$\frac12$BPS}}(\mathfrak{q}), \\
\langle \mathcal{W}_{(m,1^2)} \mathcal{W}_{\overline{(m+1,1)}}\rangle^{U(\infty)}_{\textrm{$\frac12$BPS}}(\mathfrak{q}^{-1})
&=(-\mathfrak{q})^{m+2}
\langle \mathcal{W}_{(m,1^2)} \mathcal{W}_{\overline{(2,1^{m})}}\rangle^{U(\infty)}_{\textrm{$\frac12$BPS}}(\mathfrak{q}). 
\end{align}

In the large representation limit $m\rightarrow \infty$, we find
\begin{align}
\label{large_hm11&mp2}
&
\lim_{m\rightarrow \infty}
\langle \mathcal{W}_{(m,1^2)} \mathcal{W}_{\overline{(m+2)}}\rangle^{U(\infty)}_{\textrm{$\frac12$BPS}}(\mathfrak{q})
=\frac{\mathfrak{q}^3}{(1-\mathfrak{q}) (1-\mathfrak{q}^2)} 
\prod_{n=1}^{\infty} \frac{1}{1-\mathfrak{q}^n}
\nonumber\\
&=\mathfrak{q}^3 \sum_{n=0}^{\infty}a_{\{(\infty,1^2), (\infty+2) \}}^{(H)}(n) \mathfrak{q}^{n}
\nonumber\\
&=
\mathfrak{q}^3
(1+2\mathfrak{q}+5\mathfrak{q}^2+9\mathfrak{q}^3+17\mathfrak{q}^4+28\mathfrak{q}^5+47\mathfrak{q}^6+73\mathfrak{q}^7
+114\mathfrak{q}^8+170\mathfrak{q}^9+253\mathfrak{q}^{10}+\cdots), \\
&\lim_{m\rightarrow \infty}
\langle \mathcal{W}_{(m,1^2)} \mathcal{W}_{\overline{(1^{m+2})}}\rangle^{U(\infty)}_{\textrm{$\frac12$BPS}}(\mathfrak{q})=0, \\
&
\lim_{m\rightarrow \infty}
\langle \mathcal{W}_{(m,1^2)} \mathcal{W}_{\overline{(m+1,1)}}\rangle^{U(\infty)}_{\textrm{$\frac12$BPS}}(\mathfrak{q})
=\frac{\mathfrak{q} (1-\mathfrak{q}^2+\mathfrak{q}^3)}{(1-\mathfrak{q})^2 (1-\mathfrak{q}^2)}
\prod_{n=1}^{\infty}\frac{1}{1-\mathfrak{q}^n},\\
&
\lim_{m\rightarrow \infty}
\langle \mathcal{W}_{(m,1^2)} \mathcal{W}_{\overline{(2,1^{m})}}\rangle^{U(\infty)}_{\textrm{$\frac12$BPS}}(\mathfrak{q})
=0. 
\end{align}
The function (\ref{large_hm11&mp2}) is the generating function for the numbers $g(n)$ of pairs of partitions of $n$ which differ by $1$ box \cite{MR2718277}. 
It follows that
\begin{align}
a_{\{(\infty,1^2), (\infty+2)\}}^{(H)}(n)&=g(n+2). 
\end{align}

\subsubsection{Reconstructing flavored correlators}
We can also reconstruct the flavored correlators from special limits.
From \eqref{large_um11&m11}, \eqref{large_um11&mp2}, \eqref{large_hm11&m11} and \eqref{large_hm11&mp2}, we find the following analytic expressions:
\begin{align}
&\lim_{m\rightarrow \infty}
\langle \mathcal{W}_{(m,1^2)} \mathcal{W}_{\overline{(m,1^2)}}\rangle^{U(\infty)}(t;q)\\
&=\frac{P_{\{ (\infty,1^2), (\infty,1^2)\}}(t;q)}{(1-q^{\frac{1}{2}}t^2)^2(1-q^{\frac{1}{2}}t^{-2})^2(1-qt^4)^2(1-qt^{-4})^2} \prod_{n=1}^\infty \frac{1}{(1-q^{\frac{n}{2}}t^{2n})(1-q^{\frac{n}{2}}t^{-2n})},\notag \\
&\lim_{m\rightarrow \infty}
\langle \mathcal{W}_{(m,1^2)} \mathcal{W}_{\overline{(m+2)}}\rangle^{U(\infty)}(t;q)\\
&=\frac{P_{\{ (\infty,1^2), (\infty+2)\}}(t;q)}{(1-q^{\frac{1}{2}}t^2)(1-q^{\frac{1}{2}}t^{-2})(1-qt^4)(1-qt^{-4})} \prod_{n=1}^\infty \frac{1}{(1-q^{\frac{n}{2}}t^{2n})(1-q^{\frac{n}{2}}t^{-2n})}, \notag
\end{align}
where
\begin{align}
&P_{\{ (\infty,1^2), (\infty,1^2)\}}(t;q)=1-(t^2+t^{-2})q^{\frac{1}{2}}+3q+(t^2+t^{-2})(2t^4-3+2t^{-4})q^\frac{3}{2} \\
&-(t^8+6t^4+3+6t^{-4}+t^{-8})q^2-(t^2+t^{-2})(t^4+t^{-4})(t^4-4+t^{-4})q^\frac{5}{2}\notag \\
&+(t^{12}+3t^8+2t^4+14+2t^{-4}+3t^{-8}+t^{-12})q^3\notag \\ 
&-(t^2+t^{-2})(3t^8-2t^4+10-2t^{-4}+3t^{-8})q^\frac{7}{2}-(8t^4-3+8t^{-4})q^4\notag \\
&+(t^2+t^{-2})(8t^4-3+8t^{-4})q^\frac{9}{2}-5(t^4-1+t^{-4})q^5-5(t^2+t^{-2})q^\frac{11}{2}+5q^6, \notag
\end{align}
and
\begin{align}
P_{\{ (\infty,1^2), (\infty+2)\}}(t;q)=q+(t^6+t^{-6})q^\frac{3}{2}-2(t^4+t^{-4})q^2-(t^2+t^{-2})q^\frac{5}{2}+3q^3.
\end{align}

%%%%%%%%%%%%%%%%%%%%%%%%%%%%%%%%%%%%
\subsection{$(m,1^3)$ and $(4,1^{m-1})$}
%%%%%%%%%%%%%%%%%%%%%%%%%%%%%%%%%%%%
Consider the Wilson line operators in the hook representations labeled by the Young diagram $\lambda=(m,1^3)$ and its conjugate $\lambda'=(4,1^{m-1})$ for $m>4$.  

For example, for $m=5$ we have 
\begin{align}
&
\langle \mathcal{W}_{\tiny \yng(5,1,1,1)} \mathcal{W}_{\overline{\tiny \yng(5,1,1,1)}}\rangle^{U(\infty)}
\nonumber\\
&=\frac{35}{1152} {\langle \mathcal{W}_{1} \mathcal{W}_{-1}\rangle^{U(\infty)}}^{8}
+\frac{5}{576} {\langle \mathcal{W}_{1} \mathcal{W}_{-1}\rangle^{U(\infty)}}^{6}{\langle \mathcal{W}_{2} \mathcal{W}_{-2}\rangle^{U(\infty)}}
\nonumber\\
&+\frac{25}{768} {\langle \mathcal{W}_{1} \mathcal{W}_{-1}\rangle^{U(\infty)}}^{4}{\langle \mathcal{W}_{2} \mathcal{W}_{-2}\rangle^{U(\infty)}}^{2}
+\frac{3}{256} {\langle \mathcal{W}_{1} \mathcal{W}_{-1}\rangle^{U(\infty)}}^{2}{\langle \mathcal{W}_{2} \mathcal{W}_{-2}\rangle^{U(\infty)}}^{3}
\nonumber\\
&+\frac{3}{2048} {\langle \mathcal{W}_{2} \mathcal{W}_{-2}\rangle^{U(\infty)}}^{4}
+\frac{5}{216} {\langle \mathcal{W}_{1} \mathcal{W}_{-1}\rangle^{U(\infty)}}^{5}{\langle \mathcal{W}_{3} \mathcal{W}_{-3}\rangle^{U(\infty)}}
\nonumber\\
&+\frac{1}{216} {\langle \mathcal{W}_{1} \mathcal{W}_{-1}\rangle^{U(\infty)}}^{3}{\langle \mathcal{W}_{2} \mathcal{W}_{-2}\rangle^{U(\infty)}}
{\langle \mathcal{W}_{3} \mathcal{W}_{-3}\rangle^{U(\infty)}}
\nonumber\\
&+\frac{1}{288} {\langle \mathcal{W}_{1} \mathcal{W}_{-1}\rangle^{U(\infty)}}{\langle \mathcal{W}_{2} \mathcal{W}_{-2}\rangle^{U(\infty)}}^{2}
{\langle \mathcal{W}_{3} \mathcal{W}_{-3}\rangle^{U(\infty)}}
\nonumber\\
&+\frac{1}{81} {\langle \mathcal{W}_{1} \mathcal{W}_{-1}\rangle^{U(\infty)}}^{2}{\langle \mathcal{W}_{3} \mathcal{W}_{-3}\rangle^{U(\infty)}}^{2}
+\frac{1}{162} {\langle \mathcal{W}_{2} \mathcal{W}_{-2}\rangle^{U(\infty)}}{\langle \mathcal{W}_{3} \mathcal{W}_{-3}\rangle^{U(\infty)}}^{2}
\nonumber\\
&+\frac{1}{384} {\langle \mathcal{W}_{1} \mathcal{W}_{-1}\rangle^{U(\infty)}}^{4}{\langle \mathcal{W}_{4} \mathcal{W}_{-4}\rangle^{U(\infty)}}
+\frac{1}{128} {\langle \mathcal{W}_{1} \mathcal{W}_{-1}\rangle^{U(\infty)}}^{2}{\langle \mathcal{W}_{2} \mathcal{W}_{-2}\rangle^{U(\infty)}}
{\langle \mathcal{W}_{4} \mathcal{W}_{-4}\rangle^{U(\infty)}}
\nonumber\\
&+\frac{1}{512} {\langle \mathcal{W}_{2} \mathcal{W}_{-2}\rangle^{U(\infty)}}^{4}{\langle \mathcal{W}_{4} \mathcal{W}_{-4}\rangle^{U(\infty)}}
+\frac{1}{144} {\langle \mathcal{W}_{1} \mathcal{W}_{-1}\rangle^{U(\infty)}}{\langle \mathcal{W}_{3} \mathcal{W}_{-3}\rangle^{U(\infty)}}
{\langle \mathcal{W}_{4} \mathcal{W}_{-4}\rangle^{U(\infty)}}
\nonumber\\
&+\frac{1}{512} {\langle \mathcal{W}_{4} \mathcal{W}_{-4}\rangle^{U(\infty)}}^2
+\frac{1}{64} {\langle \mathcal{W}_{8} \mathcal{W}_{-8}\rangle^{U(\infty)}}. 
\end{align}

%%%%%%%%%%%%%%%%%%%%%%%%%%%%%%%%%%%%
\subsubsection{Unflavored limit}
%%%%%%%%%%%%%%%%%%%%%%%%%%%%%%%%%%%%

For $m=5$ the large $N$ unflavored correlation functions are given by
\begin{align}
&\langle \mathcal{W}_{\tiny \yng(5,1,1,1)} \mathcal{W}_{\overline{\tiny \yng(5,1,1,1)}}\rangle^{U(\infty)}(q)
=\langle \mathcal{W}_{\tiny \yng(4,1,1,1,1)} \mathcal{W}_{\overline{\tiny \yng(4,1,1,1,1)}}\rangle^{U(\infty)}(q)
\nonumber\\
&=\frac{1}{(1-q^{\frac12})^2 (1-q)^2 (1-q^{\frac32})^2 (1-q^2) (1-q^4)}
\nonumber\\
&\times 
\Bigl(1+2q^{\frac12}+9q+26q^{\frac32}+65q^2+134q^{\frac52}+249q^{3}+402q^{\frac72}+590q^{4}+754q^{\frac92}+903q^5
\nonumber\\
&+980q^{\frac{11}{2}}+1012q^6+962q^{\frac{13}{2}}+869q^{7}+710q^{\frac{15}{2}}
+536q^{8}+358q^{\frac{17}{2}}+214q^{9}+108q^{\frac{19}{2}}
\nonumber\\
&+50q^{10}+18q^{\frac{21}{2}}+6q^{11}+2q^{\frac{23}{2}}
\Bigr). 
\end{align}

For general $m$, we can write 
\begin{align}
&\langle \mathcal{W}_{(m,1^3)} \mathcal{W}_{\overline{(m,1^3)}} \rangle^{U(\infty)}(q)
=\langle \mathcal{W}_{(4,1^{m-1})} \mathcal{W}_{\overline{(4,1^{m-1})}} \rangle^{U(\infty)}(q)
\nonumber\\
&=\frac{G_{\{ (m,1^3), (m,1^3)\}} (q)}
{(1-q^{\frac12}) (1-q) (1-q^{\frac32} (1-q^{\frac{m+3}{2}})}
\prod_{n=1}^{m-1}\frac{1}{1-q^{\frac{n}{2}}}, \\
&\langle \mathcal{W}_{(m,1^3)} \mathcal{W}_{\overline{(m+3)}} \rangle^{U(\infty)}(q)
=\langle \mathcal{W}_{(4,1^{m-1})} \mathcal{W}_{\overline{(1^{m+3})}} \rangle^{U(\infty)}(q)
\nonumber\\
&=\frac{G_{\{ (m,1^3), (m+3)\}} (q)}
{(1-q^{\frac12}) (1-q) (1-q^{\frac32} (1-q^{\frac{m+3}{2}})}
\prod_{n=1}^{m-1}\frac{1}{1-q^{\frac{n}{2}}}, 
\end{align}
where $G_{\{ (m,1^3), (m,1^3)\}} (q)$ and $G_{\{ (m,1^3), (m+3)\}} (q)$ are polynomials in $q$ with positive integer coefficients. 
We find that  
\begin{align}
&G_{\{ (m,1^3), (m,1^3)\}} (q)
\nonumber\\
&\equiv \frac{1+4q+8q^{\frac32}+10q^2+6q^{\frac52}+8q^3+10q^{\frac72}+10q^4+7q^6}
{(1-q^{\frac12}) (1-q) (1-q^{\frac32})}
\prod_{n=1}^{\infty}\frac{1}{1-q^{\frac{n}{2}}} \mod q^{\frac{m}{2}}, \\
&G_{\{ (m,1^3), (m+3)\}} (q)
\nonumber\\
&\equiv 2q^2(1+q^{\frac12}+2q)\prod_{n=1}^{\infty}\frac{1}{1-q^{\frac{n}{2}}}\mod q^{\frac{m+4}{2}}. 
\end{align}

In the large representation limit $m\rightarrow \infty$, we get
\begin{align}
&
\lim_{m\rightarrow \infty}
\langle \mathcal{W}_{(m,1^3)} \mathcal{W}_{\overline{(m,1^3)}}\rangle^{U(\infty)}(q)
\nonumber\\
\label{large_um111&m111}
&=\frac{1+4q+8q^{\frac32}+10q^2+6q^{\frac52}+8q^3+10q^{\frac72}+10q^4+7q^6}
{(1-q^{\frac12})^2 (1-q)^2 (1-q^{\frac32})^2}
\prod_{n=1}^{\infty}
\frac{1}{(1-q^{\frac{n}{2}})^2}, \\
&
\lim_{m\rightarrow \infty}
\langle \mathcal{W}_{(m,1^3)} \mathcal{W}_{\overline{(m+3)}}\rangle^{U(\infty)}(q)
\nonumber\\
\label{large_um111&mp3}
&=\frac{2q^2 (1+q^{\frac12}+2q)}
{(1-q^{\frac12}) (1-q) (1-q^{\frac32})}
\prod_{n=1}^{\infty}\frac{1}{(1-q^{\frac{n}{2}})^2}. 
\end{align}

%%%%%%%%%%%%%%%%%%%%%%%%%%%%%%%%%%%%
\subsubsection{Half-BPS limit}
%%%%%%%%%%%%%%%%%%%%%%%%%%%%%%%%%%%%
The half-BPS limit of the large $N$ 2-point function of these Wilson line operators takes the form
\begin{align}
\langle \mathcal{W}_{(m,1^3)} \mathcal{W}_{\overline{(m,1^3)}}\rangle^{U(\infty)}_{\textrm{$\frac12$BPS}}(\mathfrak{q})
&=\frac{H_{\{ (m,1^3), (m,1^3) \}} (\mathfrak{q})}{(1-\mathfrak{q}) (1-\mathfrak{q}^2) (1-\mathfrak{q}^3) (1-\mathfrak{q}^{m+3})}
\prod_{n=1}^{m-1}\frac{1}{1-\mathfrak{q}^n}
\end{align} 
where $H_{\{ (m,1^3), (m,1^3) \}} (\mathfrak{q})$ is a polynomial in $\mathfrak{q}$ with positive coefficients. 
For example, we find
\begin{align}
H_{\{ (6,1^3), (6,1^3) \}}
&=1+\mathfrak{q}^2+2\mathfrak{q}^3+3\mathfrak{q}^4+3\mathfrak{q}^5
+4\mathfrak{q}^6+4\mathfrak{q}^7+5\mathfrak{q}^8+5\mathfrak{q}^9+6\mathfrak{q}^{10}
\nonumber\\
&+5\mathfrak{q}^{11}+5\mathfrak{q}^{12}+3\mathfrak{q}^{13}+3\mathfrak{q}^{14}+2\mathfrak{q}^{15}
+2\mathfrak{q}^{16}+\mathfrak{q}^{17}+\mathfrak{q}^{18}. 
\end{align}

Besides, these line operators have the large $N$ correlation functions with the other Wilson line operators, which take simpler forms. 
We find the half-BPS limit of such large $N$ normalized 2-point functions is given by 
\begin{align}
\langle \mathcal{W}_{(m,1^3)} \mathcal{W}_{\overline{(m+3)}}\rangle^{U(\infty)}_{\textrm{$\frac12$BPS}}(\mathfrak{q})
&=\frac{\mathfrak{q}^6}
{(1-\mathfrak{q}) (1-\mathfrak{q}^2) (1-\mathfrak{q}^3) (1-\mathfrak{q}^{m+3})}
\prod_{n=1}^{m-1}
\frac{1}{1-\mathfrak{q}^n}, \\
\langle \mathcal{W}_{(m,1^3)} \mathcal{W}_{\overline{(m+2,1)}}\rangle^{U(\infty)}_{\textrm{$\frac12$BPS}}(\mathfrak{q})
&=\frac{\mathfrak{q}^3 (1-\mathfrak{q}^3+\mathfrak{q}^4-\mathfrak{q}^{m+3})}
{(1-\mathfrak{q})^2 (1-\mathfrak{q}^2) (1-\mathfrak{q}^3) (1-\mathfrak{q}^{m+3})}
\prod_{n=1}^{m-1}\frac{1}{1-\mathfrak{q}^{n}}, 
\end{align}

In the large representation limit $m\rightarrow \infty$, we obtain 
the half-BPS limit of the large $N$ normalized 2-point functions of the Wilson line operators labeled by the partition $\lambda=(m,1^3)$ 
\begin{align}
\label{large_hm111&m111}
\lim_{m\rightarrow \infty} 
\langle \mathcal{W}_{(m,1^3)} \mathcal{W}_{\overline{(m,1^3)}}\rangle^{U(\infty)}_{\textrm{$\frac12$BPS}}(\mathfrak{q})
&=\frac{H_{\{ (\infty,1^3), (\infty,1^3) \}} (\mathfrak{q})}{(1-\mathfrak{q}) (1-\mathfrak{q}^2) (1-\mathfrak{q}^3)}
\prod_{n=1}^{\infty}\frac{1}{1-\mathfrak{q}^n}, 
\end{align}
where 
\begin{align}
&
H_{\{ (\infty,1^3), (\infty,1^3) \}} (\mathfrak{q})
\nonumber\\
&=\frac{1-\mathfrak{q}+\mathfrak{q}^3+\mathfrak{q}^4-\mathfrak{q}^5-2\mathfrak{q}^6+\mathfrak{q}^7
+2\mathfrak{q}^8-\mathfrak{q}^{10}-\mathfrak{q}^{11}+\mathfrak{q}^{12}}
{(1-\mathfrak{q}) (1-\mathfrak{q}^2) (1-\mathfrak{q}^3)}
\nonumber\\
&=\sum_{n=0}^{\infty} a_{\{ (\infty,1^3), (\infty,1^3) \}}^{(H)}(n) \mathfrak{q}^n
\nonumber\\
&=1+\mathfrak{q}^2+2\mathfrak{q}^3+3\mathfrak{q}^4+3\mathfrak{q}^5+4\mathfrak{q}^6
+5\mathfrak{q}^7+7\mathfrak{q}^8+8\mathfrak{q}^9+10\mathfrak{q}^{10}
\nonumber\\
&+11\mathfrak{q}^{11}+14\mathfrak{q}^{12}+15\mathfrak{q}^{13}+18\mathfrak{q}^{14}
+20\mathfrak{q}^{15}+23\mathfrak{q}^{16}+25\mathfrak{q}^{17}+29\mathfrak{q}^{18}+31\mathfrak{q}^{19}+35\mathfrak{q}^{20}+\cdots.
\end{align}
%agrees with the generating function for the numbers of strict integer partitions of $2n$ with reverse-alternating sum being $4$. 

Also we have
\begin{align}
\label{large_hm111&mp3}
\lim_{m\rightarrow \infty}
\langle \mathcal{W}_{(m,1^3)} \mathcal{W}_{\overline{(m+3)}}\rangle^{U(\infty)}_{\textrm{$\frac12$BPS}}(\mathfrak{q})
&=\frac{\mathfrak{q}^6}{(1-\mathfrak{q}) (1-\mathfrak{q}^2) (1-\mathfrak{q}^3)}
\prod_{n=1}^{\infty}\frac{1}{1-\mathfrak{q}^n}, \\
\lim_{m\rightarrow \infty}
\langle \mathcal{W}_{(m,1^3)} \mathcal{W}_{\overline{(m+2,1)}}\rangle^{U(\infty)}_{\textrm{$\frac12$BPS}}(\mathfrak{q})
&=\frac{\mathfrak{q}^3 (1-\mathfrak{q}^3+\mathfrak{q}^4)}
{(1-\mathfrak{q})^2 (1-\mathfrak{q}^2) (1-\mathfrak{q}^3)}
\prod_{n=1}^{\infty}\frac{1}{1-\mathfrak{q}^n}. 
\end{align}

\subsubsection{Reconstructing flavored correlators}
From \eqref{large_um111&m111}, \eqref{large_um111&mp3}, \eqref{large_hm111&m111} and \eqref{large_hm111&mp3}, we can guess the following analytic expressions of the flavored correlators:
\begin{align}
&\lim_{m\rightarrow \infty}
\langle \mathcal{W}_{(m,1^3)} \mathcal{W}_{\overline{(m,1^3)}}\rangle^{U(\infty)}(t;q)\notag\\
&=\frac{P_{\{ (\infty,1^3), (\infty,1^3)\}}(t;q)}{(1-q^{\frac{1}{2}}t^2)^2(1-q^{\frac{1}{2}}t^{-2})^2(1-qt^4)^2(1-qt^{-4})^2(1-q^\frac{3}{2}t^6)^2(1-q^\frac{3}{2}t^{-6})^2}  \notag\\
&\hspace{7truecm}\times \prod_{n=1}^\infty\frac{1}{(1-q^{\frac{n}{2}}t^{2n})(1-q^{\frac{n}{2}}t^{-2n})}, \\
&\lim_{m\rightarrow \infty}
\langle \mathcal{W}_{(m,1^3)} \mathcal{W}_{\overline{(m+3)}}\rangle^{U(\infty)}(t;q)\notag\\
&=\frac{P_{\{ (\infty,1^3), (\infty+3)\}}(t;q)}{(1-q^{\frac{1}{2}}t^2)(1-q^{\frac{1}{2}}t^{-2})(1-qt^4)(1-qt^{-4})(1-q^\frac{3}{2}t^6)(1-q^\frac{3}{2}t^{-6})}  \notag\\
&\hspace{6truecm}\times \prod_{n=1}^\infty \frac{1}{(1-q^{\frac{n}{2}}t^{2n})(1-q^{\frac{n}{2}}t^{-2n})}, 
\end{align}
where $P_{\{ (\infty,1^3), (\infty,1^3)\}}(t;q)$ and $P_{\{ (\infty,1^3), (\infty+3)\}}(t;q)$ are complicated polynomials in $q^\frac{1}{2}$:
\begin{align}
&P_{\{ (\infty,1^3), (\infty,1^3)\}}(t;q)=1-(t^2+t^{-2})q^{\frac{1}{2}}+3q +%(t^6+t^{-6})q^\frac{3}{2} 
\cdots-7(t^2+t^{-2})q^\frac{13}{2}+7q^{12},
\end{align}
and
\begin{align}
P_{\{ (\infty,1^3), (\infty+3)\}}(t;q)=(t^4+t^{-4})q^2+(t^{12}-2+t^{-12})q^3+\cdots+(t^2+t^{-2})q^{11/2}-4q^6.
\end{align}

%%%%%%%%%%%%%%%%%%%%%%%%%%%%%%%%%%%%
\subsection{$(m,1^{m-1})$}
%%%%%%%%%%%%%%%%%%%%%%%%%%%%%%%%%%%%
Let us study the correlation functions of the Wilson line operators labeled by the self-conjugate partition $\lambda=(m,1^{m-1})$ of the hook shape. 
Unlike the previous cases, the both column and row grow as $m$ increases. 

For $m=2$ and $3$, we have 
\begin{align}
&
\langle \mathcal{W}_{\tiny \yng(2,1)} \mathcal{W}_{\overline{\tiny \yng(2,1)}}\rangle^{U(\infty)}
\nonumber\\
&=\frac23 {\langle \mathcal{W}_{1}\mathcal{W}_{-1}\rangle^{U(\infty)}}^{3}
+\frac19 \langle \mathcal{W}_{3}\mathcal{W}_{-3}\rangle^{U(\infty)}, \\
&
\langle \mathcal{W}_{\tiny \yng(3,1,1)} \mathcal{W}_{\overline{\tiny \yng(3,1,1)}}\rangle^{U(\infty)}
\nonumber\\
&=\frac{3}{10}{\langle \mathcal{W}_{1}\mathcal{W}_{-1}\rangle^{U(\infty)}}^{5}
+\frac{1}{8}{\langle \mathcal{W}_{1}\mathcal{W}_{-1}\rangle^{U(\infty)}}
{\langle \mathcal{W}_{2}\mathcal{W}_{-2}\rangle^{U(\infty)}}^{2}
+\frac{1}{25}{\langle \mathcal{W}_{5}\mathcal{W}_{-5}\rangle^{U(\infty)}}. 
\end{align}
Under $q\rightarrow q^{-1}$ the (un)flavored correlator transforms as
\begin{align}
\langle \mathcal{W}_{\tiny \yng(2,1)} \mathcal{W}_{\overline{\tiny \yng(2,1)}}\rangle^{U(\infty)}(t;q^{-1})
&=-\langle \mathcal{W}_{\tiny \yng(2,1)} \mathcal{W}_{\overline{\tiny \yng(2,1)}}\rangle^{U(\infty)}(t;q), \\
\langle \mathcal{W}_{\tiny \yng(3,1,1)} \mathcal{W}_{\overline{\tiny \yng(3,1,1)}}\rangle^{U(\infty)}(t;q^{-1})
&=-\langle \mathcal{W}_{\tiny \yng(3,1,1)} \mathcal{W}_{\overline{\tiny \yng(3,1,1)}}\rangle^{U(\infty)}(t;q). 
\end{align}

The Wilson line operators in the self-conjugate hook representations have correlation functions 
with those in the other hook representations including the (anti)symmetric representations. 
For example, 
\begin{align}
&
\langle \mathcal{W}_{\tiny \yng(2,1)} \mathcal{W}_{\overline{\tiny \yng(3)}}\rangle^{U(\infty)}
=\langle \mathcal{W}_{\tiny \yng(2,1)} \mathcal{W}_{\overline{\tiny \yng(1,1,1)}}\rangle^{U(\infty)}
\nonumber\\
&=\frac13{\langle \mathcal{W}_{1}\mathcal{W}_{-1}\rangle^{U(\infty)}}^{3}
-\frac{1}{9}{\langle \mathcal{W}_{3}\mathcal{W}_{-3}\rangle^{U(\infty)}}, \\
&
\langle \mathcal{W}_{\tiny \yng(3,1,1)} \mathcal{W}_{\overline{\tiny \yng(5)}}\rangle^{U(\infty)}
=\langle \mathcal{W}_{\tiny \yng(3,1,1)} \mathcal{W}_{\overline{\tiny \yng(1,1,1,1,1)}}\rangle^{U(\infty)}
\nonumber\\
&=\frac{1}{20}{\langle \mathcal{W}_{1}\mathcal{W}_{-1}\rangle^{U(\infty)}}^{5}
-\frac{1}{16}{\langle \mathcal{W}_{1}\mathcal{W}_{-1}\rangle^{U(\infty)}}
{\langle \mathcal{W}_{2}\mathcal{W}_{-2}\rangle^{U(\infty)}}^{2}
+\frac{1}{25}{\langle \mathcal{W}_{5}\mathcal{W}_{-5}\rangle^{U(\infty)}}, \\
&
\langle \mathcal{W}_{\tiny \yng(3,1,1)} \mathcal{W}_{\overline{\tiny \yng(4,1)}}\rangle^{U(\infty)}
=\langle \mathcal{W}_{\tiny \yng(3,1,1)} \mathcal{W}_{\overline{\tiny \yng(2,1,1,1)}}\rangle^{U(\infty)}
\nonumber\\
&=\frac15{\langle \mathcal{W}_{1}\mathcal{W}_{-1}\rangle^{U(\infty)}}^{5}
-\frac{1}{25}{\langle \mathcal{W}_{5}\mathcal{W}_{-5}\rangle^{U(\infty)}}. 
\end{align}
They obey 
\begin{align}
\langle \mathcal{W}_{\tiny \yng(2,1)} \mathcal{W}_{\overline{\tiny \yng(3)}}\rangle^{U(\infty)}(t;q^{-1})
&=-\langle \mathcal{W}_{\tiny \yng(2,1)} \mathcal{W}_{\overline{\tiny \yng(3)}}\rangle^{U(\infty)}(t;q), \\
\langle \mathcal{W}_{\tiny \yng(3,1,1)} \mathcal{W}_{\overline{\tiny \yng(4)}}\rangle^{U(\infty)}(t;q^{-1})
&=-\langle \mathcal{W}_{\tiny \yng(3,1,1)} \mathcal{W}_{\overline{\tiny \yng(4)}}\rangle^{U(\infty)}(t;q), \\
\langle \mathcal{W}_{\tiny \yng(3,1,1)} \mathcal{W}_{\overline{\tiny \yng(4,1)}}\rangle^{U(\infty)}(t;q^{-1})
&=-\langle \mathcal{W}_{\tiny \yng(3,1,1)} \mathcal{W}_{\overline{\tiny \yng(4,1)}}\rangle^{U(\infty)}(t;q). 
\end{align}

%%%%%%%%%%%%%%%%%%%%%%%%%%%%%%%%%%%%
\subsubsection{Unflavored limit}
%%%%%%%%%%%%%%%%%%%%%%%%%%%%%%%%%%%%
%unflavored correlators (2,1)
The unflavored large $N$ 2-point functions of the Wilson line operators indexed by the Young diagrams $\tiny \yng(2,1)$ and $\tiny \yng(3,1,1)$ are 
\begin{align}
\label{u21&21}
&
\langle \mathcal{W}_{\tiny \yng(2,1)} \mathcal{W}_{\overline{\tiny \yng(2,1)}}\rangle^{U(\infty)}(q)
\nonumber\\
&=\frac{1+2q^{\frac12}+5q+5q^{\frac32}+2q^2+q^{\frac52}}{(1-q^{\frac12})^2(1-q^{\frac32})}, \\
\label{u311&311}
&
\langle \mathcal{W}_{\tiny \yng(3,1,1)} \mathcal{W}_{\overline{\tiny \yng(3,1,1)}}\rangle^{U(\infty)}(q)
\nonumber\\
&=\frac{1+2q^{\frac12}+9q+20q^{\frac32}+29q^2+35q^{\frac52}
+35q^3+29q^{\frac72}+20q^4+9q^{\frac92}+2q^5+q^{\frac{11}{2}}}
{(1-q^{\frac12})^2 (1-q)^2 (1-q^{\frac52})}. 
\end{align}

For general $m$ the unflavored large $N$ 2-point functions of the Wilson line operators in the self-conjugate hook representations take the from
\begin{align}
\label{um_1^{m-1}&m_1^{m-1}}
\langle \mathcal{W}_{(m,1^{m-1})} \mathcal{W}_{\overline{(m,1^{m-1})}}\rangle^{U(\infty)}(q)
&=G_{\{(m,1^{m-1}),(m,1^{m-1})\}}(q)
\frac{1}{1-q^{\frac{2m-1}{2}}}
\prod_{n=1}^{m-1}\frac{1}{(1-q^{\frac{n}{2}})^2}, 
\end{align}
where $G_{\{(m,1^{m-1}),(m,1^{m-1})\}}(q)$ is a palindromic polynomial in $q$ with positive coefficients. 
For example, when $m=4$ and $m=5$ we have 
\begin{align}
&
G_{\left\{ \tiny\yng(4,1,1,1),\tiny \yng(4,1,1,1)\right\}}(q)
\nonumber\\
&=1+2q^{\frac12}+9q+26q^{\frac32}+61q^2+112q^{\frac52}+182q^3
+253q^{\frac72}+303q^4+331q^{\frac92}
\nonumber\\
&+331q^5+303q^{\frac{11}{2}}+253q^6+182q^{\frac{13}{2}}+112q^7+61q^{\frac{15}{2}}+26q^8+9q^{\frac{17}{2}}+2q^9+q^{\frac{19}{2}}, \\
&
G_{\left\{ \tiny\yng(5,1,1,1,1),\tiny \yng(5,1,1,1,1)\right\}}(q)
\nonumber\\
&=1+2q^{\frac12}+9q+26q^{\frac32}+69q^2+156q^{\frac52}+316q^3+576q^{\frac{7}{2}}
+960q^4+1445q^{\frac92}+1988q^5
\nonumber\\
&+2529q^{\frac{11}{2}}+2998q^6+3333q^{\frac{13}{2}}+3512q^{7}+3512q^{\frac{15}{2}}
+3333q^{8}+2998q^{\frac{17}{2}}+2529q^9
\nonumber\\
&+1988q^{\frac{19}{2}}+1445q^{10}+960q^{\frac{21}{2}}
+576q^{11}+316q^{\frac{23}{2}}+156q^{12}+69q^{\frac{25}{2}}+26q^{13}
\nonumber\\
&+9q^{\frac{27}{2}}+2q^{14}+q^{\frac{29}{2}}. 
\end{align}

In the large $m$ limit, it can be expanded as
\begin{align}
&
\lim_{m\rightarrow \infty}
G_{\{(m,1^{m-1}),(m,1^{m-1})\}}(q)
\nonumber\\
&
=1+2q^{\frac12}+9q+26q^{\frac32}+69q^2+166q^{\frac52}+384q^3+836q^{\frac72}+1762q^4+3570q^{\frac92}+7041q^5
\nonumber\\
&+13500q^{\frac{11}{2}}+25326q^6+46482q^{\frac{13}{2}}+83785q^7+148394q^{\frac{15}{2}}+258840q^8+\cdots, 
\end{align}
where the finite $m$ correction appears from the term with $q^{\frac{m}{2}}$. 
We find that
\begin{align}
\lim_{m\rightarrow \infty}
G_{\{(m,1^{m-1}),(m,1^{m-1})\}}(q)
&=G_{\{(m,1^{m-1}),(m,1^{m-1})\}}(q)+
(2 m)q^{\frac{m}{2}}+\cdots. 
\end{align}

%correlators with Wilson lines in other reps
The large $N$ unflavored 2-point functions of the Wilson line operators labeled by 
$\tiny \yng(2,1)$ and $\tiny \yng(3,1,1)$ and the (anti)symmetric Wilson line operators are evaluated as 
\begin{align}
\label{u21&3}
&
\langle \mathcal{W}_{\tiny \yng(2,1)} \mathcal{W}_{\overline{\tiny \yng(3)}}\rangle^{U(\infty)}(q)
=\langle \mathcal{W}_{\tiny \yng(2,1)} \mathcal{W}_{\overline{\tiny \yng(1,1,1)}}\rangle^{U(\infty)}(q)
\nonumber\\
&=\frac{2q^{\frac12} (1+q^{\frac12}+q+q^{\frac32})}{(1-q^{\frac12})^2 (1-q^{\frac32})}, \\
\label{u311&5}
&
\langle \mathcal{W}_{\tiny \yng(3,1,1)} \mathcal{W}_{\overline{\tiny \yng(5)}}\rangle^{U(\infty)}(q)
=\langle \mathcal{W}_{\tiny \yng(3,1,1)} \mathcal{W}_{\overline{\tiny \yng(1,1,1,1,1)}}\rangle^{U(\infty)}(q)
\nonumber\\
&=\frac{q(1+q^{\frac12}+5q+6q^{\frac32}+6q^2+5q^{\frac52}+4q^3+q^{\frac72})}
{(1-q^{\frac12})^2(1-q)^2(1-q^{\frac52})}, \\
\label{u311&41}
&
\langle \mathcal{W}_{\tiny \yng(3,1,1)} \mathcal{W}_{\overline{\tiny \yng(4,1)}}\rangle^{U(\infty)}(q)
=\langle \mathcal{W}_{\tiny \yng(3,1,1)} \mathcal{W}_{\overline{\tiny \yng(2,1,1,1)}}\rangle^{U(\infty)}(q)
\nonumber\\
&=\frac{2q^{\frac12}
(1+3q^{\frac12}+7q+12q^{\frac32}+16q^2+18q^{\frac52}+16q^3+12q^{\frac72}+7q^4+3q^{\frac92}+q^5)
}
{(1-q^{\frac12})^2 (1-q)^2 (1-q^{\frac52})}. 
\end{align}

%%%%%%%%%%%%%%%%%%%%%%%%%%%%%%%%%%%%
\subsubsection{Ramanujan's general theta functions}
%%%%%%%%%%%%%%%%%%%%%%%%%%%%%%%%%%%%
For general $m$ the large $N$ unflavored 2-point function of the Wilson line operator in the self-conjugate hook representation $(m,1^{m-1})$
and that in the symmetric representation $(2m-1)$ can be expressed as
\begin{align}
\langle \mathcal{W}_{(m,1^{m-1})} \mathcal{W}_{(2m-1)}\rangle^{U(\infty)}(q)
&=G_{\{(m,1^{m-1}),(2m-1)\}}(q)
\frac{1}{1-q^{\frac{2m-1}{2}}}
\prod_{n=1}^{m-1}\frac{1}{(1-q^{\frac{n}{2}})^2}. 
\end{align}
Here $G_{\{(m,1^{m-1}),(2m-1)\}}(q)$ is a palindromic polynomial in $q$. 

When $m=2l-1$ with $l=1,2,\cdots$, we have 
\begin{align}
&
G_{\{(2l-1,1^{2l-2}), (4l-3)\}}(q)
\nonumber\\
&=q^{\frac{l(l-1)}{2}}
(1+4q^{\frac12}+9q+20q^{\frac32}+42q^2+80q^{\frac52}+147q^{3}+260q^{\frac72}+445q^4+\cdots)
\nonumber\\
&\equiv 
q^{\frac{l(l-1)}{2}}
\left[
\sum_{n\in \mathbb{Z}}q^{n^2}\prod_{n=1}^{\infty}\frac{1}{(1-(-q)^{n})^2}\mod q^{\frac{l^2}{2}}
\right]. 
\end{align}
For $m=2l$ with $l=1,2,\cdots$, we find
\begin{align}
&
G_{\{(2l,1^{2l-1}), (4l-1)\}}(q)
\nonumber\\
&=2q^{\frac{l^2}{2}}
(1+2q^{\frac12}+6q+12q^{\frac32}+25q^2+46q^{\frac52}+86q^3+148q^{\frac72}+\cdots)
\nonumber\\
&\equiv 2q^{\frac{l^2}{2}}
\left[
\sum_{n=0}^{\infty}q^{n(n+1)}\prod_{n=1}^{\infty}\frac{1}{(1-(-q)^{n})^2}\mod q^{\frac{l(l+1)}{2}}
\right]. 
\end{align}

Let 
\begin{align}
f(a,b)&:=\sum_{m\in \mathbb{Z}}a^{\frac{m(m+1)}{2}}b^{\frac{m(m-1)}{2}}
\end{align}
be Ramanujan's general theta function with $|ab|<1$ \cite{MR1117903}. 
The special cases of $f(a,b)$ are defined by 
\begin{align}
f(-q)&:=f(-q,-q^2)=\sum_{m\in \mathbb{Z}}(-1)^m q^{\frac{m(3m-1)}{2}}=(q;q)_{\infty}, \\
\varphi(q)&:=f(q,q)=\sum_{m\in \mathbb{Z}}q^{m^2}
=(-q;q^2)^2_{\infty}(q^2;q^2)_{\infty}
=\frac{(-q;-q)_{\infty}}{(q;-q)_{\infty}}, \\
\psi(q)&:=f(q,q^3)=\sum_{m=0}^{\infty}q^{\frac{m(m+1)}{2}}
=\frac{(q^2;q^2)_{\infty}}{(q;q^2)_{\infty}}. 
\end{align}

Then we have 
\begin{align}
G_{\{(2l-1,1^{2l-2}), (4l-3)\}}(q)&\equiv q^{\frac{l(l-1)}{2}} \left[ \frac{\varphi(q^{\frac12})}{f(-q^{\frac12})^{2}} \mod q^{\frac{l^2}{2}}\right], \\
G_{\{(2l,1^{2l-1}), (4l-1)\}}(q)&\equiv 2q^{\frac{l^2}{2}}\left[ \frac{\psi(q)}{f(-q^{\frac12})^2} \mod q^{\frac{l(l+1)}{2}}\right]. 
\end{align}
Therefore in the large representation limit $l\rightarrow \infty$, 
the normalized polynomial $q^{-\frac{l(l-1)}{2}}$ $G_{\{(2l-1,1^{2l-2}), (4l-3)\}}(q)$ 
and $2^{-1}q^{-\frac{l^2}{2}}$ $G_{\{(2l,1^{2l-1}), (4l-1)\}}(q)$ are given by the Ramanujan's general theta functions. 
%%%%%%%%%%%%%%%%%%%%%%%%%%%%%%%%%%%%
\subsubsection{Half-BPS limit}
%%%%%%%%%%%%%%%%%%%%%%%%%%%%%%%%%%%%
In the half-BPS limit the large $N$ correlation functions of the Wilson line operators indexed by 
$\tiny \yng(2,1)$ and $\tiny \yng(3,1,1)$ are given by
%1/2BPS correlators (2,1) (3,1,1)
\begin{align}
\label{h21&21}
\langle \mathcal{W}_{\tiny \yng(2,1)} \mathcal{W}_{\overline{\tiny \yng(2,1)}}\rangle^{U(\infty)}_{\textrm{$\frac12$BPS}}(\mathfrak{q})
&=\frac{1+\mathfrak{q}^2}{(1-\mathfrak{q})^2 (1-\mathfrak{q}^3)}, \\
\label{h311&311}
\langle \mathcal{W}_{\tiny \yng(3,1,1)} \mathcal{W}_{\overline{\tiny \yng(3,1,1)}}\rangle^{U(\infty)}_{\textrm{$\frac12$BPS}}(\mathfrak{q})
&=\frac{1+\mathfrak{q}^2+2\mathfrak{q}^3+\mathfrak{q}^4+\mathfrak{q}^5}
{(1-\mathfrak{q})^2 (1-\mathfrak{q}^2)^2 (1-\mathfrak{q}^5)}. 
\end{align}

For general $m$, 
the half-BPS limit of the large $N$ 2-point functions of the Wilson line operators labeled by the self-conjugate hook Young diagrams can be written as
\begin{align}
\langle \mathcal{W}_{(m,1^{m-1})} \mathcal{W}_{\overline{(m,1^{m-1})}}\rangle_{\textrm{$\frac12$BPS}}^{U(\infty)}(\mathfrak{q})
&=H_{\{(m,1^{m-1}),  (m,1^{m-1})\}}(\mathfrak{q})
\prod_{n=1}^{\infty}
\frac{1}{(1-\mathfrak{q})^2}, 
\end{align}
where $H_{\{(m,1^{m-1}),  (m,1^{m-1})\}}(\mathfrak{q})$ is a palindromic polynomial in $\mathfrak{q}$. 
%H
For example, we have 
\begin{align}
&
H_{\left\{ \tiny{\yng(4,1,1,1)}, \tiny{\yng(4,1,1,1)}\right\}}(\mathfrak{q})
\nonumber\\
&=1+\mathfrak{q}^2+2\mathfrak{q}^3+3\mathfrak{q}^4+2\mathfrak{q}^5+2\mathfrak{q}^6
+2\mathfrak{q}^7+3\mathfrak{q}^8+2\mathfrak{q}^9+\mathfrak{q}^{10}+\mathfrak{q}^{12}, \\
&
H_{\left\{ \tiny{\yng(5,1,1,1,1)}, \tiny{\yng(5,1,1,1,1)}\right\}}(\mathfrak{q})
\nonumber\\
&=1+\mathfrak{q}^2+2\mathfrak{q}^3+3\mathfrak{q}^4+4\mathfrak{q}^5+4\mathfrak{q}^6
+4\mathfrak{q}^7+6\mathfrak{q}^8+6\mathfrak{q}^9+8\mathfrak{q}^{10}
\nonumber\\
&+6\mathfrak{q}^{11}+6\mathfrak{q}^{12}+4\mathfrak{q}^{13}+4\mathfrak{q}^{14}+4\mathfrak{q}^{15}
+3\mathfrak{q}^{16}+2\mathfrak{q}^{17}+\mathfrak{q}^{18}+\mathfrak{q}^{20}. 
\end{align}
We find 
\begin{align}
\label{H_m_1^m-1}
&H_{\{(m, 1^{m-1}),(m, 1^{m-1})\}}(\mathfrak{q})
\nonumber\\
&\equiv \prod_{n=1}^{\infty} \frac{1}{1-\mathfrak{q}^n} \sum_{n=0}^{\infty}(-1)^n \mathfrak{q}^{\frac{n(n+1)}{2}}\mod \mathfrak{q}^{m+1}
\nonumber\\
&\equiv \sum_{n=0}^{\infty}\frac{\mathfrak{q}^{n^2+n}}{(\mathfrak{q};\mathfrak{q})_n^2}\mod \mathfrak{q}^{m+1}. 
\end{align}

%%%%%%%%%%%%%%%%%%%%%%%%%%%%%%%%%%%%
\subsubsection{Non-negative crank partitions}
%%%%%%%%%%%%%%%%%%%%%%%%%%%%%%%%%%%%
In the large representation limit $m\rightarrow \infty$, we find that 
the half-BPS limit of the large $N$ 2-point functions of the Wilson line operators labeled by the partition $\lambda=(m,1^{m-1})$ is given by
\begin{align}
&
\lim_{m\rightarrow \infty}
\langle \mathcal{W}_{(m,1^{m-1})} \mathcal{W}_{\overline{(m,1^{m-1})}}\rangle^{U(\infty)}_{\textrm{$\frac12$BPS}}(\mathfrak{q})
=H_{\{(\infty, 1^{\infty-1}),(\infty, 1^{\infty-1})\}}(\mathfrak{q})\prod_{n=1}^{\infty} \frac{1}{(1-\mathfrak{q}^n)^2}
\nonumber\\
&=\sum_{n=0}^{\infty} a_{\{(\infty,1^{\infty-1}),(\infty,1^{\infty-1})\}}^{(H)}(n)\mathfrak{q}^n
\nonumber\\
&=1+2\mathfrak{q}+6\mathfrak{q}^2+14\mathfrak{q}^3+32\mathfrak{q}^4+66\mathfrak{q}^5
+134\mathfrak{q}^6+256\mathfrak{q}^7+480\mathfrak{q}^8+868\mathfrak{q}^9+\cdots, 
\end{align}
where 
\begin{align}
\label{H_{infty,1^{infty-1}}}
&
H_{\{(\infty,1^{\infty-1}),(\infty,1^{\infty-1})\}}(\mathfrak{q})
=\prod_{n=1}^{\infty} \frac{1}{1-\mathfrak{q}^n} \sum_{n=0}^{\infty}(-1)^n \mathfrak{q}^{\frac{n(n+1)}{2}}
=\sum_{n=0}^{\infty}\frac{\mathfrak{q}^{n^2+n}}{(\mathfrak{q};\mathfrak{q})_{n}^2}
\nonumber\\
&=\sum_{n=0}^{\infty} b_{\{(\infty,1^{\infty-1}),(\infty,1^{\infty-1})\}}^{(H)}(n)\mathfrak{q}^n
\nonumber\\
&=1+\mathfrak{q}^2+2\mathfrak{q}^3+3\mathfrak{q}^4+4\mathfrak{q}^5
+6\mathfrak{q}^6+8\mathfrak{q}^7+12\mathfrak{q}^{8}+16\mathfrak{q}^{9}+23\mathfrak{q}^{10}+\cdots. 
\end{align}
The function (\ref{H_{infty,1^{infty-1}}}) exactly coincides with the generating function for the numbers of partitions with non-negative crank \cite{MR3803977,MR4072556}. 
Here the crank of partition $\lambda$ $=$ $(\lambda_1,\cdots,\lambda_r)$ is defined by \cite{MR3077150,MR929094}
\begin{align}
\textrm{crank}(\lambda)
&=\begin{cases}
\lambda_1&\textrm{if $\omega(\lambda)=0$}\cr
\mu(\lambda)-\omega(\lambda)&\textrm{if $\omega(\lambda)>0$}\cr
\end{cases}, 
\end{align}
where $\omega(\lambda)$ is the number of ones in $\lambda$ 
and $\mu(\lambda)$ is the number of parts of $\lambda$ greater than $\omega(\lambda)$. 
In other words, it is equal to the largest part of the partition if there are no ones as parts 
and otherwise to the number of parts larger than the number of ones minus the number of ones. 
Therefore the coefficient $b_{\{(\infty,1^{\infty-1}),(\infty,1^{\infty-1})\}}(n)$ is equal to the number of partitions of $n$ with non-negative crank. 
\footnote{It can be also viewed as the number of partitions of $n$ with smallest missing part (mex) odd \cite{MR4072556,MR4115761}. }
It is given by \cite{MR4098158}
\begin{align}
b_{\{(\infty,1^{\infty-1}),(\infty,1^{\infty-1})\}}^{(H)}(n)&=\sum_{k=0}^{\infty}(-1)^k p(n-T_k), 
\end{align}
where $p(n)$ is the number of partitions of $n$ and $T_k$ $=$ $k(k+1)/2$ is the triangular number. 
The asymptotic expression for $b_{\{(\infty,1^{\infty-1}),(\infty,1^{\infty-1})\}}(n)$ takes the form \cite{MR45147}
\begin{align}
b_{\{(\infty,1^{\infty-1}),(\infty,1^{\infty-1})\}}^{(H)}(n)&\sim 
\frac{1}{8\sqrt{3}n}\exp
\left(
\sqrt{\frac{2}{3}}\pi n^{\frac12}
\right). 
\end{align}
Making use of the convolution theorem \cite{MR3043606}, we obtain the following asymptotic growth of 
the degeneracy $a_{\{(\infty,1^{\infty-1}),(\infty,1^{\infty-1})\}}^{(H)}(n)$ for the half-BPS limit of the normalized correlator:
\begin{align}
a_{\{(\infty,1^{\infty-1}),(\infty,1^{\infty-1})\}}^{(H)}(n)
&\sim \frac{1}{16\sqrt{2} n^{\frac32}}
\exp\Bigl(
\sqrt{2}\pi n^{\frac12}
\Bigr). 
\end{align}
The actual coefficients and the analytic values are evaluated as
\begin{align}
\begin{array}{c|c|c}
n&a_{\{(\infty,1^{\infty-1}),(\infty,1^{\infty-1})\}}^{(H)}(n)&a_{\{(\infty,1^{\infty-1}),(\infty,1^{\infty-1})\} \textrm{asym.}}^{(H)}(n)\\ \hline 
10&1540&1766.21 \\
100&8.37149\times 10^{14}&8.72088\times 10^{14} \\
1000&1.43424\times 10^{55}&1.45255\times 10^{55} \\
10000&3.94083\times 10^{185}&3.95657\times 10^{185} \\
\end{array}. 
\end{align}

Under $\mathfrak{q}$ $\rightarrow$ $\mathfrak{q}^{-1}$ the correlation function transforms as
\begin{align}
\langle \mathcal{W}_{(m,1^{m-1})} \mathcal{W}_{\overline{(m,1^{m-1})}}\rangle_{\textrm{$\frac12$BPS}}^{U(\infty)}(\mathfrak{q}^{-1})
&=-\mathfrak{q}^{2m-1}
\langle \mathcal{W}_{(m,1^{m-1})} \mathcal{W}_{\overline{(m,1^{m-1})}}\rangle_{\textrm{$\frac12$BPS}}^{U(\infty)}(\mathfrak{q}). 
\end{align}

%1/2BPS correlators (2,1)&(3), (3,1,1)&(5), (3,1,1)&(4,1)
The half-BPS limit of the correlation functions with the Wilson line operators in other representations can be also evaluated. 
For example, for $\tiny \yng(2,1)$ and $\tiny \yng(3,1,1)$, we obtain
\begin{align}
\label{h21&3}
\langle \mathcal{W}_{\tiny \yng(2,1)} \mathcal{W}_{\overline{\tiny \yng(3)}}\rangle^{U(\infty)}_{\textrm{$\frac12$BPS}}(\mathfrak{q})
&=\frac{\mathfrak{q}}{(1-\mathfrak{q})^2 (1-\mathfrak{q}^3)}, \\
\label{h311&5}
\langle \mathcal{W}_{\tiny \yng(3,1,1)} \mathcal{W}_{\overline{\tiny \yng(5)}}\rangle^{U(\infty)}_{\textrm{$\frac12$BPS}}(\mathfrak{q})
&=\frac{\mathfrak{q}^3}{(1-\mathfrak{q})^2 (1-\mathfrak{q}^2)^2 (1-\mathfrak{q}^5)}, \\
\label{h311&41}
\langle \mathcal{W}_{\tiny \yng(3,1,1)} \mathcal{W}_{\overline{\tiny \yng(4,1)}}\rangle^{U(\infty)}_{\textrm{$\frac12$BPS}}(\mathfrak{q})
&=\frac{\mathfrak{q}+\mathfrak{q}^2+\mathfrak{q}^4+\mathfrak{q}^5}
{(1-\mathfrak{q})^2 (1-\mathfrak{q}^2)^2 (1-\mathfrak{q}^5)}. 
\end{align}

For general $m$, we get
\begin{align}
\langle \mathcal{W}_{(m,1^{m-1})} \mathcal{W}_{\overline{(2m-1)}}\rangle^{U(\infty)}_{\textrm{$\frac12$BPS}}(\mathfrak{q})
&=\frac{\mathfrak{q}^{\frac{m(m-1)}{2}}}{1-\mathfrak{q}^{2m-1}}
\prod_{n=1}^{m-1}\frac{1}{(1-\mathfrak{q})^2}, \\
\langle \mathcal{W}_{(m,1^{m-1})} \mathcal{W}_{\overline{(2m-2,1)}}\rangle^{U(\infty)}_{\textrm{$\frac12$BPS}}(\mathfrak{q})
&=
\frac{\mathfrak{q}^{\frac{(m-1) (m-2)}{2}} (1+\mathfrak{q}^m) (1-\mathfrak{q}^{m-1})}
{(1-\mathfrak{q})(1-\mathfrak{q}^{2m-1})}
\prod_{n=1}^{m-1}\frac{1}{(1-\mathfrak{q})^2}. 
\end{align}

%%%%%%%%%%%%%%%%%%%%%%%%%%%%%%%%%%%%
%%%%%%%%%%%%%%%%%%%%%%%%%%%%%%%%%%%%
\section{Rectangular representations}
\label{sec_rect}
%%%%%%%%%%%%%%%%%%%%%%%%%%%%%%%%%%%%
%%%%%%%%%%%%%%%%%%%%%%%%%%%%%%%%%%%%
In this section, we study the large $N$ correlation functions of the Wilson line operators in the representations 
indexed by the Young diagrams of rectangular shapes. 
According to the Jacobi-Trudi identities 
\begin{align}
s_{\lambda}&=\det (h_{\lambda_i+j-i}),& 
s_{\lambda}&=\det (e_{\lambda'_i+j-i}), 
\end{align}
and Newton's identities (\ref{Newton1})-(\ref{Newton2}), 
the Schur functions indexed by the rectangular Young diagrams are expressible in terms of the power sum symmetric functions. 

%%%%%%%%%%%%%%%%%%%%%%%%%%%%%%%%%%%%
\subsection{$(m^2)$ and $(2^m)$}
\label{sec_m^2&2^m}
%%%%%%%%%%%%%%%%%%%%%%%%%%%%%%%%%%%%
Consider the Wilson line operators in the rectangular representations labeled by the partitions $(m^2)$ and $(2^m)$ with $m>2$. 
Simple examples are the cases with $m=3$ and $4$, that is $\tiny \yng(3,3)$, $\tiny \yng(4,4)$, $\tiny \yng(2,2,2)$ and $\tiny \yng(2,2,2,2)$. 
The Schur functions indexed by them are expanded with respect to the power sum symmetric functions as
\begin{align}
\label{schur_3&3}
s_{\tiny \yng(3,3)}
&=\frac{1}{144}p_1^6
+\frac{1}{48}p_1^4p_2
+\frac{1}{16}p_1^2p_2^2
-\frac{1}{16}p_2^3
-\frac{1}{18}p_1^3p_3
\nonumber\\
&+\frac{1}{6}p_1p_2p_3
+\frac{1}{9}p_3^2
-\frac{1}{8}p_1^2p_4
-\frac{1}{8}p_2p_4, \\
\label{schur_3&3}
s_{\tiny \yng(2,2,2)}
&=\frac{1}{144}p_1^6
-\frac{1}{48}p_1^4p_2
+\frac{1}{16}p_1^2p_2^2
+\frac{1}{16}p_2^3
-\frac{1}{18}p_1^3p_3
\nonumber\\
&-\frac{1}{6}p_1p_2p_3
+\frac{1}{9}p_3^2
+\frac{1}{8}p_1^2p_4
-\frac{1}{8}p_2p_4, 
\end{align}
\begin{align}
\label{schur_4&4}
s_{\tiny \yng(4,4)}&=\frac{1}{2880}p_1^8+\frac{1}{360}p_1^6p_2+\frac{1}{96}p_1^4p_2^2
+\frac{1}{64}p_2^4-\frac{1}{360}p_1^5p_3+\frac{1}{36}p_1^3p_2p_3
\nonumber\\
&-\frac{1}{24}p_1p_2^2p_3+\frac{1}{18}p_1^2p_3^2-\frac{1}{18}p_2p_3^2-\frac{1}{48}p_1^4p_4+\frac{1}{16}p_2^2p_4+\frac{1}{12}p_1p_3p_4
\nonumber\\
&+\frac{1}{16}p_4^2-\frac{1}{30}p_1^3p_5-\frac{1}{10}p_1p_2p_5-\frac{1}{15}p_3p_5, \\
\label{schur_2&2&2&2}
s_{\tiny \yng(2,2,2,2)}
&=\frac{1}{2880}p_1^8
-\frac{1}{360}p_1^6p_2
+\frac{1}{96}p_1^4p_2^2
+\frac{1}{64}p_2^4
-\frac{1}{360}p_1^5p_3
-\frac{1}{36}p_1^3p_2p_3
\nonumber\\
&-\frac{1}{24}p_1p_2^2p_3
+\frac{1}{18}p_1^2p_3^2
+\frac{1}{18}p_2p_3^2
+\frac{1}{48}p_1^4p_4
-\frac{1}{16}p_2^2p_4
-\frac{1}{12}p_1p_3p_4
\nonumber\\
&+\frac{1}{16}p_4^2
-\frac{1}{30}p_1^3p_5
+\frac{1}{10}p_1p_2p_5
-\frac{1}{15}p_3p_5. 
\end{align}
Accordingly, we obtain the large $N$ normalized 2-point functions of the Wilson line operators associated with the rectangular Young diagrams $\tiny \yng(3,3)$ and $\tiny \yng(2,2,2)$
\begin{align}
&
\langle \mathcal{W}_{\tiny \yng(3,3)} \mathcal{W}_{\overline{\tiny \yng(3,3)}}\rangle^{U(\infty)}
=\langle \mathcal{W}_{\tiny \yng(2,2,2)} \mathcal{W}_{\overline{\tiny \yng(2,2,2)}}\rangle^{U(\infty)}
\nonumber\\
&=\frac{5}{144}{\langle \mathcal{W}_{1}\mathcal{W}_{-1}\rangle^{U(\infty)}}^{6}
+\frac{1}{96}{\langle \mathcal{W}_{1}\mathcal{W}_{-1}\rangle^{U(\infty)}}^{4}{\langle \mathcal{W}_{2}\mathcal{W}_{-2}\rangle^{U(\infty)}}
+\frac{1}{64}{\langle \mathcal{W}_{1}\mathcal{W}_{-1}\rangle^{U(\infty)}}^{2}{\langle \mathcal{W}_{2}\mathcal{W}_{-2}\rangle^{U(\infty)}}^{2}
\nonumber\\
&+\frac{3}{128}{\langle \mathcal{W}_{2}\mathcal{W}_{-2}\rangle^{U(\infty)}}^{3}
+\frac{1}{54}{\langle \mathcal{W}_{1}\mathcal{W}_{-1}\rangle^{U(\infty)}}^{3}{\langle \mathcal{W}_{3}\mathcal{W}_{-3}\rangle^{U(\infty)}}
\nonumber\\
&+\frac{1}{36}{\langle \mathcal{W}_{1}\mathcal{W}_{-1}\rangle^{U(\infty)}}{\langle \mathcal{W}_{2}\mathcal{W}_{-2}\rangle^{U(\infty)}}{\langle \mathcal{W}_{3}\mathcal{W}_{-3}\rangle^{U(\infty)}}
+\frac{2}{81}{\langle \mathcal{W}_{3}\mathcal{W}_{-3}\rangle^{U(\infty)}}^{2}
\nonumber\\
&+\frac{1}{32}{\langle \mathcal{W}_{1}\mathcal{W}_{-1}\rangle^{U(\infty)}}^{2}{\langle \mathcal{W}_{4}\mathcal{W}_{-4}\rangle^{U(\infty)}}
+\frac{1}{64}{\langle \mathcal{W}_{2}\mathcal{W}_{-2}\rangle^{U(\infty)}}^{2}{\langle \mathcal{W}_{4}\mathcal{W}_{-4}\rangle^{U(\infty)}}, \\
&
\langle \mathcal{W}_{\tiny \yng(3,3)} \mathcal{W}_{\overline{\tiny \yng(2,2,2)}}\rangle^{U(\infty)}
\nonumber\\
&=\frac{5}{144}{\langle \mathcal{W}_{1}\mathcal{W}_{-1}\rangle^{U(\infty)}}^{6}
-\frac{1}{96}{\langle \mathcal{W}_{1}\mathcal{W}_{-1}\rangle^{U(\infty)}}^{4}{\langle \mathcal{W}_{2}\mathcal{W}_{-2}\rangle^{U(\infty)}}
+\frac{1}{64}{\langle \mathcal{W}_{1}\mathcal{W}_{-1}\rangle^{U(\infty)}}^{2}{\langle \mathcal{W}_{2}\mathcal{W}_{-2}\rangle^{U(\infty)}}^{2}
\nonumber\\
&-\frac{3}{128}{\langle \mathcal{W}_{2}\mathcal{W}_{-2}\rangle^{U(\infty)}}^{3}
+\frac{1}{54}{\langle \mathcal{W}_{1}\mathcal{W}_{-1}\rangle^{U(\infty)}}^{3}{\langle \mathcal{W}_{3}\mathcal{W}_{-3}\rangle^{U(\infty)}}
\nonumber\\
&-\frac{1}{36}{\langle \mathcal{W}_{1}\mathcal{W}_{-1}\rangle^{U(\infty)}}{\langle \mathcal{W}_{2}\mathcal{W}_{-2}\rangle^{U(\infty)}}
{\langle \mathcal{W}_{3}\mathcal{W}_{-3}\rangle^{U(\infty)}}
+\frac{2}{81}{\langle \mathcal{W}_{3}\mathcal{W}_{-3}\rangle^{U(\infty)}}^{2}
\nonumber\\
&-\frac{1}{32}{\langle \mathcal{W}_{1}\mathcal{W}_{-1}\rangle^{U(\infty)}}^{2}{\langle \mathcal{W}_{4}\mathcal{W}_{-4}\rangle^{U(\infty)}}
+\frac{1}{64}{\langle \mathcal{W}_{2}\mathcal{W}_{-2}\rangle^{U(\infty)}}^{2}{\langle \mathcal{W}_{4}\mathcal{W}_{-4}\rangle^{U(\infty)}}. 
\end{align}
They satisfy 
\begin{align}
\langle \mathcal{W}_{\tiny \yng(3,3)} \mathcal{W}_{\overline{\tiny \yng(3,3)}}\rangle^{U(\infty)}(t;q^{-1})
&=\langle \mathcal{W}_{\tiny \yng(3,3)} \mathcal{W}_{\overline{\tiny \yng(2,2,2)}}\rangle^{U(\infty)}(t;q). 
\end{align}

Similarly, for the rectangular diagrams $\tiny \yng(4,4)$ and $\tiny \yng(2,2,2,2)$, 
the large $N$ normalized 2-point functions are expressed as
\begin{align}
&
\langle \mathcal{W}_{\tiny \yng(4,4)} \mathcal{W}_{\overline{\tiny \yng(4,4)}}\rangle^{U(\infty)}
=\langle \mathcal{W}_{\tiny \yng(2,2,2,2)} \mathcal{W}_{\overline{\tiny \yng(2,2,2,2)}}\rangle^{U(\infty)}
\nonumber\\
&=\frac{7}{1440}{\langle \mathcal{W}_{1}\mathcal{W}_{-1}\rangle^{U(\infty)}}^{8}
+\frac{1}{180}{\langle \mathcal{W}_{1}\mathcal{W}_{-1}\rangle^{U(\infty)}}^{6}{\langle \mathcal{W}_{2}\mathcal{W}_{-2}\rangle^{U(\infty)}}
+\frac{1}{192}{\langle \mathcal{W}_{1}\mathcal{W}_{-1}\rangle^{U(\infty)}}^{4}{\langle \mathcal{W}_{2}\mathcal{W}_{-2}\rangle^{U(\infty)}}^{2}
\nonumber\\
&+\frac{3}{512}{\langle \mathcal{W}_{2}\mathcal{W}_{-2}\rangle^{U(\infty)}}^{4}
+\frac{1}{1080}{\langle \mathcal{W}_{1}\mathcal{W}_{-1}\rangle^{U(\infty)}}^{5}{\langle \mathcal{W}_{3}\mathcal{W}_{-3}\rangle^{U(\infty)}}
\nonumber\\
&+\frac{1}{216}{\langle \mathcal{W}_{1}\mathcal{W}_{-1}\rangle^{U(\infty)}}^{3}{\langle \mathcal{W}_{2}\mathcal{W}_{-2}\rangle^{U(\infty)}}
{\langle \mathcal{W}_{3}\mathcal{W}_{-3}\rangle^{U(\infty)}}
+\frac{1}{288}{\langle \mathcal{W}_{1}\mathcal{W}_{-1}\rangle^{U(\infty)}}
{\langle \mathcal{W}_{2}\mathcal{W}_{-2}\rangle^{U(\infty)}}^{2}{\langle \mathcal{W}_{3}\mathcal{W}_{-3}\rangle^{U(\infty)}}
\nonumber\\
&+\frac{1}{81}{\langle \mathcal{W}_{1}\mathcal{W}_{-1}\rangle^{U(\infty)}}^{2}{\langle \mathcal{W}_{3}\mathcal{W}_{-3}\rangle^{U(\infty)}}^{2}
+\frac{1}{162}{\langle \mathcal{W}_{2}\mathcal{W}_{-2}\rangle^{U(\infty)}}{\langle \mathcal{W}_{3}\mathcal{W}_{-3}\rangle^{U(\infty)}}^{2}
\nonumber\\
&+\frac{1}{96}{\langle \mathcal{W}_{1}\mathcal{W}_{-1}\rangle^{U(\infty)}}^{4}{\langle \mathcal{W}_{4}\mathcal{W}_{-4}\rangle^{U(\infty)}}
+\frac{1}{128}{\langle \mathcal{W}_{2}\mathcal{W}_{-2}\rangle^{U(\infty)}}^{2}{\langle \mathcal{W}_{4}\mathcal{W}_{-4}\rangle^{U(\infty)}}
\nonumber\\
&+\frac{1}{144}{\langle \mathcal{W}_{1}\mathcal{W}_{-1}\rangle^{U(\infty)}}{\langle \mathcal{W}_{3}\mathcal{W}_{-3}\rangle^{U(\infty)}}
{\langle \mathcal{W}_{4}\mathcal{W}_{-4}\rangle^{U(\infty)}}
+\frac{1}{128}{\langle \mathcal{W}_{4}\mathcal{W}_{-4}\rangle^{U(\infty)}}^{2}
\nonumber\\
&+\frac{1}{150}{\langle \mathcal{W}_{1}\mathcal{W}_{-1}\rangle^{U(\infty)}}^{3}{\langle \mathcal{W}_{5}\mathcal{W}_{-5}\rangle^{U(\infty)}}
+\frac{1}{100}{\langle \mathcal{W}_{1}\mathcal{W}_{-1}\rangle^{U(\infty)}}{\langle \mathcal{W}_{2}\mathcal{W}_{-2}\rangle^{U(\infty)}}
{\langle \mathcal{W}_{4}\mathcal{W}_{-4}\rangle^{U(\infty)}}
\nonumber\\
&+\frac{1}{225}{\langle \mathcal{W}_{3}\mathcal{W}_{-3}\rangle^{U(\infty)}}{\langle \mathcal{W}_{5}\mathcal{W}_{-5}\rangle^{U(\infty)}}, 
\end{align}
\begin{align}
&
\langle \mathcal{W}_{\tiny \yng(4,4)} \mathcal{W}_{\overline{\tiny \yng(2,2,2,2)}}\rangle^{U(\infty)}
\nonumber\\
&=\frac{7}{1440}{\langle \mathcal{W}_{1}\mathcal{W}_{-1}\rangle^{U(\infty)}}^{8}
-\frac{1}{180}{\langle \mathcal{W}_{1}\mathcal{W}_{-1}\rangle^{U(\infty)}}^{6}{\langle \mathcal{W}_{2}\mathcal{W}_{-2}\rangle^{U(\infty)}}
+\frac{1}{192}{\langle \mathcal{W}_{1}\mathcal{W}_{-1}\rangle^{U(\infty)}}^{4}{\langle \mathcal{W}_{2}\mathcal{W}_{-2}\rangle^{U(\infty)}}^{2}
\nonumber\\
&+\frac{3}{512}{\langle \mathcal{W}_{2}\mathcal{W}_{-2}\rangle^{U(\infty)}}^{4}
+\frac{1}{1080}{\langle \mathcal{W}_{1}\mathcal{W}_{-1}\rangle^{U(\infty)}}^{5}{\langle \mathcal{W}_{3}\mathcal{W}_{-3}\rangle^{U(\infty)}}
\nonumber\\
&-\frac{1}{216}{\langle \mathcal{W}_{1}\mathcal{W}_{-1}\rangle^{U(\infty)}}^{3}{\langle \mathcal{W}_{2}\mathcal{W}_{-2}\rangle^{U(\infty)}}
{\langle \mathcal{W}_{3}\mathcal{W}_{-3}\rangle^{U(\infty)}}
+\frac{1}{288}{\langle \mathcal{W}_{1}\mathcal{W}_{-1}\rangle^{U(\infty)}}
{\langle \mathcal{W}_{2}\mathcal{W}_{-2}\rangle^{U(\infty)}}^{2}{\langle \mathcal{W}_{3}\mathcal{W}_{-3}\rangle^{U(\infty)}}
\nonumber\\
&+\frac{1}{81}{\langle \mathcal{W}_{1}\mathcal{W}_{-1}\rangle^{U(\infty)}}^{2}{\langle \mathcal{W}_{3}\mathcal{W}_{-3}\rangle^{U(\infty)}}^{2}
-\frac{1}{162}{\langle \mathcal{W}_{2}\mathcal{W}_{-2}\rangle^{U(\infty)}}{\langle \mathcal{W}_{3}\mathcal{W}_{-3}\rangle^{U(\infty)}}^{2}
\nonumber\\
&-\frac{1}{96}{\langle \mathcal{W}_{1}\mathcal{W}_{-1}\rangle^{U(\infty)}}^{4}{\langle \mathcal{W}_{4}\mathcal{W}_{-4}\rangle^{U(\infty)}}
-\frac{1}{128}{\langle \mathcal{W}_{2}\mathcal{W}_{-2}\rangle^{U(\infty)}}^{2}{\langle \mathcal{W}_{4}\mathcal{W}_{-4}\rangle^{U(\infty)}}
\nonumber\\
&-\frac{1}{144}{\langle \mathcal{W}_{1}\mathcal{W}_{-1}\rangle^{U(\infty)}}{\langle \mathcal{W}_{3}\mathcal{W}_{-3}\rangle^{U(\infty)}}
{\langle \mathcal{W}_{4}\mathcal{W}_{-4}\rangle^{U(\infty)}}
+\frac{1}{128}{\langle \mathcal{W}_{4}\mathcal{W}_{-4}\rangle^{U(\infty)}}^{2}
\nonumber\\
&+\frac{1}{150}{\langle \mathcal{W}_{1}\mathcal{W}_{-1}\rangle^{U(\infty)}}^{3}{\langle \mathcal{W}_{5}\mathcal{W}_{-5}\rangle^{U(\infty)}}
-\frac{1}{100}{\langle \mathcal{W}_{1}\mathcal{W}_{-1}\rangle^{U(\infty)}}{\langle \mathcal{W}_{2}\mathcal{W}_{-2}\rangle^{U(\infty)}}
{\langle \mathcal{W}_{4}\mathcal{W}_{-4}\rangle^{U(\infty)}}
\nonumber\\
&+\frac{1}{225}{\langle \mathcal{W}_{3}\mathcal{W}_{-3}\rangle^{U(\infty)}}{\langle \mathcal{W}_{5}\mathcal{W}_{-5}\rangle^{U(\infty)}}. 
\end{align}
Under $q$ $\rightarrow$ $q^{-1}$, they transform as
\begin{align}
\langle \mathcal{W}_{\tiny \yng(4,4)} \mathcal{W}_{\overline{\tiny \yng(4,4)}}\rangle^{U(\infty)}(t;q^{-1})
&=\langle \mathcal{W}_{\tiny \yng(4,4)} \mathcal{W}_{\overline{\tiny \yng(2,2,2,2)}}\rangle^{U(\infty)}(t;q). 
\end{align}

More generally, we have
\begin{align}
\langle \mathcal{W}_{(m^2)} \mathcal{W}_{\overline{(m^2)}}\rangle^{U(\infty)}(t;q^{-1})
&=\langle \mathcal{W}_{(m^2)} \mathcal{W}_{\overline{(2^m)}}\rangle^{U(\infty)}(t;q). 
\end{align}

The Wilson line operators in the rectangular representations $(m^2)$ or $(2^m)$ also have correlators with those in the other representations. 
For example, the large $N$ normalized 2-point functions of the Wilson line operators labeled by $\tiny \yng(3,3)$, $\tiny \yng(2,2,2)$ and those in the (anti)symmetric representations can be calculated from the relation
\begin{align}
&
\langle \mathcal{W}_{\tiny \yng(3,3)} \mathcal{W}_{\overline{\tiny \yng(6)}}\rangle^{U(\infty)}(t;q)
=\langle \mathcal{W}_{\tiny \yng(2,2,2)} \mathcal{W}_{\overline{\tiny \yng(1,1,1,1,1,1)}}\rangle^{U(\infty)}(t;q)
\nonumber\\
&=\frac{1}{144}{\langle \mathcal{W}_{1}\mathcal{W}_{-1}\rangle^{U(\infty)}}^{6}
+\frac{1}{96}{\langle \mathcal{W}_{1}\mathcal{W}_{-1}\rangle^{U(\infty)}}^{4}{\langle \mathcal{W}_{2}\mathcal{W}_{-2}\rangle^{U(\infty)}}
+\frac{1}{64}{\langle \mathcal{W}_{1}\mathcal{W}_{-1}\rangle^{U(\infty)}}^2{\langle \mathcal{W}_{2}\mathcal{W}_{-2}\rangle^{U(\infty)}}^2
\nonumber\\
&-\frac{1}{128}{\langle \mathcal{W}_{2}\mathcal{W}_{-2}\rangle^{U(\infty)}}^3
-\frac{1}{54}{\langle \mathcal{W}_{1}\mathcal{W}_{-1}\rangle^{U(\infty)}}^3{\langle \mathcal{W}_{3}\mathcal{W}_{-3}\rangle^{U(\infty)}}
\nonumber\\
&+\frac{1}{36}{\langle \mathcal{W}_{1}\mathcal{W}_{-1}\rangle^{U(\infty)}}{\langle \mathcal{W}_{2}\mathcal{W}_{-2}\rangle^{U(\infty)}}
{\langle \mathcal{W}_{3}\mathcal{W}_{-3}\rangle^{U(\infty)}}+\frac{1}{81}{\langle \mathcal{W}_{3}\mathcal{W}_{-3}\rangle^{U(\infty)}}^2
\nonumber\\
&-\frac{1}{32}{\langle \mathcal{W}_{1}\mathcal{W}_{-1}\rangle^{U(\infty)}}^{2}{\langle \mathcal{W}_{4}\mathcal{W}_{-4}\rangle^{U(\infty)}}
-\frac{1}{64}{\langle \mathcal{W}_{2}\mathcal{W}_{-2}\rangle^{U(\infty)}}{\langle \mathcal{W}_{4}\mathcal{W}_{-4}\rangle^{U(\infty)}}. 
\end{align}
It follows that 
\begin{align}
\langle \mathcal{W}_{(m^2)} \mathcal{W}_{\overline{(2m)}}\rangle^{U(\infty)}(t;q^{-1})
&=\langle \mathcal{W}_{(m^2)} \mathcal{W}_{\overline{(1^{2m})}}\rangle^{U(\infty)}(t;q). 
\end{align}
More generally, we have
\begin{align}
\langle \mathcal{W}_{(m^2)} \mathcal{W}_{\overline{\lambda}}\rangle^{U(\infty)}(t;q^{-1})
&=\langle \mathcal{W}_{(m^2)} \mathcal{W}_{\overline{\lambda'}}\rangle^{U(\infty)}(t;q), 
\end{align}
where $\lambda$ is the partition with $|\lambda|=2m$. 

%%%%%%%%%%%%%%%%%%%%%%%%%%%%%%%%%%%%
\subsubsection{Unflavored limit}
%%%%%%%%%%%%%%%%%%%%%%%%%%%%%%%%%%%%
The large $N$ unflavored 2-point functions of the Wilson line operators in the rectangular representations 
$\tiny \yng(3,3)$ and $\tiny \yng(2,2,2)$ are evaluated as
\begin{align}
&
\langle \mathcal{W}_{\tiny \yng(3,3)} \mathcal{W}_{\overline{\tiny \yng(3,3)}}\rangle^{U(\infty)}(q)
=\langle \mathcal{W}_{\tiny \yng(2,2,2)} \mathcal{W}_{\overline{\tiny \yng(2,2,2)}}\rangle^{U(\infty)}(q)
\nonumber\\
&=\frac{1}{(1-q^{\frac12}) (1-q)^2 (1-q^{\frac32})^2 (1-q^2)}
\Bigl(
1+q^{\frac12}+4q+8q^{\frac32}+19q^2+29q^{\frac52}+52q^3+52q^{\frac72}
\nonumber\\
&+56q^4+42q^{\frac92}+31q^5+15q^{\frac{11}{2}}+9q^6+q^{\frac{13}{2}}
\Bigr). 
\end{align}
For the partitions $\tiny \yng(4,4)$ and $\tiny \yng(2,2,2,2)$ we find
\begin{align}
&
\langle \mathcal{W}_{\tiny \yng(4,4)} \mathcal{W}_{\overline{\tiny \yng(4,4)}}\rangle^{U(\infty)}(q)
=\langle \mathcal{W}_{\tiny \yng(2,2,2,2)} \mathcal{W}_{\overline{\tiny \yng(2,2,2,2)}}\rangle^{U(\infty)}(q)
\nonumber\\
&=\frac{1}{(1-q^{\frac12}) (1-q)^2 (1-q^{\frac32})^2 (1-q^2)^2 (1-q^{\frac52})}
\Bigl(
1+q^{\frac12}+4q+8q^{\frac32}+23q^2+42q^{\frac52}+92q^3
\nonumber\\
&+151q^{\frac72}+256q^4+335q^{\frac92}+432q^5+464q^{\frac{11}{2}}+479q^6
+414q^{\frac{13}{2}}+348q^7+237q^{\frac{15}{2}}
\nonumber\\
&+159q^8+80q^{\frac{17}{2}}+40q^9+12q^{\frac{19}{2}}+6q^{10}
\Bigr).
\end{align}

We find that the large $N$ unflavored 2-point function of the Wilson line operators 
in the rectangular representation labeled by the partitions $(m^2)$ or $(2^m)$ can be expressed as
\begin{align}
&
\langle \mathcal{W}_{(m^2)} \mathcal{W}_{\overline{(m^2)}}\rangle^{U(\infty)}(q)
=\langle \mathcal{W}_{(2^m)} \mathcal{W}_{\overline{(2^m)}}\rangle^{U(\infty)}(q)
\nonumber\\
&=\frac{G_{\{(m^2), (m^2)\}}(q)}{(1-q^{\frac12}) (1-q^{\frac{m+1}{2}})}
\prod_{n=2}^{m}\frac{1}{(1-q^{\frac{n}{2}})^2}, 
\end{align}
where $G_{\{(m^2), (m^2)\}}(q)$ is a polynomial in $q$ with positive integer coefficients. 
In the large $m$ limit, it has the $q$-series expansion
\begin{align}
&
\lim_{m\rightarrow \infty} 
G_{\{(m^2), (m^2)\}}(q)
\nonumber\\
&=1+q^{\frac12}+4q+8q^{\frac32}+23q^2+47q^{\frac52}+115q^3+235q^{\frac72}+514q^4+1030q^{\frac92}+2104q^5
\nonumber\\
&+4084q^{\frac{11}{2}}+7957q^6+14989q^{\frac{13}{2}}+28114q^7+51538q^{\frac{15}{2}}
+93754q^8+167638q^{\frac{17}{2}}
\nonumber\\
&+297209q^9+519749q^{\frac{19}{2}}+901145q^{10}+\cdots
\end{align}
and the finite $m$ correction starts from the term with $q^{\frac{m+1}{2}}$
\begin{align}
\lim_{m\rightarrow \infty} 
G_{\{(m^2), (m^2)\}}(q)
=G_{\{(m^2), (m^2)\}}(q)
+(m+1)q^{\frac{m+1}{2}}+\cdots. 
\end{align}

%(3,3)&(6), (4,4)&8
The large $N$ correlators with the Wilson line operators in the (anti)symmetric representations can be evaluated as
\begin{align}
&
\langle \mathcal{W}_{\tiny \yng(3,3)} \mathcal{W}_{\overline{\tiny \yng(6)}}\rangle^{U(\infty)}(q)
\nonumber\\
&=\frac{q^{\frac32}(4+5q^{\frac12}+9q+10q^{\frac32}+12q^2+10q^{\frac52}+10q^3+3q^{\frac72}+q^4)}{(1-q^{\frac12}) (1-q)^2 (1-q^{\frac32})^2 (1-q^2)}, \\
&
\langle \mathcal{W}_{\tiny \yng(4,4)} \mathcal{W}_{\overline{\tiny \yng(8)}}\rangle^{U(\infty)}(q)
\nonumber\\
&=\frac{q^2(5+7q^{\frac12}+14q+20q^{\frac32}+28q^2+31q^{\frac52}+38q^{2}+35q^{\frac72}+34q^4+22q^{\frac92}+14q^5+5q^{\frac{11}{2}}+3q^6)}
{(1-q^{\frac12}) (1-q)^2 (1-q^{\frac32})^2 (1-q^2)^2 (1-q^{\frac52})}. 
\end{align}

In general, the large $N$ unflavored 2-point function of the Wilson line operators in the rectangular representation 
$(m^2)$ or $(2^m)$ and those in the (anti)symmetric representation takes the form
\begin{align}
\langle \mathcal{W}_{(m^2)} \mathcal{W}_{(2m)}\rangle^{U(\infty)}(q)
&=q^{\frac{m}{2}}
\frac{G_{\{(m^2), (2m)\}}(q)}
{(1-q^{\frac12}) (1-q^{\frac{m+1}{2}})}\prod_{n=2}^{m}\frac{1}{(1-q^{\frac{n}{2}})^2}. 
\end{align}
Here $G_{\{(m^2), (2m)\}}(q)$ is a polynomial in $q$ with positive coefficients. 

We find that it obeys
\begin{align}
G_{\{(m^2), (2m)\}}(q)-G_{\{((m-1)^2), (2(m-1))\}}(q)
&\equiv \prod_{n=1}^{\infty}\frac{1}{(1-q^{\frac{n}{2}})^2}\mod q^{\frac{m}{2}}. 
\end{align}

%%%%%%%%%%%%%%%%%%%%%%%%%%%%%%%%%%%%
\subsubsection{Half-BPS limit}
%%%%%%%%%%%%%%%%%%%%%%%%%%%%%%%%%%%%
In the half-BPS limit, the large $N$ normalized 2-point functions of the Wilson line operators labeled by 
the partitions $\tiny \yng(3,3)$ and $\tiny \yng(2,2,2)$ are given by
\begin{align}
&
\langle \mathcal{W}_{\tiny \yng(3,3)} \mathcal{W}_{\overline{\tiny \yng(3,3)}}\rangle^{U(\infty)}_{\textrm{$\frac12$BPS}} (\mathfrak{q})
=\langle \mathcal{W}_{\tiny \yng(2,2,2)} \mathcal{W}_{\overline{\tiny \yng(2,2,2)}}\rangle^{U(\infty)}_{\textrm{$\frac12$BPS}} (\mathfrak{q})
\nonumber\\
&=\frac{1+\mathfrak{q}^4+\mathfrak{q}^5+2\mathfrak{q}^6}
{(1-\mathfrak{q}) (1-\mathfrak{q}^2)^2 (1-\mathfrak{q}^3)^2 (1-\mathfrak{q}^4)}. 
\end{align}
Under $\mathfrak{q}\rightarrow \mathfrak{q}^{-1}$, they transform as 
\begin{align}
\langle \mathcal{W}_{\tiny \yng(3,3)} \mathcal{W}_{\overline{\tiny \yng(3,3)}}\rangle^{U(\infty)}_{\textrm{$\frac12$BPS}} (\mathfrak{q}^{-1})
&=\mathfrak{q}^6 \langle \mathcal{W}_{\tiny \yng(3,3)} \mathcal{W}_{\overline{\tiny \yng(2,2,2)}}\rangle^{U(\infty)}_{\textrm{$\frac12$BPS}} (\mathfrak{q}). 
\end{align}
For the partitions $\tiny \yng(4,4)$ and $\tiny \yng(2,2,2,2)$ we obtain
\begin{align}
&
\langle \mathcal{W}_{\tiny \yng(4,4)} \mathcal{W}_{\overline{\tiny \yng(4,4)}}\rangle^{U(\infty)}_{\textrm{$\frac12$BPS}} (\mathfrak{q})
=\langle \mathcal{W}_{\tiny \yng(2,2,2,2)} \mathcal{W}_{\overline{\tiny \yng(2,2,2,2)}}\rangle^{U(\infty)}_{\textrm{$\frac12$BPS}} (\mathfrak{q})
\nonumber\\
&=\frac{1+\mathfrak{q}^4+\mathfrak{q}^5+3\mathfrak{q}^6+2\mathfrak{q}^7+3\mathfrak{q}^8+\mathfrak{q}^9+\mathfrak{q}^{10}+\mathfrak{q}^{12}}
{(1-\mathfrak{q}) (1-\mathfrak{q}^2)^2 (1-\mathfrak{q}^3)^2 (1-\mathfrak{q}^4)^2 (1-\mathfrak{q}^5)}. 
\end{align}

In general, the half-BPS limit of the large $N$ 2-point function of the Wilson line operators transforming in the rectangular representations $(m^2)$ and $(2^m)$ takes the form
\begin{align}
&
\langle \mathcal{W}_{(m^2)} \mathcal{W}_{\overline{(m^2)}}\rangle^{U(\infty)}_{\textrm{$\frac12$BPS}} (\mathfrak{q})
=\langle \mathcal{W}_{(2^m)} \mathcal{W}_{\overline{(2^m)}}\rangle^{U(\infty)}_{\textrm{$\frac12$BPS}} (\mathfrak{q})
\nonumber\\
&=\frac{H_{\{(m^2), (m^2)\}}(\mathfrak{q}) }{(1-\mathfrak{q}) (1-\mathfrak{q}^{m+1})}
\prod_{n=2}^{m}\frac{1}{(1-\mathfrak{q}^n)^2}, 
\end{align}
where $H_{\{(m^2), (m^2)\}}(\mathfrak{q})$ is a polynomial in $\mathfrak{q}$ with positive coefficients. 
In the large $m$ limit, it has the $q$-series expansion
\begin{align}
&
\lim_{m\rightarrow \infty}
H_{\{(m^2), (m^2)\}}(\mathfrak{q})
\nonumber\\
&=1+\mathfrak{q}^4+\mathfrak{q}^5+3\mathfrak{q}^6+3\mathfrak{q}^7+6\mathfrak{q}^8
+7\mathfrak{q}^9+12\mathfrak{q}^{10}+14\mathfrak{q}^{11}+24\mathfrak{q}^{12}
+29\mathfrak{q}^{13}+45\mathfrak{q}^{14}
\nonumber\\
&+58\mathfrak{q}^{15}+86\mathfrak{q}^{16}+110\mathfrak{q}^{17}+160\mathfrak{q}^{18}
+206\mathfrak{q}^{19}+290\mathfrak{q}^{20}+\cdots. 
\end{align}

The half-BPS limit of the large $N$ correlation functions 
of the Wilson line operators in the rectangular representation $\tiny \yng(3,3)$ or $\tiny \yng(4,4)$ and those in the (anti)symmetric representations is
\begin{align}
&
\langle \mathcal{W}_{\tiny \yng(3,3)} \mathcal{W}_{\overline{\tiny \yng(6)}}\rangle^{U(\infty)}_{\textrm{$\frac12$BPS}} (\mathfrak{q})
=\langle \mathcal{W}_{\tiny \yng(2,2,2)} \mathcal{W}_{\overline{\tiny \yng(1,1,1,1,1,1)}}\rangle^{U(\infty)}_{\textrm{$\frac12$BPS}} (\mathfrak{q})
\nonumber\\
&=\frac{\mathfrak{q}^3}
{(1-\mathfrak{q}) (1-\mathfrak{q}^2)^2 (1-\mathfrak{q}^3)^2 (1-\mathfrak{q}^4)}, \\
&
\langle \mathcal{W}_{\tiny \yng(4,4)} \mathcal{W}_{\overline{\tiny \yng(8)}}\rangle^{U(\infty)}_{\textrm{$\frac12$BPS}} (\mathfrak{q})
=\langle \mathcal{W}_{\tiny \yng(2,2,2,2)} \mathcal{W}_{\overline{\tiny \yng(1,1,1,1,1,1,1,1)}}\rangle^{U(\infty)}_{\textrm{$\frac12$BPS}} (\mathfrak{q})
\nonumber\\
&=\frac{\mathfrak{q}^4}
{(1-\mathfrak{q}) (1-\mathfrak{q}^2)^2 (1-\mathfrak{q}^3)^2 (1-\mathfrak{q}^4)^2 (1-\mathfrak{q}^5)}. 
\end{align}

More generally, the half-BPS limit of the large $N$ correlation functions  
of the Wilson line operators in the rectangular representations $(m^2)$ and $(2^m)$ 
and those in the (anti)symmetric representations can be expressed as
\begin{align}
&
\langle \mathcal{W}_{(m^2)} \mathcal{W}_{\overline{(2m)}}\rangle^{U(\infty)}_{\textrm{$\frac12$BPS}} (\mathfrak{q})
=\langle \mathcal{W}_{(2^m)} \mathcal{W}_{\overline{(1^{2m})}}\rangle^{U(\infty)}_{\textrm{$\frac12$BPS}} (\mathfrak{q})
\nonumber\\
&=\frac{\mathfrak{q}^{m}}
{(1-\mathfrak{q}) (1-\mathfrak{q}^{m+1})}
\prod_{n=2}^{m}\frac{1}{(1-\mathfrak{q}^{n})^2}. 
\end{align}
One can check that 
the correlator transforms as
\begin{align}
\langle \mathcal{W}_{(m^2)} \mathcal{W}_{\overline{(2m)}}\rangle^{U(\infty)}_{\textrm{$\frac12$BPS}} (\mathfrak{q}^{-1})
&=\mathfrak{q}^{2m} \langle \mathcal{W}_{(m^2)} \mathcal{W}_{\overline{(1^{2m})}}\rangle^{U(\infty)}_{\textrm{$\frac12$BPS}} (\mathfrak{q})
\end{align}
under $\mathfrak{q}\rightarrow \mathfrak{q}^{-1}$. 

%%%%%%%%%%%%%%%%%%%%%%%%%%%%%%%%%%%%
\subsection{$(m^3)$ and $(3^m)$}
\label{sec_m^3&3^m}
%%%%%%%%%%%%%%%%%%%%%%%%%%%%%%%%%%%%
Next we consider the large $N$ correlation functions involving the Wilson line operators 
in the rectangular representations labeled by the Young diagrams $(m^3)$ and $(3^m)$ with $m>3$. 

The calculation is similar to that in subsection \ref{sec_m^2&2^m}. 
It can be shown that the large $N$ normalized 2-point functions of the Wilson line operators in these representations satisfy 
\begin{align}
\langle \mathcal{W}_{(m^3)} \mathcal{W}_{\overline{(m^3)}}\rangle^{U(\infty)}(t;q^{-1})
&=(-1)^m \langle \mathcal{W}_{(m^3)} \mathcal{W}_{\overline{(3^m)}}\rangle^{U(\infty)}(t;q). 
\end{align}
Also the large $N$ normalized 2-point functions of these Wilson line operators and those in the other representations associated with the Young diagram $\lambda$ obey
\begin{align}
\langle \mathcal{W}_{(m^3)} \mathcal{W}_{\overline{\lambda}}\rangle^{U(\infty)}(t;q^{-1})
&=(-1)^m \langle \mathcal{W}_{(m^3)} \mathcal{W}_{\overline{\lambda'}}\rangle^{U(\infty)}(t;q), 
\end{align}
where $\lambda$ is the partition with $|\lambda|=3m$. 

%%%%%%%%%%%%%%%%%%%%%%%%%%%%%%%%%%%%
\subsubsection{Unflavored limit}
%%%%%%%%%%%%%%%%%%%%%%%%%%%%%%%%%%%%
The large $N$ unflavored 2-point functions of the Wilson line operators 
in the rectangular representations labeled by the partitions $\tiny \yng(4,4,4)$ and $\tiny \yng(3,3,3,3)$ are evaluated as
\begin{align}
&
\langle \mathcal{W}_{\tiny \yng(4,4,4)} \mathcal{W}_{\overline{\tiny \yng(4,4,4)}}\rangle^{U(\infty)}(q)
=\langle \mathcal{W}_{\tiny \yng(3,3,3,3)} \mathcal{W}_{\overline{\tiny \yng(3,3,3,3)}}\rangle^{U(\infty)}(q)
\nonumber\\
&=\frac{1}{(1-q^{\frac12}) (1-q)^2 (1-q^{\frac32})^3 (1-q^2)^3 (1-q^{\frac52})^2 (1-q^3)}
\Bigl(
1+q^{\frac12}+4q+11q^{\frac32}+32q^2
\nonumber\\
&+79q^{\frac52}+216q^3+507q^{\frac72}+1205q^4+2606q^{\frac92}
+5367q^5+10179q^{\frac{11}{2}}+18220q^6
\nonumber\\
&+30057q^{\frac{13}{2}}+46734q^7+67723q^{\frac{15}{2}}+92562q^8+118637q^{\frac{17}{2}}
+143910q^9+164217q^{\frac{19}{2}}
\nonumber\\
&+177626q^{10}+181113q^{\frac{21}{2}}+175022q^{11}
+159405q^{\frac{23}{2}}+137578q^{12}+111603q^{\frac{25}{2}}
+85655q^{13}
\nonumber\\
&+61610q^{\frac{27}{2}}+41801q^{14}+26405q^{\frac{29}{2}}
+15722q^{15}+8613q^{\frac{31}{2}}+4442q^{16}+2077q^{\frac{33}{2}}
\nonumber\\
&+908q^{17}+343q^{\frac{35}{2}}+125q^{18}+30q^{\frac{37}{2}}+6q^{19}
\Bigr).
\end{align}

For higher $m$ the explicit form is rather expensive, 
however, for general $m$ the large $N$ unflavored 2-point function can be written as
\begin{align}
&
\langle \mathcal{W}_{(m^3)} \mathcal{W}_{\overline{(m^3)}}\rangle^{U(\infty)}(q)
\nonumber\\
&=\frac{G_{\{(m^3), (m^3)\}}(q)}
{(1-q^{\frac12}) (1-q)^2 (1-q^{\frac{m+1}{2}})^2 (1-q^{\frac{m+2}{2}})}
\prod_{n=3}^{m}
\frac{1}{(1-q^{\frac{n}{2}})^3}, 
\end{align}
where $G_{\{(m^3), (m^3)\}}(q)$ is a polynomial in $q$ with positive integers. 
In the large $m$ limit, it can be expanded as
\begin{align}
&
\lim_{m\rightarrow \infty}G_{\{(m^3), (m^3)\}}(q)
\nonumber\\
&=1+q^{\frac12}+4q+11q^{\frac32}+84q^{\frac52}+239q^3+614q^{\frac72}+1617q^{4}+4101q^{\frac92}+10282q^5
\nonumber\\
&+25188q^{\frac{11}{2}}+60933q^6+144486q^{\frac{13}{2}}+337974q^7+778039q^{\frac{15}{2}}+\cdots. 
\end{align}
We find that the finite $m$ correction appears from the term with $q^{\frac{m+1}{2}}$ 
\begin{align}
\lim_{m\rightarrow \infty}
G_{\{(m^3), (m^3)\}}(q)
&=G_{\{(m^3), (m^3)\}}(q)+(m+1)q^{\frac{m+1}{2}}+\cdots. 
\end{align}

%%%%%%%%%%%%%%%%%%%%%%%%%%%%%%%%%%%%
\subsubsection{Half-BPS limit}
%%%%%%%%%%%%%%%%%%%%%%%%%%%%%%%%%%%%
In the half-BPS limit, the large $N$ normalized 2-point functions of the Wilson line operators associated with $\tiny \yng(4,4,4)$ is 
\begin{align}
&
\langle \mathcal{W}_{\tiny \yng(4,4,4)} \mathcal{W}_{\overline{\tiny \yng(4,4,4)}}\rangle^{U(\infty)}_{\textrm{$\frac12$BPS}} (\mathfrak{q})
\nonumber\\
&=\frac{1}
{(1-\mathfrak{q}) (1-\mathfrak{q})^2 (1-\mathfrak{q}^3)^3 (1-\mathfrak{q}^4)^3 (1-\mathfrak{q}^5)^2 (1-\mathfrak{q}^6)}
\Bigl(
1+\mathfrak{q}^4+2\mathfrak{q}^5+6\mathfrak{q}^6+7\mathfrak{q}^7+12\mathfrak{q}^8
\nonumber\\
&+15\mathfrak{q}^9+21\mathfrak{q}^{10}+26\mathfrak{q}^{11}+41\mathfrak{q}^{12}
+42\mathfrak{q}^{13}+50\mathfrak{q}^{14}+48\mathfrak{q}^{15}+46\mathfrak{q}^{16}+34\mathfrak{q}^{17}
+31\mathfrak{q}^{18}+20\mathfrak{q}^{19}
\nonumber\\
&+19\mathfrak{q}^{20}+14\mathfrak{q}^{21}+12\mathfrak{q}^{22}+7\mathfrak{q}^{23}+6\mathfrak{q}^{24}+\mathfrak{q}^{25}
\Bigr). 
\end{align}

For general $m$ the large $N$ unflavored 2-point function takes the form
\begin{align}
&
\langle \mathcal{W}_{(m^3)} \mathcal{W}_{\overline{(m^3)}}\rangle^{U(\infty)}_{\textrm{$\frac12$BPS}} (\mathfrak{q})
\nonumber\\
&=\frac{H_{\{(m^3), (m^3)\}}(\mathfrak{q})}
{(1-\mathfrak{q}) (1-\mathfrak{q}^2)^2 (1-\mathfrak{q}^{m+1})^2 (1-\mathfrak{q}^{m+2})}
\prod_{n=3}^{m}
\frac{1}{(1-\mathfrak{q}^n)^3}, 
\end{align}
where $H_{\{(m^3), (m^3)\}}(\mathfrak{q})$ is a polynomial in $\mathfrak{q}$ with positive coefficients. 
In the large $m$ limit, it has the $q$-series expansion
\begin{align}
&
\lim_{m\rightarrow \infty}
H_{\{(m^3), (m^3)\}}(\mathfrak{q})
\nonumber\\
&=1+\mathfrak{q}^4+2\mathfrak{q}^5+6\mathfrak{q}^6+8\mathfrak{q}^7+16\mathfrak{q}^8
+26\mathfrak{q}^9+45\mathfrak{q}^{10}+72\mathfrak{q}^{11}+131\mathfrak{q}^{12}
+208\mathfrak{q}^{13}
\nonumber\\
&+356\mathfrak{q}^{14}+582\mathfrak{q}^{15}+\cdots. 
\end{align}

%%%%%%%%%%%%%%%%%%%%%%%%%%%%%%%%%%%%
\subsection{$(m^m)$}
%%%%%%%%%%%%%%%%%%%%%%%%%%%%%%%%%%%%
Let us study the large $N$ correlation functions of the Wilson line operators transforming in the representations 
labeled by the $m$-squares $(m^m)$, i.e. the self-conjugate Young diagrams of the rectangular shapes with $m$ rows each of length $m$. 

The Schur functions indexed by the $m$-square $(m^m)$ can be expanded with respect to the power sum symmetric functions. 
For example, for $\tiny \yng(2,2)$ and $\tiny \yng(3,3,3)$ one finds
\begin{align}
s_{\tiny \yng(2,2)}&=
\frac{1}{12}p_1^4+\frac14 p_2^2-\frac13 p_1 p_3, \\
\label{schur3^3_power}
s_{\tiny \yng(3,3,3)}
&=\frac{1}{8640}p_1^9+\frac{1}{480}p_1^5p_2^2-\frac{1}{64}p_1p_2^4-\frac{1}{360}p_1^6p_3
+\frac{1}{24}p_1^2p_2^2p_3+\frac{1}{27}p_3^3-\frac{1}{24}p_1^3p_2p_4
\nonumber\\
&-\frac{1}{12}p_2p_3p_4+\frac{1}{16}p_1p_4^2+\frac{1}{60}p_1^4p_5
+\frac{1}{20}p_2^2p_5-\frac{1}{15}p_1p_3p_5. 
\end{align}

The large $N$ normalized 2-point functions of the Wilson line operators in the rectangular representations 
$\tiny \yng(2,2)$ and $\tiny \yng(3,3,3)$ are evaluated as
\begin{align}
\label{f22&22}
&
\langle \mathcal{W}_{\tiny \yng(2,2)} \mathcal{W}_{\overline{\tiny \yng(2,2)}}\rangle^{U(\infty)}
\nonumber\\
&=\frac16 {\langle W_{1}W_{-1}\rangle^{U(\infty)}}^{4}
+\frac18 {\langle W_{2}W_{-2}\rangle^{U(\infty)}}^{2}
+\frac19 {\langle W_{1}W_{-1}\rangle^{U(\infty)}}{\langle W_{3}W_{-3}\rangle^{U(\infty)}},
\end{align}
\begin{align}
\label{f333&333}
&
\langle \mathcal{W}_{\tiny \yng(3,3,3)} \mathcal{W}_{\overline{\tiny \yng(3,3,3)}}\rangle^{U(\infty)}
\nonumber\\
&=\frac{7}{1440}{\langle W_{1}W_{-1}\rangle^{U(\infty)}}^{9}
+\frac{1}{960}{\langle W_{1}W_{-1}\rangle^{U(\infty)}}^{5}{\langle W_{2}W_{-2}\rangle^{U(\infty)}}^{2}
\nonumber\\
&+\frac{3}{512}{\langle W_{1}W_{-1}\rangle^{U(\infty)}}{\langle W_{2}W_{-2}\rangle^{U(\infty)}}^{4}
+\frac{1}{180}{\langle W_{1}W_{-1}\rangle^{U(\infty)}}^6{\langle W_{3}W_{-3}\rangle^{U(\infty)}}
\nonumber\\
&+\frac{1}{144}{\langle W_{1}W_{-1}\rangle^{U(\infty)}}^2{\langle W_{3}W_{-2}\rangle^{U(\infty)}}^2
{\langle W_{3}W_{-3}\rangle^{U(\infty)}}
+\frac{2}{243}{\langle W_{3}W_{-3}\rangle^{U(\infty)}}^3
\nonumber\\
&+\frac{1}{96}{\langle W_{1}W_{-1}\rangle^{U(\infty)}}^3{\langle W_{2}W_{-2}\rangle^{U(\infty)}}
{\langle W_{4}W_{-4}\rangle^{U(\infty)}}
+\frac{1}{144}{\langle W_{2}W_{-2}\rangle^{U(\infty)}}{\langle W_{3}W_{-3}\rangle^{U(\infty)}}
{\langle W_{4}W_{-4}\rangle^{U(\infty)}}
\nonumber\\
&+\frac{1}{128}{\langle W_{1}W_{-1}\rangle^{U(\infty)}}{\langle W_{4}W_{-4}\rangle^{U(\infty)}}^2
+\frac{1}{150}{\langle W_{1}W_{-1}\rangle^{U(\infty)}}^4{\langle W_{5}W_{-5}\rangle^{U(\infty)}}
\nonumber\\
&+\frac{1}{200}{\langle W_{2}W_{-2}\rangle^{U(\infty)}}^2{\langle W_{5}W_{-5}\rangle^{U(\infty)}}
+\frac{1}{225}{\langle W_{1}W_{-1}\rangle^{U(\infty)}}{\langle W_{3}W_{-3}\rangle^{U(\infty)}}
{\langle W_{5}W_{-5}\rangle^{U(\infty)}}. 
\end{align}
Under $q\rightarrow q^{-1}$ the correlators (\ref{f22&22}) and (\ref{f333&333}) obey the transformation laws
\begin{align}
\langle \mathcal{W}_{\tiny \yng(2,2)} \mathcal{W}_{\overline{\tiny \yng(2,2)}}\rangle^{U(\infty)}(t;q^{-1})
&=\langle \mathcal{W}_{\tiny \yng(2,2)} \mathcal{W}_{\overline{\tiny \yng(2,2)}}\rangle^{U(\infty)}(t;q), \\
\langle \mathcal{W}_{\tiny \yng(3,3,3)} \mathcal{W}_{\overline{\tiny \yng(3,3,3)}}\rangle^{U(\infty)}(t;q^{-1})
&=- \langle \mathcal{W}_{\tiny \yng(3,3,3)} \mathcal{W}_{\overline{\tiny \yng(3,3,3)}}\rangle^{U(\infty)}(t;q). 
\end{align}
For general $m$, we have
\begin{align}
\langle \mathcal{W}_{(m^m)} \mathcal{W}_{\overline{(m^m)}}\rangle^{U(\infty)}(t;q^{-1})
&=(-1)^m \langle \mathcal{W}_{(m^m)} \mathcal{W}_{\overline{(m^m)}}\rangle^{U(\infty)}(t;q). 
\end{align}

We can proceed the calculation of the large $N$ normalized 2-point functions of the Wilson line operators labeled by 
the $m$-square Young diagrams and those in the other representations. 
For example, the large $N$ normalized 2-point functions of the Wilson line operators in the rectangular representations 
$\tiny \yng(2,2)$ and $\tiny \yng(3,3,3)$ and those in the (anti)symmetric representations are given by
\begin{align}
\label{f22&4}
&
\langle \mathcal{W}_{\tiny \yng(2,2)} \mathcal{W}_{\overline{\tiny \yng(4)}}\rangle^{U(\infty)}
=\langle \mathcal{W}_{\tiny \yng(2,2)} \mathcal{W}_{\overline{\tiny \yng(1,1,1,1)}}\rangle^{U(\infty)}
\nonumber\\
&=\frac{1}{12} {\langle W_{1}W_{-1}\rangle^{U(\infty)}}^{4}
+\frac{1}{16} {\langle W_{2}W_{-2}\rangle^{U(\infty)}}^{2}
-\frac19 {\langle W_{1}W_{-1}\rangle^{U(\infty)}}{\langle W_{3}W_{-3}\rangle^{U(\infty)}}, 
\end{align}
\begin{align}
\label{f333&9}
&
\langle \mathcal{W}_{\tiny \yng(3,3,3)} \mathcal{W}_{\overline{\tiny \yng(9)}}\rangle^{U(\infty)}
=\langle \mathcal{W}_{\tiny \yng(3,3,3)} \mathcal{W}_{\overline{\tiny \yng(1,1,1,1,1,1,1,1,1)}}\rangle^{U(\infty)}
\nonumber\\
&=\frac{1}{8640} {\langle W_{1}W_{-1}\rangle^{U(\infty)}}^{9}
+\frac{1}{1920} {\langle W_{1}W_{-1}\rangle^{U(\infty)}}^{5}{\langle W_{2}W_{-2}\rangle^{U(\infty)}}^{2}
\nonumber\\
&-\frac{1}{1024} {\langle W_{1}W_{-1}\rangle^{U(\infty)}}{\langle W_{2}W_{-2}\rangle^{U(\infty)}}^{4}
-\frac{1}{1080} {\langle W_{1}W_{-1}\rangle^{U(\infty)}}^{6}{\langle W_{2}W_{-2}\rangle^{U(\infty)}}^{3}
\nonumber\\
&+\frac{1}{288} {\langle W_{1}W_{-1}\rangle^{U(\infty)}}^{2}{\langle W_{2}W_{-2}\rangle^{U(\infty)}}^{2}{\langle W_{3}W_{-3}\rangle^{U(\infty)}}
+\frac{1}{729}{\langle W_{3}W_{-3}\rangle^{U(\infty)}}^3
\nonumber\\
&-\frac{1}{192} {\langle W_{1}W_{-1}\rangle^{U(\infty)}}^{3}{\langle W_{2}W_{-2}\rangle^{U(\infty)}}{\langle W_{4}W_{-4}\rangle^{U(\infty)}}
\nonumber\\
&-\frac{1}{288} {\langle W_{2}W_{-2}\rangle^{U(\infty)}}{\langle W_{3}W_{-3}\rangle^{U(\infty)}}{\langle W_{4}W_{-4}\rangle^{U(\infty)}}
+\frac{1}{256} {\langle W_{1}W_{-1}\rangle^{U(\infty)}}{\langle W_{4}W_{-4}\rangle^{U(\infty)}}^2
\nonumber\\
&+\frac{1}{300} {\langle W_{1}W_{-1}\rangle^{U(\infty)}}^{4}{\langle W_{5}W_{-5}\rangle^{U(\infty)}}
+\frac{1}{400} {\langle W_{2}W_{-2}\rangle^{U(\infty)}}^{2}{\langle W_{5}W_{-5}\rangle^{U(\infty)}}
\nonumber\\
&-\frac{1}{225} {\langle W_{1}W_{-1}\rangle^{U(\infty)}}{\langle W_{3}W_{-3}\rangle^{U(\infty)}}{\langle W_{5}W_{-5}\rangle^{U(\infty)}}. 
\end{align}
They obey
\begin{align}
\langle \mathcal{W}_{\tiny \yng(2,2)} \mathcal{W}_{\overline{\tiny \yng(4)}}\rangle^{U(\infty)}(t;q^{-1})
&=\langle \mathcal{W}_{\tiny \yng(2,2)} \mathcal{W}_{\overline{\tiny \yng(4)}}\rangle^{U(\infty)}(t;q), \\
\langle \mathcal{W}_{\tiny \yng(3,3,3)} \mathcal{W}_{\overline{\tiny \yng(9)}}\rangle^{U(\infty)}(t;q^{-1})
&=- \langle \mathcal{W}_{\tiny \yng(3,3,3)} \mathcal{W}_{\overline{\tiny \yng(9)}}\rangle^{U(\infty)}(t;q). 
\end{align}
More generally, we find that
\begin{align}
\langle \mathcal{W}_{(m^m)} \mathcal{W}_{\overline{\lambda}}\rangle^{U(\infty)}(t;q^{-1})
&=(-1)^m \langle \mathcal{W}_{(m^m)} \mathcal{W}_{\overline{\lambda}}\rangle^{U(\infty)}(t;q), 
\end{align}
where $\lambda$ is the partition with $|\lambda|=m^2$. 

%%%%%%%%%%%%%%%%%%%%%%%%%%%%%%%%%%%%
\subsubsection{Unflavored limit}
%%%%%%%%%%%%%%%%%%%%%%%%%%%%%%%%%%%%
In the unflavored limit, the large $N$ normalized 2-point function of the Wilson line operators in the rectangular representations 
$\tiny \yng(2,2)$ and $\tiny \yng(3,3,3)$ are evaluated as
\begin{align}
&\langle \mathcal{W}_{\tiny \yng(2,2)} \mathcal{W}_{\overline{\tiny \yng(2,2)}}\rangle^{U(\infty)}(q)
=\frac{1+q^{\frac12}+4q+5q^{\frac32}+10q^2+5q^{\frac52}+4q^3+q^{\frac72}+q^4}
{(1-q^{\frac12}) (1-q)^2 (1-q^{\frac32})}
\nonumber\\
&=1+2q^{\frac12}+8q+16q^{\frac32}+38q^2+62q^{\frac52}+110q^3+162q^{\frac72}
+246q^4+336q^{\frac92}+\cdots, 
\end{align}
\begin{align}
&
\langle \mathcal{W}_{\tiny \yng(3,3,3)} \mathcal{W}_{\overline{\tiny \yng(3,3,3)}}\rangle^{U(\infty)}(q)
=\frac{1}
{(1-q^{\frac12}) (1-q)^2 (1-q^{\frac32})^3 (1-q^{2})^2 (1-q^{\frac52})}
\nonumber\\
&\times 
\Bigl(
1+q^{\frac12}+4q+11q^{\frac32}+28q^2+66q^{\frac52}+158q^3+316q^{\frac72}+595q^4+1004q^{\frac92}
\nonumber\\
&+1487q^5+1995q^{\frac{11}{2}}+2426q^6+2660q^{\frac{13}{2}}+2660q^7+2426q^{\frac{15}{2}}+1995q^{8}
\nonumber\\
&+1487q^{\frac{17}{2}}+1004q^9+595q^{\frac{19}{2}}+316q^{10}+158q^{\frac{21}{2}}+66q^{11}+28q^{\frac{23}{2}}
\nonumber\\
&+11q^{12}+4q^{\frac{25}{2}}+q^{13}+q^{\frac{27}{2}}
\Bigr)
\nonumber\\
&=1+2q^{\frac12}+8q+24q^{\frac32}+68q^2+180q^{\frac52}+468q^3+1114q^{\frac72}
+2542q^4+5468q^{\frac92}+\cdots. 
\end{align}

For higher $m$, calculating the exact form of the normalized correlators is computationally expensive whereas it is straightforward. 
For general $m$, the large $N$ unflavored 2-point function takes the form
\begin{align}
\langle \mathcal{W}_{(m^m)} \mathcal{W}_{\overline{(m^m)}}\rangle^{U(\infty)}(q)
&=
G_{\{(m^m), (m^m)\}}(q)
\prod_{n=1}^{m}
\frac{1}{(1-q^{\frac{n}{2}})^{n} (1-q^{\frac{m+n}{2}})^{m-n}}, 
\end{align}
where $G_{\{(m^m), (m^m)\}}(q)$ is a polynomial in $q$ with positive coefficients. 

For $m=4,5,6,7,8$, the correlation functions can be expanded as
\begin{align}
&
\langle \mathcal{W}_{(4^4)} \mathcal{W}_{\overline{(4^4)}}\rangle^{U(\infty)}(q)
\nonumber\\
&=1+2q^{\frac12}+8q+24q^{\frac32}+78q^2+220q^{\frac52}+656q^3+1832q^{\frac72}+5128q^{4}+\cdots, 
\\
&
\langle \mathcal{W}_{(5^5)} \mathcal{W}_{\overline{(5^5)}}\rangle^{U(\infty)}(q)
\nonumber\\
&=1+2q^{\frac12}+8q+24q^{\frac32}+78q^2+232q^{\frac52}+706q^3+2076q^{\frac72}+6140q^{4}+\cdots, 
\\
&
\langle \mathcal{W}_{(6^6)} \mathcal{W}_{\overline{(6^6)}}\rangle^{U(\infty)}(q)
\nonumber\\
&=1+2q^{\frac12}+8q+24q^{\frac32}+78q^2+232q^{\frac52}+720q^3+2136q^{\frac72}+6440q^{4}+\cdots, 
\\
&
\langle \mathcal{W}_{(7^7)} \mathcal{W}_{\overline{(7^7)}}\rangle^{U(\infty)}(q)
\nonumber\\
&=1+2q^{\frac12}+8q+24q^{\frac32}+78q^2+232q^{\frac52}+720q^3+2152q^{\frac72}+6510q^{4}+\cdots, 
\\
&
\langle \mathcal{W}_{(8^8)} \mathcal{W}_{\overline{(8^8)}}\rangle^{U(\infty)}(q)
\nonumber\\
&=1+2q^{\frac12}+8q+24q^{\frac32}+78q^2+232q^{\frac52}+720q^3+2152q^{\frac72}+6528q^{4}+\cdots. 
\end{align}
In the large $m$ limit, it has the $q$-series expansion
\begin{align}
\label{large_um^m&m^m}
&
\lim_{m\rightarrow \infty}
\langle \mathcal{W}_{(m^m)} \mathcal{W}_{\overline{(m^m)}}\rangle^{U(\infty)}(q)
\nonumber\\
&=\sum_{n=0}^{\infty} a_{\{ (\infty^{\infty}), (\infty^{\infty}) \}}^{(G)} (n)q^{\frac{n}{2}}
\nonumber\\
&=1+2q^{\frac12}+8q+24q^{\frac32}+78q^2+232q^{\frac52}+720q^3+2152q^{\frac72}+6528q^4
+\cdots. 
\end{align}
The finite $m$ correction appears from the term with $q^{\frac{m+1}{2}}$. 
We find that 
\begin{align}
&
\langle \mathcal{W}_{((m+1)^{m+1})} \mathcal{W}_{\overline{((m+1)^{m+1})}}\rangle^{U(\infty)}(q)
\nonumber\\
&=\langle \mathcal{W}_{(m^m)} \mathcal{W}_{\overline{(m^m)}}\rangle^{U(\infty)}(q)
+(2m+4)q^{\frac{m+1}{2}}
+(10m+10)q^{\frac{m+2}{2}}
+\cdots. 
\end{align}
The $q$-series expansion (\ref{large_um^m&m^m}) of the large $m$ unflavored 2-point function precisely coincides with 
\begin{align}
\label{GLn3_conj}
\prod_{n=1}^{\infty}
\frac{1-q^{\frac{n}{2}}}
{1-3q^{\frac{n}{2}}}. 
\end{align}
This function plays a role of the generating function for the number $c(n,3)$ of conjugacy classes of $GL(n,3)$, i.e. the general linear group of degree $n$ over a finite field with $3$ elements \cite{MR109810,MR615131}. 
Hence we have the correspondence between the emergent BPS local operators 
due to the Wilson line operators in the large rectangular representation in the large $N$ limit and the conjugacy classes of $GL(n,3)$: 
\begin{align}
a_{\{ (\infty^{\infty}), (\infty^{\infty}) \}}^{(G)}(n)&=c(n,3). 
\end{align}

The function (\ref{GLn3_conj}) can be rewritten as
\begin{align}
\sum_{m=0}^{\infty}3^m q^{\frac{m}{2}}
\prod_{l=m+1}^{\infty}(1-q^{\frac{l}{2}}). 
\end{align}
As $n\rightarrow \infty$, the asymptotic expression for $a_{\{ (\infty^{\infty}), (\infty^{\infty}) \}}(n)$ is given by \cite{MR937520}
\begin{align}
\label{asymu_rect}
a_{\{ (\infty^{\infty}), (\infty^{\infty}) \}}^{(G)}(n)&\sim 
\exp \Bigl[
(\log 3) n
\Bigr]. 
\end{align}
The actual coefficients and the analytic values are given by
\begin{align}
\begin{array}{c|c|c}
n&a_{\{ (\infty^{\infty}), (\infty^{\infty}) \}}^{(G)}(n)&a_{\{ (\infty^{\infty}), (\infty^{\infty}) \} \textrm{asym.}}^{(G)}(n)\\ \hline 
10&58944&59049 \\
100&5.15378\times 10^{47}&5.15378\times 10^{47} \\
1000&1.32207\times 10^{477}&1.32207\times 10^{477} \\
10000&1.63135\times 10^{4771}&1.63135\times 10^{4771} \\
\end{array}. 
\end{align}
Holographically, the degeneracy $a_{\{ (\infty^{\infty}), (\infty^{\infty}) \}}(n)$ will be understood as 
the fluctuation modes for the dual bubbling geometry with genus one surface. 
Curiously, the asymptotic behavior (\ref{asymu_rect}) grows much faster than 
that for the $p$-branes of the form 
$\exp (\alpha n^{\frac{p}{p+1}})$ \cite{Fubini:1972mf, Dethlefsen:1974dr,Strumia:1975rd,Alvarez:1991qs,Harms:1992jt} with some constant $\alpha$. 

%%%%%%%%%%%%%%%%%%%%%%%%%%%%%%%%%%%%
\subsubsection{Half-BPS limit}
%%%%%%%%%%%%%%%%%%%%%%%%%%%%%%%%%%%%
In the half-BPS limit the large $N$ 2-point functions for the Wilson line operators in the rectangular representations 
$\tiny \yng(2,2)$ and $\tiny \yng(3,3,3)$ are given by
\begin{align}
&
\langle \mathcal{W}_{\tiny \yng(2,2)} \mathcal{W}_{\overline{\tiny \yng(2,2)}}\rangle^{U(\infty)}_{\textrm{$\frac12$BPS}} (\mathfrak{q})
=\frac{1+\mathfrak{q}^4}
{(1-\mathfrak{q}) (1-\mathfrak{q}^2)^2 (1-\mathfrak{q}^3)}
\nonumber\\
&=1+\mathfrak{q}+3\mathfrak{q}^2+4\mathfrak{q}^3+8\mathfrak{q}^4+10\mathfrak{q}^5+17\mathfrak{q}^6+21\mathfrak{q}^7+31\mathfrak{q}^8+\cdots, 
\end{align}
\begin{align}
&
\langle \mathcal{W}_{\tiny \yng(3,3,3)} \mathcal{W}_{\overline{\tiny \yng(3,3,3)}}\rangle^{U(\infty)}_{\textrm{$\frac12$BPS}} (\mathfrak{q})
=\frac{1}{(1-\mathfrak{q}) (1-\mathfrak{q}^2)^2 (1-\mathfrak{q}^3)^3 (1-\mathfrak{q}^4)^2 (1-\mathfrak{q}^5)}
\nonumber\\
&\Bigl(
1+\mathfrak{q}^4+2\mathfrak{q}^5+5\mathfrak{q}^6+4\mathfrak{q}^7
+5\mathfrak{q}^8+6\mathfrak{q}^9+5\mathfrak{q}^{10}+4\mathfrak{q}^{11}+5\mathfrak{q}^{12}+2\mathfrak{q}^{13}+\mathfrak{q}^{14}+\mathfrak{q}^{18}
\Bigr)
\nonumber\\
&=1+\mathfrak{q}+3\mathfrak{q}^2+6\mathfrak{q}^3+12\mathfrak{q}^4+21\mathfrak{q}^5+42\mathfrak{q}^6+70\mathfrak{q}^7+125\mathfrak{q}^8+\cdots. 
\end{align}
Under $\mathfrak{q}\rightarrow \mathfrak{q}^{-1}$ they transform as
\begin{align}
\langle \mathcal{W}_{\tiny \yng(2,2)} \mathcal{W}_{\overline{\tiny \yng(2,2)}}\rangle^{U(\infty)}_{\textrm{$\frac12$BPS}} (\mathfrak{q}^{-1})
&=\mathfrak{q}^4 \langle \mathcal{W}_{\tiny \yng(2,2)} \mathcal{W}_{\overline{\tiny \yng(2,2)}}\rangle^{U(\infty)}_{\textrm{$\frac12$BPS}} (\mathfrak{q}), \nonumber\\
\langle \mathcal{W}_{\tiny \yng(3,3,3)} \mathcal{W}_{\overline{\tiny \yng(3,3,3)}}\rangle^{U(\infty)}_{\textrm{$\frac12$BPS}} (\mathfrak{q}^{-1})
&=-\mathfrak{q}^9 
\langle \mathcal{W}_{\tiny \yng(3,3,3)} \mathcal{W}_{\overline{\tiny \yng(3,3,3)}}\rangle^{U(\infty)}_{\textrm{$\frac12$BPS}} (\mathfrak{q}). 
\end{align}

Again, the exact expressions of the normalized correlators are computationally expensive for higher $m$. 
For general $m$, the large $N$ 2-point function can be written as
\begin{align}
\langle \mathcal{W}_{(m^m)} \mathcal{W}_{\overline{(m^m)}}\rangle^{U(\infty)}_{\textrm{$\frac12$BPS}} (\mathfrak{q})
&=H_{\{(m^m), (m^m)\}}(\mathfrak{q})
\prod_{n=1}^{m}
\frac{1}{(1-\mathfrak{q}^n)^n (1-\mathfrak{q}^{m+n})^{m-n}},
\end{align}
where $H_{\{(m^m), (m^m)\}}(\mathfrak{q})$ is a polynomial in $\mathfrak{q}$ with positive coefficients. 
It follows that 
\begin{align}
\langle \mathcal{W}_{(m^m)} \mathcal{W}_{\overline{(m^m)}}\rangle^{U(\infty)}_{\textrm{$\frac12$BPS}} (\mathfrak{q}^{-1})
&=(-1)^m \mathfrak{q}^{m^2}
\langle \mathcal{W}_{(m^m)} \mathcal{W}_{\overline{(m^m)}}\rangle^{U(\infty)}_{\textrm{$\frac12$BPS}} (\mathfrak{q}). 
\end{align}

For $m=4,5,6,7, 8$, we can expand the correlators as
\begin{align}
&
\langle \mathcal{W}_{(4^4)} \mathcal{W}_{\overline{(4^4)}}\rangle^{U(\infty)}_{\textrm{$\frac12$BPS}} (\mathfrak{q})
\nonumber\\
&=1+\mathfrak{q}+3\mathfrak{q}^2+6\mathfrak{q}^3+14\mathfrak{q}^4+25\mathfrak{q}^5+54\mathfrak{q}^6+99\mathfrak{q}^7+198\mathfrak{q}^8+\cdots, \\
&
\langle \mathcal{W}_{(5^5)} \mathcal{W}_{\overline{(5^5)}}\rangle^{U(\infty)}_{\textrm{$\frac12$BPS}} (\mathfrak{q})
\nonumber\\
&=1+\mathfrak{q}+3\mathfrak{q}^2+6\mathfrak{q}^3+14\mathfrak{q}^4+27\mathfrak{q}^5+58\mathfrak{q}^6+111\mathfrak{q}^7+228\mathfrak{q}^8+\cdots, \\
&
\langle \mathcal{W}_{(6^6)} \mathcal{W}_{\overline{(6^6)}}\rangle^{U(\infty)}_{\textrm{$\frac12$BPS}} (\mathfrak{q})
\nonumber\\
&=1+\mathfrak{q}+3\mathfrak{q}^2+6\mathfrak{q}^3+14\mathfrak{q}^4+27\mathfrak{q}^5+60\mathfrak{q}^6+115\mathfrak{q}^7+240\mathfrak{q}^8+\cdots, \\
&
\langle \mathcal{W}_{(7^7)} \mathcal{W}_{\overline{(7^7)}}\rangle^{U(\infty)}_{\textrm{$\frac12$BPS}} (\mathfrak{q})
\nonumber\\
&=1+\mathfrak{q}+3\mathfrak{q}^2+6\mathfrak{q}^3+14\mathfrak{q}^4+27\mathfrak{q}^5+60\mathfrak{q}^6+117\mathfrak{q}^7+244\mathfrak{q}^8+\cdots, \\
&
\langle \mathcal{W}_{(8^8)} \mathcal{W}_{\overline{(8^8)}}\rangle^{U(\infty)}_{\textrm{$\frac12$BPS}} (\mathfrak{q})
\nonumber\\
&=1+\mathfrak{q}+3\mathfrak{q}^2+6\mathfrak{q}^3+14\mathfrak{q}^4+27\mathfrak{q}^5+60\mathfrak{q}^6+117\mathfrak{q}^7+246\mathfrak{q}^8+\cdots. 
\end{align}

In the large $m$ limit, we find 
\begin{align}
\label{large_hm^m&m^m}
&
\lim_{m\rightarrow \infty}
\langle \mathcal{W}_{(m^m)} \mathcal{W}_{\overline{(m^m)}}\rangle^{U(\infty)}_{\textrm{$\frac12$BPS}} (\mathfrak{q})
\nonumber\\
&=\sum_{n=0}^{\infty}a_{\{ (\infty^{\infty}), (\infty^{\infty}) \}}^{(H)}(n)\mathfrak{q}^n
\nonumber\\
&=1+\mathfrak{q}+3\mathfrak{q}^2+6\mathfrak{q}^3+14\mathfrak{q}^4+27\mathfrak{q}^5+60\mathfrak{q}^6+117\mathfrak{q}^7
+246\mathfrak{q}^8+\cdots.
\end{align}
The finite $m$ correction starts from the term with $\mathfrak{q}^{m+1}$. 
It follows that 
\begin{align}
&
\langle \mathcal{W}_{((m+1)^{m+1})} \mathcal{W}_{\overline{((m+1)^{m+1})}}\rangle^{U(\infty)}_{\textrm{$\frac12$BPS}} (\mathfrak{q})
\nonumber\\
&=
\langle \mathcal{W}_{(m^m)} \mathcal{W}_{\overline{(m^m)}}\rangle^{U(\infty)}_{\textrm{$\frac12$BPS}} (\mathfrak{q})
+2\mathfrak{q}^{m+1}+4\mathfrak{q}^{m+2}
+\cdots. 
\end{align}
The $q$-series expansion (\ref{large_hm^m&m^m}) of the half-BPS limit of the large $m$ correlation function of the Wilson line operators 
associated with the $m$-square $(m^m)$ agrees with 
\begin{align}
\label{GLn2_conj}
\prod_{n=1}^{\infty}\frac{1-\mathfrak{q}^n}{1-2\mathfrak{q}^n}. 
\end{align}
This is similar to the function (\ref{GLn3_conj}). 
It is known as the generating function for the number $c(n,2)$ of conjugacy classes of $GL(n,2)$, 
that is the general linear group of degree $n$ over a finite field with $2$ elements \cite{MR109810,MR615131}. 
So we have 
\begin{align}
a_{\{ (\infty^{\infty}), (\infty^{\infty}) \}}^{(H)}(n)&=c(n,2). 
\end{align}

Again one can express the function (\ref{GLn2_conj}) as 
\begin{align}
\sum_{m=0}^{\infty}2^m \mathfrak{q}^m
\prod_{l=m+1}^{\infty}(1-\mathfrak{q}^{l}). 
\end{align}
The asymptotic degeneracy behaves as \cite{MR937520}
\begin{align}
\label{asymh_rect}
a_{\{ (\infty^{\infty}), (\infty^{\infty}) \}}^{(H)}(n)
&\sim \exp\Bigl
[(\log 2)n
\Bigr]. 
\end{align}
The actual coefficients and the analytic values are 
\begin{align}
\begin{array}{c|c|c}
n&a_{\{ (\infty^{\infty}), (\infty^{\infty}) \}}^{(H)}(n)&a_{\{ (\infty^{\infty}), (\infty^{\infty}) \} \textrm{asym.}}^{(H)}(n)\\ \hline 
10&1002&1024 \\
100&1.26765\times 10^{30}&1.26765\times 10^{30} \\
1000&1.07151\times 10^{301}&1.07151\times 10^{301} \\
10000&1.99506\times 10^{3010}&1.99506\times 10^{3010} \\
\end{array}. 
\end{align}
Again we observe that the asymptotic degeneracy (\ref{asymh_rect}) is much larger than that for the $p$-branes 
\cite{Fubini:1972mf, Dethlefsen:1974dr,Strumia:1975rd,Alvarez:1991qs,Harms:1992jt}. 

%%%%%%%%%%%%%%%%%%%%%%%%%%%%%%%%%%%%
\subsubsection{$q$-MZVs}
%%%%%%%%%%%%%%%%%%%%%%%%%%%%%%%%%%%%
We have found that in the large representation limit $m\rightarrow\infty$, 
the unflavored limit and the half-BPS limit of the large $N$ normalized 2-point function 
of the Wilson line operators in the representation labeled by the $m$-square Young diagram becomes 
the generating functions (\ref{GLn3_conj}) and (\ref{GLn2_conj}) for the conjugacy classes of $GL(n,3)$ and $GL(n,2)$ respectively. 
It may be worth pointing out that they are expressible as \cite{brindle2021dualities} 
\begin{align}
\prod_{n=1}^{\infty}
\frac{1-q^n}{1-cq^n}
&=\sum_{k\ge 0}(c-1)^k \zeta_q(\{1\}^k), 
\end{align}
where 
\begin{align}
\zeta_q(k_1,\cdots, k_r)&:=
\sum_{m_1>\cdots>m_k>0}
\frac{q^{m_1}}{(1-q^{m_1})^{k_1} (1-q^{m_2})^{k_2}\cdots (1-q^{m_r})^{k_r}}
\end{align}
is the $q$-analogues of multiple zeta values ($q$-MZVs) \cite{MR2843304}. 
There are several versions of $q$-MZVs in the mathematical literature \cite{schlesinger2001some,MR2069738,MR2111222,MR1992130,MR2341851,MR2322731,
MR2843304,MR3141529,okounkov2014hilbert,MR3338962,MR3473421,MR3522085,milas2022generalized} where their algebraic structures have been examined. 
They are closely related to the multiple Kronecker theta series \cite{Hatsuda:2023iwi} which plays a role of the building blocks of the Schur line defect correlators. 
We leave it future work to investigate further connection between the Schur line defect correlators and the $q$-MZVs. 

%%%%%%%%%%%%%%%%%%%%%%%%%%%%%%%%%%%%
%%%%%%%%%%%%%%%%%%%%%%%%%%%%%%%%%%%%
\section{Conjectural properties}
\label{sec_conj}
%%%%%%%%%%%%%%%%%%%%%%%%%%%%%%%%%%%%
%%%%%%%%%%%%%%%%%%%%%%%%%%%%%%%%%%%%
We have seen that the large $N$ correlation functions satisfy certain transformation laws 
and that they can be expanded in specific forms by computing various examples. 
In this section, we summarize several conjectural properties for the large $N$ Schur line defect correlation functions. 

%%%%%%%%%%%%%%%%%%%%%%%%%%%%%%%%%%%%
\subsection{Conjugation}
%%%%%%%%%%%%%%%%%%%%%%%%%%%%%%%%%%%%
Consider the large $N$ correlation function of the Wilson line operators 
in the representations labeled by a set of the Young diagrams $\{\lambda_i\}_{i=1}^{k}$ 
associated with the positive powers of gauge fugacities 
and those labeled by the other set of diagrams $\{\mu\}_{j=1}^{l}$ for the negative powers of gauge fugacities. 
The correlator is non-trivial when the total sums of the boxes are equal, $\sum_{i=1}^{k}|\lambda_i|$ $=$ $\sum_{j=1}^{l}|\mu_j|$.  
Then we find that 
\begin{align}
\label{CONJ_1}
\langle \mathcal{W}_{\lambda_1}\cdots \mathcal{W}_{\lambda_k} 
\mathcal{W}_{\overline{\mu_1}} \cdots \mathcal{W}_{\overline{\mu_l}} \rangle^{U(\infty)}(t;q)
&=
\langle \mathcal{W}_{\lambda_1'}\cdots \mathcal{W}_{\lambda_k'} 
\mathcal{W}_{\overline{\mu_1'}} \cdots \mathcal{W}_{\overline{\mu_l'}} \rangle^{U(\infty)}(t;q), 
\end{align}
where $\lambda_i'$ and $\mu_j'$ are conjugate to $\lambda_i$ and $\mu_j$ respectively. 

For example, when $k=l=1$ and $\lambda_1=\mu_1=(1^m)$, 
the relation (\ref{CONJ_1}) results in the observation \cite{Hatsuda:2023iwi} 
that the large $N$ 2-point function of the Wilson line operators in the rank-$m$ antisymmetric representation 
is equal to that of the Wilson line operators in the rank-$m$ symmetric representation 
\footnote{For the unflavored 2-point function, it was observed in \cite{Drukker:2015spa}. }
\begin{align}
\label{CONJ_2}
\langle \mathcal{W}_{(1^m)} \mathcal{W}_{\overline{(1^m)}}\rangle^{U(\infty)}(t;q)
&=\langle \mathcal{W}_{(m)} \mathcal{W}_{\overline{(m)}}\rangle^{U(\infty)}(t;q). 
\end{align}

%%%%%%%%%%%%%%%%%%%%%%%%%%%%%%%%%%%%
\subsection{Automorphy}
%%%%%%%%%%%%%%%%%%%%%%%%%%%%%%%%%%%%
While the large $N$ correlation functions of the Wilson line operators are invariant 
when one replaces all the partitions labeling the representations of the Wilson line operators with their conjugate, 
they also have a nice transformation law under the replacement of one of the set of partitions with its conjugate. 
We find that 
\begin{align}
\label{AUT_1}
&
\langle \mathcal{W}_{\lambda_1}\cdots \mathcal{W}_{\lambda_k} 
\mathcal{W}_{\overline{\mu}_1}\cdots \mathcal{W}_{\overline{\mu}_l} \rangle^{U(\infty)}(t;q^{-1})
\nonumber\\
&=(-1)^{\sum_{i=1}^{k}|\lambda_i|} 
\langle \mathcal{W}_{\lambda_1'}\cdots \mathcal{W}_{\lambda_k'} 
\mathcal{W}_{\overline{\mu}_1}\cdots \mathcal{W}_{\overline{\mu}_l} \rangle^{U(\infty)}(t;q). 
\end{align}
In other words, the replacement of one of the sets of partitions, e.g. $\{\lambda_i\}_{i=1}^{k}$ with $\{\lambda_i'\}_{i=1}^{k}$, 
relates the transformation $q\rightarrow q^{-1}$ up to the phase factor $(-1)^{\sum_{i=1}^{k}|\lambda_i|} $. 

For example, it follows that
\begin{align}
&
\langle \mathcal{W}_{\tiny \yng(2)} \mathcal{W}_{\tiny \yng(3)} \mathcal{W}_{\overline{\tiny \yng(5)}}\rangle^{U(\infty)}(t;q^{-1})
\nonumber\\
&=-\langle \mathcal{W}_{\tiny \yng(1,1)} \mathcal{W}_{\tiny \yng(1,1,1)} \mathcal{W}_{\overline{\tiny \yng(5)}}\rangle^{U(\infty)}(t;q)
\nonumber\\
&=-\langle \mathcal{W}_{\tiny \yng(2)} \mathcal{W}_{\tiny \yng(3)} \mathcal{W}_{\overline{\tiny \yng(1,1,1,1,1)}}\rangle^{U(\infty)}(t;q). 
\end{align}

For the special case where one of the sets of Young diagrams is self-conjugate, e.g. $\{\lambda_i'\}=\{\lambda_i\}$, 
the large $N$ correlator satisfies the automorphy
\begin{align}
\label{AUT2}
&
\langle \mathcal{W}_{\lambda_1} \cdots \mathcal{W}_{\lambda_k} 
\mathcal{W}_{\overline{\mu_1}} \cdots \mathcal{W}_{\overline{\mu_l}}
\rangle^{U(\infty)}(t;q^{-1})
\nonumber\\
&=(-1)^{\sum_{i=1}^{k}|\lambda_i|} 
\langle \mathcal{W}_{\lambda_1} \cdots \mathcal{W}_{\lambda_k} 
\mathcal{W}_{\overline{\mu_1}} \cdots \mathcal{W}_{\overline{\mu_l}}
\rangle^{U(\infty)}(t;q). 
\end{align}

For the half-BPS limit of the large $N$ normalized correlation functions we have
\begin{align}
\label{AUT3}
&
\langle \mathcal{W}_{\lambda_1} \cdots \mathcal{W}_{\lambda_k}
\mathcal{W}_{\overline{\mu_1}} \cdots \mathcal{W}_{\overline{\mu_l}}
\rangle^{U(\infty)}_{\textrm{$\frac12$BPS}}(\mathfrak{q}^{-1})
\nonumber\\
&=(-\mathfrak{q})^{\sum_{i=1}^{k}|\lambda_i|} 
\langle \mathcal{W}_{\lambda_1'} \cdots \mathcal{W}_{\lambda_k'}
\mathcal{W}_{\overline{\mu_1}} \cdots \mathcal{W}_{\overline{\mu_l}}
\rangle^{U(\infty)}_{\textrm{$\frac12$BPS}}(\mathfrak{q}). 
\end{align}
When $\{\lambda_i'\}=\{\lambda_i\}$ or $\{\mu_j'\}=\{\mu_j\}$, 
the half-BPS limit of the large $N$ normalized 2-point function satisfies the automorphy
\begin{align}
\label{AUT4}
&
\langle \mathcal{W}_{\lambda_1} \cdots \mathcal{W}_{\lambda_k}
\mathcal{W}_{\overline{\mu_1}} \cdots \mathcal{W}_{\overline{\mu_l}}
\rangle^{U(\infty)}_{\textrm{$\frac12$BPS}}(\mathfrak{q}^{-1})
\nonumber\\
&=(-\mathfrak{q})^{\sum_{i=1}^{k}|\lambda_i|} 
\langle \mathcal{W}_{\lambda_1} \cdots \mathcal{W}_{\lambda_k}
\mathcal{W}_{\overline{\mu_1}} \cdots \mathcal{W}_{\overline{\mu_l}}
\rangle^{U(\infty)}_{\textrm{$\frac12$BPS}}(\mathfrak{q}). 
\end{align}

%%%%%%%%%%%%%%%%%%%%%%%%%%%%%%%%%%%%
\subsection{Hook-length exapansion}
%%%%%%%%%%%%%%%%%%%%%%%%%%%%%%%%%%%%
Suppose that $\lambda$ and $\mu$ are the Young diagrams 
which describe the representations of the Wilson line operators have a finite number of boxes. 
Let $h(b)$ be the hook-length of a box $b$ in the Young diagram. 
It follows that 
the large $N$ normalized unflavored 2-point functions of the Wilson line operators $W_{\lambda}$ and $W_{\mu}$ can be factorized as
\begin{align}
\langle \mathcal{W}_{\lambda} \mathcal{W}_{\overline{\mu}}\rangle^{U(\infty)}(q)
&=
G_{\{\lambda, \mu\}}(q)
\prod_{b\in \lambda}
\frac{1}{1-q^{\frac{h(b)}{2}}}, 
\end{align}
where $G_{\{(\lambda), (\mu)\}}(q)$ is some polynomial whose coefficients are positive integers. 

Similarly, we find that the half-BPS limit of the large $N$ normalized 2-point functions of the Wilson line operators can be written as
\begin{align}
\langle \mathcal{W}_{\lambda} \mathcal{W}_{\overline{\mu}}\rangle^{U(\infty)}_{\textrm{$\frac12$BPS}}(\mathfrak{q})
&=
H_{\{\lambda, \mu\}}(\mathfrak{q})
\prod_{b\in \lambda}
\frac{1}{1-\mathfrak{q}^{h(b)}}, 
\end{align}
where $H_{\{(\lambda), (\mu)\}}(\mathfrak{q})$ is some polynomial whose coefficients are positive integers. 

Note that when either $\lambda$ or $\mu$ is self-conjugate, 
the large $N$ normalized 2-point functions have the automorphy in that they obey the relations (\ref{AUT2}) or (\ref{AUT4}). 
In that case, the polynomials $G_{\{(\lambda), (\mu)\}}(q)$ and $H_{\{(\lambda), (\mu)\}}(\mathfrak{q})$ are palindromic. 

%%%%%%%%%%%%%%%%%%%%%%%%%%%%%%%%%%%
\subsection*{Acknowledgements}
The authors would like to thank Kimyeong Lee, Hai Lin and Masatoshi Noumi for useful discussions and comments. 
The work of Y.H. was supported in part by JSPS KAKENHI Grant No. 22K03641. 
The work of T.O. was supported by the Startup Funding no. 4007012317 of the Southeast University. 
%%%%%%%%%%%%%%%%%%%%%%%%%%%%%%%%%%%

\appendix

\bibliographystyle{utphys}
\bibliography{ref}

\end{document}